\documentclass[11pt,a4paper]{article}
\usepackage[utf8]{inputenc}
\usepackage{jheppub}
\usepackage{mathrsfs}
\usepackage{comment}
\usepackage{graphicx}
\allowdisplaybreaks
\def\di{\mathrm{d}}

\def\Res_#1{\operatorname*{Res}_{#1}}

\def\braket#1{\langle #1 \rangle}

\newcommand{\<}{\langle}
\renewcommand{\>}{\rangle}
\def\be{\begin{equation}\begin{aligned}}
\def\fe{\end{aligned}\end{equation}}

\def\eqn#1{eq.~\eqref{#1}}

\def\fig#1{figure~{\ref{#1}}}
\def\figs#1#2{figures~{\ref{#1}} and~{\ref{#2}}}

\def\sec#1{section~{\ref{#1}}}

\def\Sec#1{Section~{\ref{#1}}}

\def\app#1{appendix~{\ref{#1}}}

\title{Inverse KLT kernel of AdS string integrals from the intersection matrix}
\author[a,b]{Yi-Xiao Tao}
\author[c]{and Shijie Zhang}
\affiliation[a]{Department of Mathematical Sciences, Tsinghua University, Beijing 100084, China}
\affiliation[b]{Institute for Theoretical and Mathematical Physics, Lomonosov Moscow State University, Moscow 119991, Russia}
\affiliation[c]{Center for Fundamental Physics, School of Physical Science and Technology, \\
ShanghaiTech University, 393 Middle Huaxia Road, Shanghai 201210, China}
\emailAdd{tyixiao@itmp.msu.ru}
\emailAdd{zhangshj2025@shanghaitech.edu.cn}

\abstract{We develop a topological and combinatorial
interpretation of the Kawai--Lewellen--Tye (KLT)
relations for $n$-point AdS string-amplitude
building blocks, which are disk integrals with multivariable multiple polylogarithm insertions and sphere integrals with
their single-valued counterparts.
Building on the framework of twisted de Rham theory over a noncommutative ring, we reformulate the problem in terms of associahedral chambers of the moduli space loaded with solutions of the
Knizhnik--Zamolodchikov equation.
We construct regularized twisted cycles and express
their intersection pairings through local monodromies and Drinfeld associators, which control
transport between fusion-tree normalizations.
Organizing the intersection contributions by common
boundary faces yields 
the intersection matrix, which is identified with the inverse KLT kernel by the twisted period relation. The field-theory limit of the inverse KLT kernel is also given.}

\begin{document}
\maketitle


\section{Introduction}
\label{intro}
In perturbative string theory, one of the most important relations between string amplitudes is the Kawai-Lewellen-Tye (KLT) relation \cite{Kawai:1985xq}. The KLT relation connects open-string amplitudes and closed-string amplitudes at tree level, and also has a one-loop generalization \cite{Stieberger:2022lss,Stieberger:2023nol}. At tree level, it reads
\begin{equation}
A_{\rm closed}(1,2,\cdots,n)=\sum_{\alpha,\beta\in S_{n-3}}A_{\rm open}(1,\alpha,n,n-1)S[\alpha|\beta]\tilde{A}_{\rm open}(1,\beta,n-1,n)
\end{equation}
where $S[\alpha|\beta]$ denotes the KLT kernel,
and $S_{n-3}$ is the permutation group acting on
$\{2,3,\ldots,n-2\}$. This relation not only has deep physical meaning, but also gives some mathematical insight. It was shown in \cite{Mizera:2017cqs} that KLT relations at tree level can be understood using twisted de Rham theory, where the inverse of the KLT kernel can be obtained from intersection numbers. Beyond genus zero, twisted de rham theory has been applied to
open-string monodromy relations at arbitrary genus
\cite{Casali:2019ihm} and to one-loop string double-copy
relations \cite{Mazloumi:2024wys}.
Related double-copy relations have also been established
for generalized hypergeometric integrals at genus one
and higher genus \cite{Bhardwaj:2023vvm,Pokraka:2025zlh}. Given the fundamental role of KLT relations in flat-space string amplitudes, it is natural to investigate whether similar structures persist in curved spacetime. This question provides one of the main motivations for the present work.

However, calculating string amplitudes in curved spacetime presents significant challenges. Even in Anti-de Sitter (AdS) spacetime, which is perhaps the most familiar curved spacetime, developing systematic approaches to computing string amplitudes remains an active area of research. In recent years, a new method based on holography, called "worldsheet bootstrap" \cite{Alday:2022uxp,Alday:2022xwz,Alday:2023jdk,Alday:2023mvu,Alday:2024yax,Alday:2024ksp,Alday:2024rjs,Alday:2024xpq,Jiang:2025oar,Chester:2024wnb,Chester:2024esn,Alday:2026frs,Wang:2025pjo,Wang:2025owf,Ren:2026zxs}, gives us a new perspective on computing string amplitudes with AdS curvature corrections. It allows us to bootstrap the AdS string amplitudes based on holography techniques. A key physical motivation for this approach comes from
expanding the AdS metric around flat space. This suggests
interpreting curvature corrections in terms of additional
soft-graviton insertions in flat-space worldsheet correlators
\cite{Alday:2023jdk,Alday:2024ksp}, which gives the building blocks of AdS string amplitudes. The curvature-corrected open-string integrals at four points are defined as follows \cite{Alday:2024yax}:
\[P_w(s,t)=\int_0^1 z^{s-1}(1-z)^{t-1}G_w(z)\,dz,\]
where \(w\) labels the word associated with the corresponding multiple polylogarithm (MPL) \(G_w(z)\). The generating series of these integrals defines the basic building blocks \cite{Alday:2025bjp}
\begin{equation}
    P(e_\ell;s,t) =
\int_0^1 z^{s-1}(1-z)^{t-1}
G(e_\ell;z)dz .
\end{equation}
where $G(e_\ell;z)$, with $\ell=0,1$, denotes the
generating series of MPLs. The closed-string counterpart is given by a sphere integral with an insertion of  single-valued
multiple polylogarithms (SVMPLs)
\begin{equation}
I(e_\ell;s,t)=
\int_{\mathbb C}
|z|^{2s-2}|1-z|^{2t-2}
\operatorname{sv}(G)(e_\ell;z,\bar{z})\,
dz\,d\bar z .
\end{equation}
where $\operatorname{sv}(G)(e_\ell;z,\bar z)$
denotes the single-valued counterpart of the MPL
generating series \cite{Brown:2004ugm}.

In view of the double-copy structure of flat-space worldsheet integrals, the close analogy between the Koba--Nielsen factor and the MPL generating series naturally motivates the search for a corresponding double-copy structure of the AdS string amplitude building blocks. Indeed, at four points, the AdS building blocks
satisfy a KLT-type relation\cite{Alday:2025bjp,Kakkad:2025klm}:
\begin{equation}
\label{KLT}
    I(e_{\ell};s,t) = P(e_{\ell};s,t)  K(e_{\ell};s,t) P^\vee(e'_{\ell};s,t), 
\end{equation}
where the KLT kernel is given by
\begin{equation}
    K(e_{\ell};s,t) =-\left( 1 +\frac{1}{e^{2\pi i s}M_0-1}+\frac{1}{e^{2\pi i t}M_1-1}\right)^{-1}
\end{equation}
with the monodromy of \(G(e_{\ell};z)\) at \(z=0\) and \(z=1\):
\begin{equation}
    M_0 = e^{2\pi i e_0},
    \qquad
    M_1 = \Phi(e_1,e_0)\,
          e^{2\pi i e_1}\,
          \Phi(e_0,e_1).
\end{equation}
The general $n$-point KLT matrix is also given in \cite{Nocchi:2026yms}, which will be reviewed later.

To generalize this construction to higher multiplicities, it is useful to recall the
disk integrals $Z$ and sphere integrals $J$ underlying flat-space string
amplitudes \cite{Mafra:2011nw,Schlotterer:2018zce}.\footnote{In this notation, the
four-point open- and closed-string generating integrals denoted by $P$
and $I$ above belong to the $Z$- and $J$-type families, respectively.} These integrals encode the full string-theoretic
dependence of the amplitudes, including all
$\alpha'$ corrections. 
They are defined by
\begin{align}
Z(\tau|\rho)
&=\int_{\gamma(\tau)}d\mu_{\mathbb R}\,
\mathrm{KN}_{\tau}(z)\,\mathrm{PT}_{\rho}(z),
\label{eq:intro-flat-Z}\\
J(\tau|\rho)
&=
\int_{\mathcal M_{0,n}(\mathbb C)}d\mu_{\mathbb C}\,
\prod_{i<j}|z_{ij}|^{2s_{ij}}\,
\mathrm{PT}_{\rho}(z)\,\mathrm{PT}_{\tau}(\bar z),
\label{eq:intro-flat-J}
\end{align}
where $z_{ij}=z_i-z_j$, the dimensionless $s_{ij}$ include the appropriate power of
inverse string tension $\alpha'$, and
\[
d\mu_{\mathbb R}
=\frac{\prod_{i=1}^{n}dz_i}{\mathrm{vol}\,\mathrm{SL}(2,\mathbb R)},
\qquad
d\mu_{\mathbb C}
=\frac{\prod_{i=1}^{n}d^2z_i}{\mathrm{vol}\,\mathrm{SL}(2,\mathbb C)},
\qquad
\mathrm{PT}_{\rho}(z)
=\frac{1}{z_{\rho(1)\rho(2)}\cdots z_{\rho(n)\rho(1)}}.
\]
The quotient measures are implemented by fixing three punctures and
including the corresponding Jacobians. In the disk integral $Z(\tau|\rho)$, the ordering $\tau$ specifies
a real chamber $\gamma(\tau)$, on which the chosen branch
of the Koba--Nielsen factor $\mathrm{KN}_{\tau}$
takes the form $\prod_{i<j}|z_{ij}|^{s_{ij}}$
\cite{Koba:1969rw}, while $\rho$ determines the
Parke--Taylor (PT) factor \cite{Parke:1986gb}.
In contrast, both orderings in the sphere integral $J(\tau|\rho)$
label Parke--Taylor factors.

Contour deformation gives a KLT relation directly between these
integrals \cite{Kawai:1985xq,Schlotterer:2018zce}:
\begin{equation}
J(\tau|\rho)
=\sum_{\alpha,\beta\in S_{n-3}}
Z(\alpha,1,n-1,n|\rho)\,
S_{\mathrm{flat}}[\alpha|\beta]\,
Z(1,\beta,n-1,n|\tau),
\label{eq:intro-flat-KLT}
\end{equation}
The string momentum kernel $S_{\mathrm{flat}}$ in these bases contains
the sine factors arising from Koba--Nielsen monodromy and is independent
of the Parke--Taylor orderings $\tau,\rho$
\cite{BjerrumBohr:2010hn}.
{The reduction to an $(n-3)!$-element basis follows from
open-string monodromy relations
\cite{BjerrumBohr:2009rd,Stieberger:2009hq,BjerrumBohr:2010zs}.}
Equivalently, at the level of their low-energy expansions, the sphere
integrals obey $J(\tau|\rho)=\operatorname{sv}Z(\tau|\rho)$, where the
single-valued map acts on the multiple zeta values in the disk
integrals \cite{Stieberger:2013wea,Stieberger:2014hba,Schlotterer:2018zce,Brown:2019wna}.
Twisted de Rham theory relates the bilinear factorization
\eqref{eq:intro-flat-KLT} to period pairings and identifies its inverse
kernel with twisted-cycle intersections, with the corresponding
normalization factors \cite{Mizera:2017cqs}.

In general, the $n$-point AdS string amplitude
building blocks \cite{Nocchi:2026yms} can be expressed as $Z$- and
$J$-type integrals with insertions of multivariable
multiple polylogarithms (MMPLs) and their
single-valued counterparts, respectively. In a fixed gauge, inserting the
chamber-normalized solutions of the Knizhnik–Zamolodchikov (KZ) equation $G_\tau$ in
\eqref{eq:intro-flat-Z} defines $Z^{(\mathrm{AdS})}(\tau|\rho)$:
\begin{equation}
    Z^{(\mathrm{AdS})}(\tau|\rho)=\int_{\gamma(\tau)}d\mu_{\mathbb R}\,
\mathrm{KN}_{\tau}(z)\,\mathrm{PT}_{\rho}(z) G_{\tau}(z;e_{ij}).
\end{equation}
and inserting its single-valued counterpart $\operatorname{sv}G_{\mathbb I_n}$ in
\eqref{eq:intro-flat-J} defines $J^{(\mathrm{AdS})}(\tau|\rho)$.
\begin{equation}
    J^{(\mathrm{AdS})}(\tau | \rho) = \int_{\mathcal M_{0,n}(\mathbb C)}d\mu_{\mathbb C}\,
\prod_{i<j}|z_{ij}|^{2s_{ij}}\,
\mathrm{PT}_{\rho}(z)\,\mathrm{PT}_{\tau}(\bar z)\operatorname{sv}G_{\mathbb I_n}(z,\bar{z};e_{ij}),
\label{eq:intro-ads-J} 
\end{equation}
Here $\mathbb I_n=(1,\ldots,n)$ fixes the reference normalization. The $e_{ij}$ are noncommuting generators defining the differential equation, and the corresponding solutions
$G_\tau$ are formal power series in these generators,
whose coefficients are MMPLs.
Setting all $e_{ij}=0$ reduces the KZ solutions to unity,
recovering the flat-space integrals $Z$ and $J$.

Their KLT relation takes the ordered form \cite{Nocchi:2026yms}
\begin{equation}
J^{(\mathrm{AdS})}(\tau|\rho)
=\sum_{\alpha,\beta\in S_{n-3}}
Z^{(\mathrm{AdS})}(\alpha,1,n-1,n|\rho)\,
S_{\mathrm{AdS}}[\alpha|\beta]\,
\widetilde Z^{(\mathrm{AdS})}(1,\beta,n-1,n|\tau).
\label{eq:intro-AdS-KLT}
\end{equation}
The dual integral $\widetilde Z^{(\mathrm{AdS})}$ is matched to the
same single-valued normalization. 
The kernel combines the shifted
residues $E_{ij}=s_{ij}\mathbf1+e_{ij}$ with Drinfeld associators
relating different normalizations of KZ solutions. The definitions and basis conventions are reviewed in \sec{2.5}.
This noncommutative monodromy provides the starting point for a
geometric description of the inverse AdS KLT kernel.

Building on twisted de Rham theory 
over a noncommutative ring framework developed in \cite{Kakkad:2025klm},
we extend the approach of \cite{Mizera:2017cqs}
to derive the KLT relations for AdS string
amplitude building blocks. We also figure out the intersection theory for the inverse KLT kernel in this case. Furthermore, we discuss the field-theory limit of the inverse KLT kernel and define the bi-adjoint scalar amplitude building blocks. Such building blocks establish connections between the string amplitudes and the field amplitudes in the language of the worldsheet bootstrap.

The paper is organized as follows. In section \ref{KZ}, we will first review some basic concepts of KZ equations and multivariable MPLs, which are the main objects of the building blocks, and then introduce the associahedra with effects from MPLs. In section \ref{derham}, we will review the twisted de Rham theory and show how to express the AdS stringy KLT relations in this language. In section \ref{inter}, we will show how the inverse KLT kernel for AdS building blocks can be obtained from the intersection numbers. Finally, we will show the field-theory limit of the inverse KLT kernel. Until we discuss the field-theory limit,
we work in units where $\alpha'=1$.

\section{KZ equations, associahedra, and KLT relations}
\label{KZ}
\subsection{KZ equations and fusion trees}
\label{sec:2.1}
The natural generalization of the four-point AdS string building block \cite{Alday:2025cxr,Alday:2025bjp,Kakkad:2025klm,Nocchi:2026yms} to higher multiplicities is obtained by replacing the single-variable MPL insertion with its multivariable counterpart. MMPLs arise naturally as coefficients of fundamental solutions to the KZ equation \cite{Knizhnik:1984nr} associated with a flat connection:
\begin{equation}
    \di G=\Omega G,
\end{equation}
where \(\Omega\) is the KZ connection
\begin{equation}
\Omega=
\sum_{i<j}e_{ij}\,\di\log(z_i-z_j).
\end{equation}
The generators \(e_{ij}=e_{ji}\) of the
\textit{Drinfeld--Kohno Lie algebra}, also known as the
infinitesimal pure braid Lie algebra
\cite{Kohno:1987monodromy,Drinfeld:1991quasi},
obey the infinitesimal braid relations
\begin{equation}
\label{braid}
\begin{aligned}
    [e_{ij},e_{kl}]=0
\quad\text{for pairwise distinct }i,j,k,l,
\\
[e_{ij},e_{ik}+e_{jk}]=0
\quad\text{for pairwise distinct }i,j,k.
\end{aligned}
\end{equation}
which ensure the flatness of the KZ connection:
\begin{equation}
    \di\Omega - \Omega\wedge\Omega=0.
\end{equation}

For the stringy integrals of interest, we consider the moduli space of \(n\) ordered marked points on the Riemann sphere
\begin{equation}
    {\cal M}_{0,n}=
\operatorname{Conf}_n(\mathbb P^1)/\text{PSL}(2,\mathbb C).
\end{equation}
The open-string integration domains are real chambers of this moduli space \({\cal M}_{0,n}\). In this setting, the Drinfeld--Kohno Lie algebra is replaced by its spherical counterpart. In addition to the infinitesimal braid relations, the generators satisfy the spherical relations
\begin{equation}
\label{sph}
    \sum_{j\neq i} e_{ij}=0,
\qquad i=1,\ldots,n.
\end{equation}
These relations eliminate the redundancy associated with global Möbius transformations,
\[
z_i\mapsto z_i'=\frac{az_i+b}{cz_i+d}.
\]
Indeed, under such a transformation one finds
\[
\di\log(z_i-z_j)
\mapsto
\di\log(z_i-z_j)
-\di\log(cz_i+d)
-\di\log(cz_j+d),
\]
and consequently
\[
\Omega
\mapsto
\Omega-
\sum_i
\left(
\sum_{j\neq i}e_{ij}
\right)
\di\log(cz_i+d).
\]
The additional term therefore vanishes upon imposing the spherical relations \eqn{sph}, so that the KZ connection is invariant under the global \(\text{PSL}(2,\mathbb C)\) action and hence descends naturally to the moduli space \(\mathcal M_{0,n}\).

The solution of a differential equation is fixed by appropriate boundary data, which, for the KZ equations, are formulated as asymptotic normalization conditions near singular loci. A choice of asymptotic normalization defines an \textit{asymptotic zone} and singles out a unique fundamental solution with prescribed behavior near the corresponding singular limit \cite{etingofLecturesRepresentationTheory1998}. Different asymptotic regions are labelled by trees encoding the successive fusion of marked points, which we refer to as \textit{fusion trees}, or equivalently by parenthesizations of the ordered marked points. We distinguish the marked point \(n\) as the root of the tree.
The remaining \(n-1\) labels are the inputs of the binary
tree. In our notation, the root label is appended to the
parenthesization of these inputs \cite{Brown:2009qja}. 

{A fusion tree is a planar rooted binary tree whose leaves are labelled by the ordered marked points. Each internal vertex represents the fusion of two subsets of marked points, so that the tree records the hierarchical order in which the corresponding collision limits are taken. Equivalently, a fusion tree is encoded by a complete parenthesization of the ordered marked points.
In the KZ equation, such a hierarchy specifies an asymptotic regime in which different relative separations approach zero at parametrically different rates. A canonical fundamental solution is then defined by prescribing its leading behavior in this asymptotic zone.}

The simplest example is the single-variable KZ equation for the generating series of multiple polylogarithms, which is used at four-point level \cite{Alday:2025cxr,Alday:2025bjp}
\begin{equation}
    \di G
    =
    \left(
        e_0\,\frac{\di z}{z}
        +
        e_1\,\frac{\di z}{z-1}
    \right)G .
\end{equation}
On the real chamber \(0<z<1\), its two boundary limits \(z\to0\) and \(z\to1\) define two canonical fundamental solutions\footnote{
The iterated-integral description of these solutions uses the
formalism in \cite{Chen:1977oja} and the theory of multiple polylogarithms
\cite{Goncharov:1998kja,Goncharov:2001iea}.}$G_0$ and $G_1$, respectively, with the standard asymptotic
normalizations \cite{Drinfeld:1991quasi,Brown:2009qja}
\[
    G_0(z)\sim z^{e_0},
    \qquad z\to0,
\]
and
\[
    G_1(z)\sim (1-z)^{e_1},
    \qquad z\to1,
\]
respectively. 
Equivalently, the asymptotic normalization is specified by requiring the regular part of the solution at each singular point to be the identity,
\[
    \lim_{z\to0} G_0(z)\,z^{-e_0}=1,
    \qquad
    \lim_{z\to1} G_1(z)\,(1-z)^{-e_1}=1 .
\]
Since \(G_0\) and \(G_1\) solve the same differential equation, their ratio is independent of \(z\), which is the Drinfeld associator \cite{Drinfeld:1991quasi}:
\begin{equation}
    G_{0}(z) = G_{1}(z)\Phi(e_0,e_1).
\end{equation}
More generally, for a fixed ordering of the marked points, we use parenthesizations to label the canonical KZ solutions associated with different fusion trees. For \(n=4\), after fixing the \(\text{PSL}(2,\mathbb C)\) redundancy by
\[
z_1=0,\qquad z_3=1,\qquad z_4=\infty ,
\]the moduli space reduces to
\[
\mathcal M_{0,4}\cong \mathbb P^1\setminus\{0,1,\infty\}.
\]
For the real chamber \(0<z_2<1\), the two asymptotic normalizations correspond to the two possible maximal parenthesizations compatible with the fixed ordering,
\[
    ((12)3)4,
    \qquad
    (1(23))4.
\]
Equivalently, the two asymptotic limits \(z_2\to0\) and \(z_2\to1\) are represented by two planar binary fusion trees, encoding two distinct hierarchical collision patterns. The first, \(z_2\to0\), corresponds to the fusion of \(z_1\) and \(z_2\), characterized by
\[
|z_2-z_1|\ll |z_3-z_2|
\quad\leftrightarrow\quad
((12)3)4,
\]whereas the second, \(z_2\to1\), corresponds to the fusion of \(z_2\) and \(z_3\),
\[
|z_3-z_2|\ll |z_2-z_1|
\quad\leftrightarrow\quad
(1(23))4.
\]
Identifying the generators of the single-variable KZ equation with
\begin{equation}
    e_0\longrightarrow e_{12},
    \qquad
    e_1\longrightarrow e_{23},
\end{equation}
the canonical KZ solutions associated with the two fusion trees are related by the Drinfeld associator,
\begin{equation}
    G_{((12)3)4}
    =
    G_{(1(23))4}\,
    \Phi(e_{12},e_{23}) .
\end{equation}
Thus, the Drinfeld associator can be interpreted as the transition matrix between canonical KZ solutions associated with different fusion trees.

In the following discussions, we will denote the solution associated with the left-comb tree by
\[
G_{1234}:= G_{((12)3)4}.
\]
More generally, for an ordering
\(\alpha=(i_1,\ldots,i_{n-1},n)\) with fixed root label \(n\),
we write
\begin{equation}
   G_{\alpha}
\equiv G_{i_1\cdots i_{n-1}n}
:=
G_{(((i_1i_2)i_3)\cdots i_{n-1})n}. 
\end{equation}
Thus, an ordering without explicit parentheses denotes the
canonical KZ solution normalized at the corresponding
left-comb tree.

In the four point case, the monodromy around \(z=0\) is associated with the current fusion tree and therefore acts directly:
\[
G_{1234}M_0=G_{1234}e^{2\pi i e_{12}},
\]where \(e_{12}\) is identified with \(e_0\) in the standard four-point gauge fixing.
The monodromy around \(z=1\), however, corresponds to the fusion of \(2\) and \(3\). Therefore, one has to first transport the solution to the alternative fusion tree \(G_{(1(23))4}\) using the Drinfeld associator and then transport it back to the original tree:
\[
G_{1234}M_1
    =
G_{1234}
\Phi(e_{23},e_{12})
e^{2\pi i e_{23}}
\Phi(e_{12},e_{23}).
\]This is the familiar associator-conjugated monodromy representation \cite{Alday:2025bjp,Kakkad:2025klm}
\[
M_1=\Phi(e_{1},e_{0})e^{2\pi i e_1}\Phi(e_{0},e_{1})
\]in the standard single-variable MPL notation.

\subsection{Associahedra and boundaries }
In the previous subsection, we introduced the asymptotic normalization and fusion-tree structure of solutions to the KZ equations, using the single-variable case as an illustrative example. We now turn to the general multivariable setting.
We use Brown's description of associahedral boundary strata by stable
partitions and compatible chord systems
\cite{Brown:2009qja},
together with Bar--Natan's cabling formalism for parenthesized braids
and chord diagrams
\cite{barnatan1998}. {Throughout this paper, the KZ solutions are understood
as formal series in the generators \(e_{ij}\).}

The fusion hierarchies introduced in the previous subsection
have a natural interpretation in the boundary geometry of
\(\mathcal M_{0,n}\). In a naive coordinate compactification,
the collision divisors need not meet with normal crossings.
The Deligne--Mumford compactification
\(\overline{\mathcal M}_{0,n}\) provides a geometric description
of these degenerations with additional components that retain the relative configurations within the colliding clusters \cite{Knudsen1983,Keel1992}.  
Its boundary is a simple normal-crossing divisor
\cite{Keel1992,Brown:2009qja,Mizera:2017cqs}.
Fix a cyclic ordering \(\alpha\), and let \(C_\alpha\) be the
corresponding real chamber of \(\mathcal M_{0,n}(\mathbb R)\).
Its closure
\begin{equation}
    K_{n-1}(\alpha):=\overline{C_n}(\alpha)
\subset\overline{\mathcal M}_{0,n}(\mathbb R)
\end{equation}
is an \((n-3)\)-dimensional associahedron \cite{10.2307/1993608,devadoss2000tessellationsmodulispacesmosaic}.
We describe its facets and their intersections using stable
\(\alpha\)-planar splits, and relate them to the fusion trees introduced above.
The same boundary geometry will underlie the regularization
of twisted cycles discussed in Section~\ref{inter}.

Fix a cyclic ordering \(\alpha\) of \(L_n=\{1,\ldots,n\}\)
and choose \(n\) as the root label. We consider subsets
\[
I\subset L_n\setminus\{n\},
\qquad 2\leq |I|\leq n-2,
\]
whose elements form a cyclically consecutive block in \(\alpha\).
Each such subset represents a stable \(\alpha\)-planar split
\(I\mid I^c\), with the representative uniquely chosen by the
condition \(n\notin I\).
The associahedron \(K_\alpha\) is an \((n-3)\)-dimensional
simple polytope characterized combinatorially as follows.
Its facets \(F_I\) are indexed by these subsets, and 
geometrically, \(K_{n-1}(\alpha)\) is realized as the closure of the
real chamber with cyclic ordering \(\alpha\) in
\(\overline{\mathcal M}_{0,n}(\mathbb R)\), and its facets are
\begin{equation}
    F_I(\alpha)=K_{n-1}(\alpha)\cap D_I,
\qquad D_I=D_{I^c},
\end{equation}
where \(D_I\) denotes the boundary divisor corresponding to
the unordered stable split \(I\mid I^c\). Since exchanging
the two sides gives the same split, \(D_I=D_{I^c}\).

\begin{figure}[t]
\centering
\includegraphics[width=0.8\textwidth]{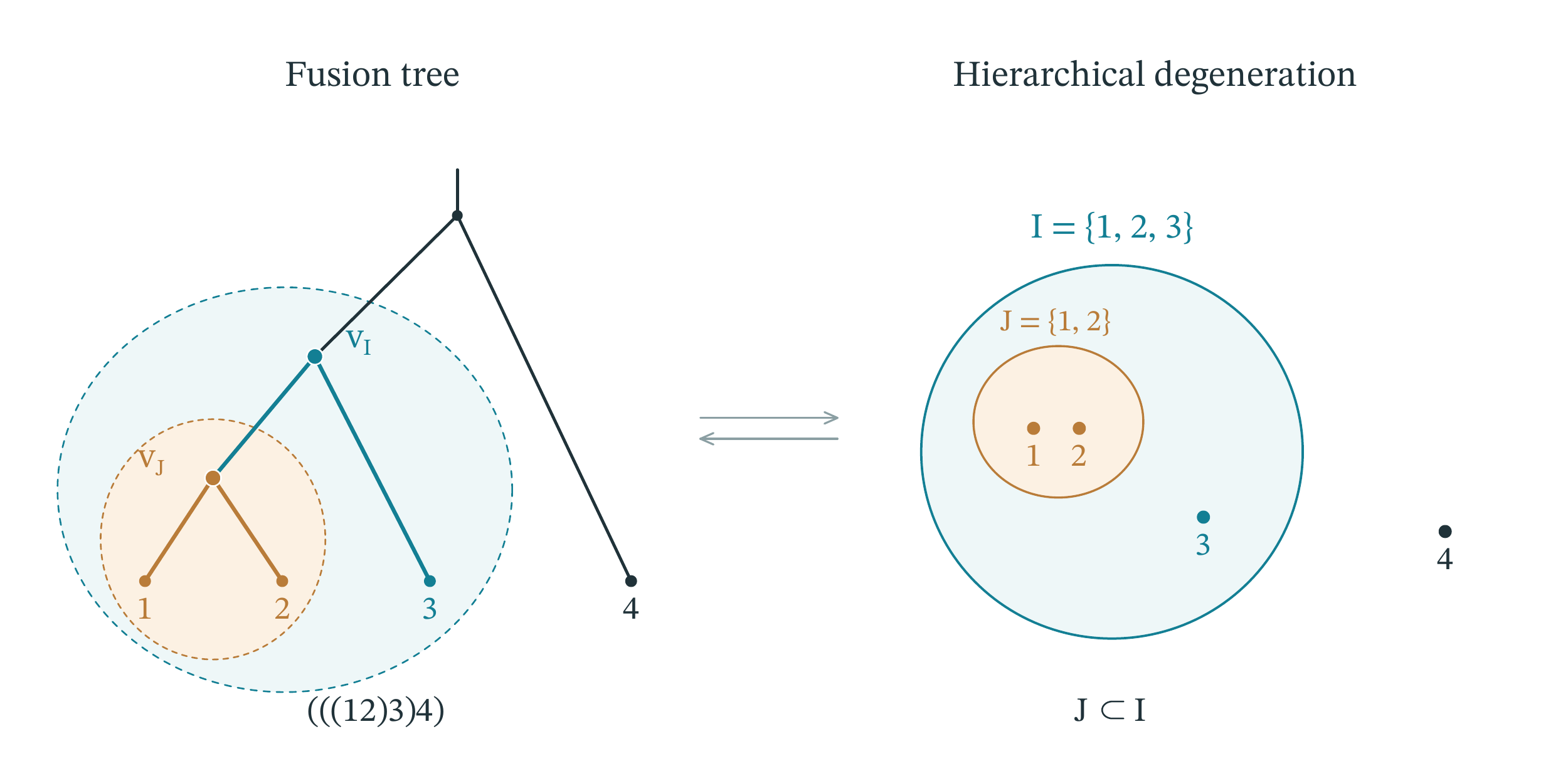}
\vspace{-5pt}
\caption{Fusion trees and clusters}
\label{fig:fusion_tree_clusters}
\end{figure}

A \textit{cluster} is a subset of marked points corresponding to the leaves descending from a common internal vertex of a fusion tree. Equivalently, it represents a collection of marked points that approach one another at the same stage of the associated hierarchical degeneration. See \fig{fig:fusion_tree_clusters} as an example

\begin{figure}[t]
\centering
\includegraphics[width=0.8\textwidth]{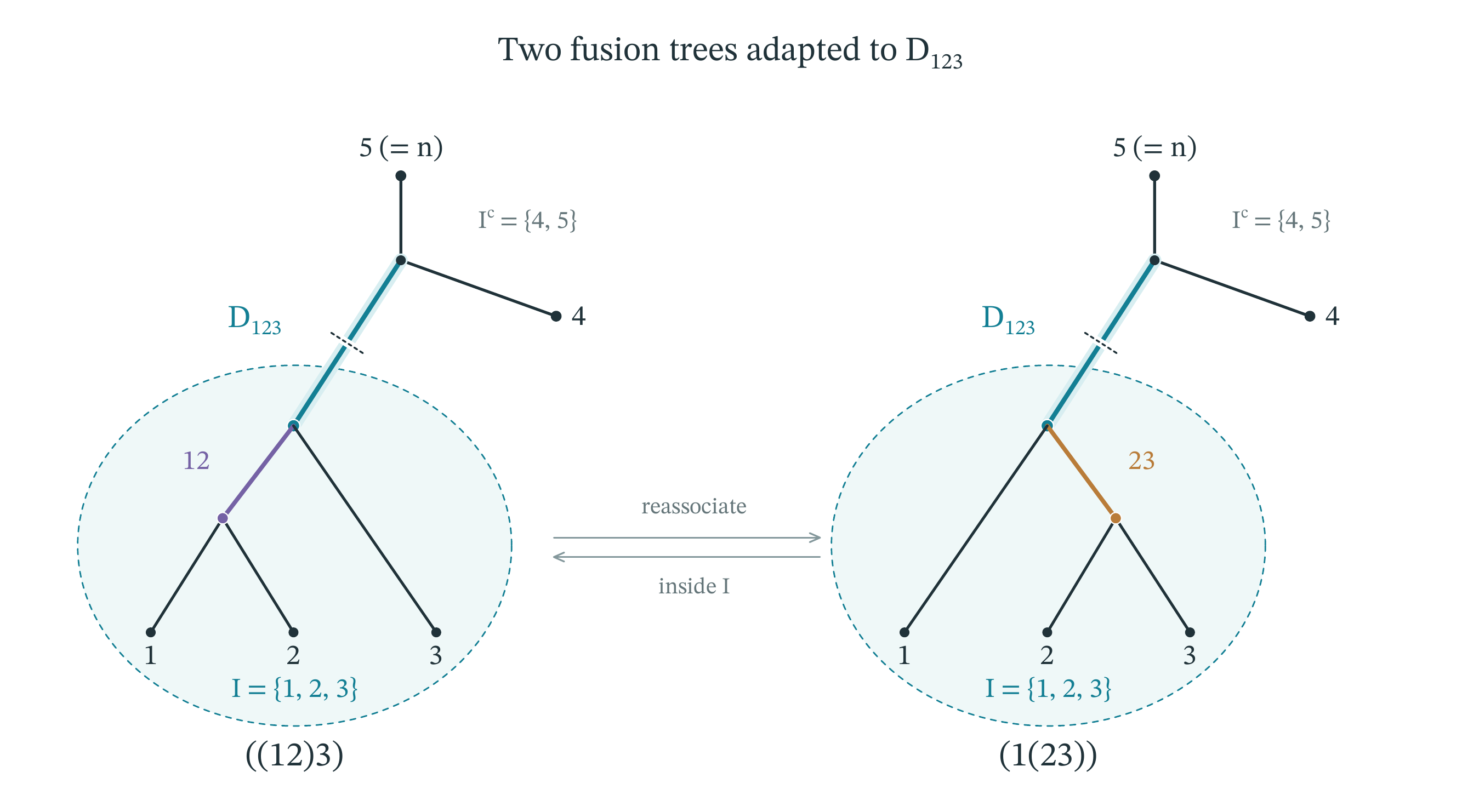}
\vspace{-5pt}
\caption{An example of the stable split and cluster adapted fusion tree}
\label{fig:stable_split_D123}
\end{figure}

{In the fusion-tree description, a stable split \(I\mid I^c\)
is represented by an internal edge: removing this edge separates
the labelled leaves into the two sets \(I\) and \(I^c\).
With the tree rooted at the leaf \(n\), the side not containing
\(n\) forms a subtree whose leaf set is \(I\). We choose \(I\) as the representative of the split \(I \mid I^c \), which is precisely a cluster in the sense defined earlier.  

A fusion tree \textit{adapted} to \(I\) therefore contains \(I\) as a
cluster, while the remaining edges specify further refinements
of the degeneration.
For example, a fusion tree adapted to \(\{123\}\) contains a
subtree with leaves \(1,2,3\). This subtree may be bracketed as
\(((12)3)\) or \(1(23)\), introducing the additional channel
\(12\) or \(23\), respectively. Both choices retain the cluster
\(123\): the divisor \(D_{123}\) specifies the separation of
this cluster from its complement, but does not fix its internal
bracketing as illustrated in \fig{fig:stable_split_D123}. }

For a cluster \(I\), we can assign the generator $e_I$ to each divisor $D_I$. Let us first define the internal generator $e_I$ by
\begin{equation}
e_I
:=
\sum_{\{a,b\}\subset I}e_{ab}.
\label{eq:internal-cluster-generator}
\end{equation}
Geometrically one always has
\[
D_I=D_{I^c}.
\]
Algebraically, however, the equality
\[
e_I=e_{I^c}
\]
does not hold in the ordinary Drinfeld--Kohno Lie algebra. But for a KZ system descending to \(\overline{\mathcal M}_{0,n}\), using the spherical relations introduced above
\[
\sum_{j\neq i}e_{ij}=0,
\]
and summing these equations over \(i\in I\) will give
\[
2e_I+e_{I,I^c}=0,
\]
where 
\[e_{I,I^c}:=\sum_{a\in I,\ b\in I^c}e_{ab}.\]
And if we sum over \(i\in I^c\), it will give
\[
2e_{I^c}+e_{I,I^c}=0,
\]
which leads to
\begin{equation}
e_I=e_{I^c}.
\label{eq:complement-residue}
\end{equation}
Thus, in the spherical quotient, this assignment is valid. And before taking that quotient, one must retain the
chosen rooted representative \(I\).

\paragraph{Faces.} Having introduced the facets of \(K_{n-1}(\alpha)\), we now turn to its higher-codimension faces, focusing on codimension-\(k\) faces with \(2\leq k\leq n-3\).
Let
\[
\mathcal N=\{I_1,\ldots,I_k\}
\]
be a collection of distinct stable \(\alpha\)-planar clusters, namely the set of \(k\) facets of \(K_{n-1}(\alpha)\).
The facets have nonempty intersection if and only if the clusters are pairwise compatible \cite{Keel1992,Brown:2009qja}: for every
\(a\neq b\),
\begin{equation}
\label{eq:compatiblity}
I_a\subset I_b,\qquad
I_b\subset I_a,\qquad\text{or}\qquad
I_a\cap I_b=\varnothing.
\end{equation}
Such a collection is called a \textit{planar nested set}.
It determines the face \(F_{\cal N}\) of \(K_{n-1}\)
\begin{equation}
F_{\mathcal N}(\alpha)
=
\bigcap_{I\in\mathcal N}F_I(\alpha)
=
K_{n-1}(\alpha)\cap\bigcap_{I\in\mathcal N}D_I.
\label{eq:face-from-nested-set}
\end{equation}
Since \(K_{n-1}(\alpha)\) is a  polytope of dimension \(n-3\),
\[
\operatorname{codim}_{K_{n-1}}F_{\mathcal N}=k,
\qquad
\dim F_{\mathcal N}=n-3-k.
\]
The empty nested set corresponds to \(K_{n-1}(\alpha)\) itself.

\begin{figure}[t]
\centering
\includegraphics[width=0.5\textwidth]{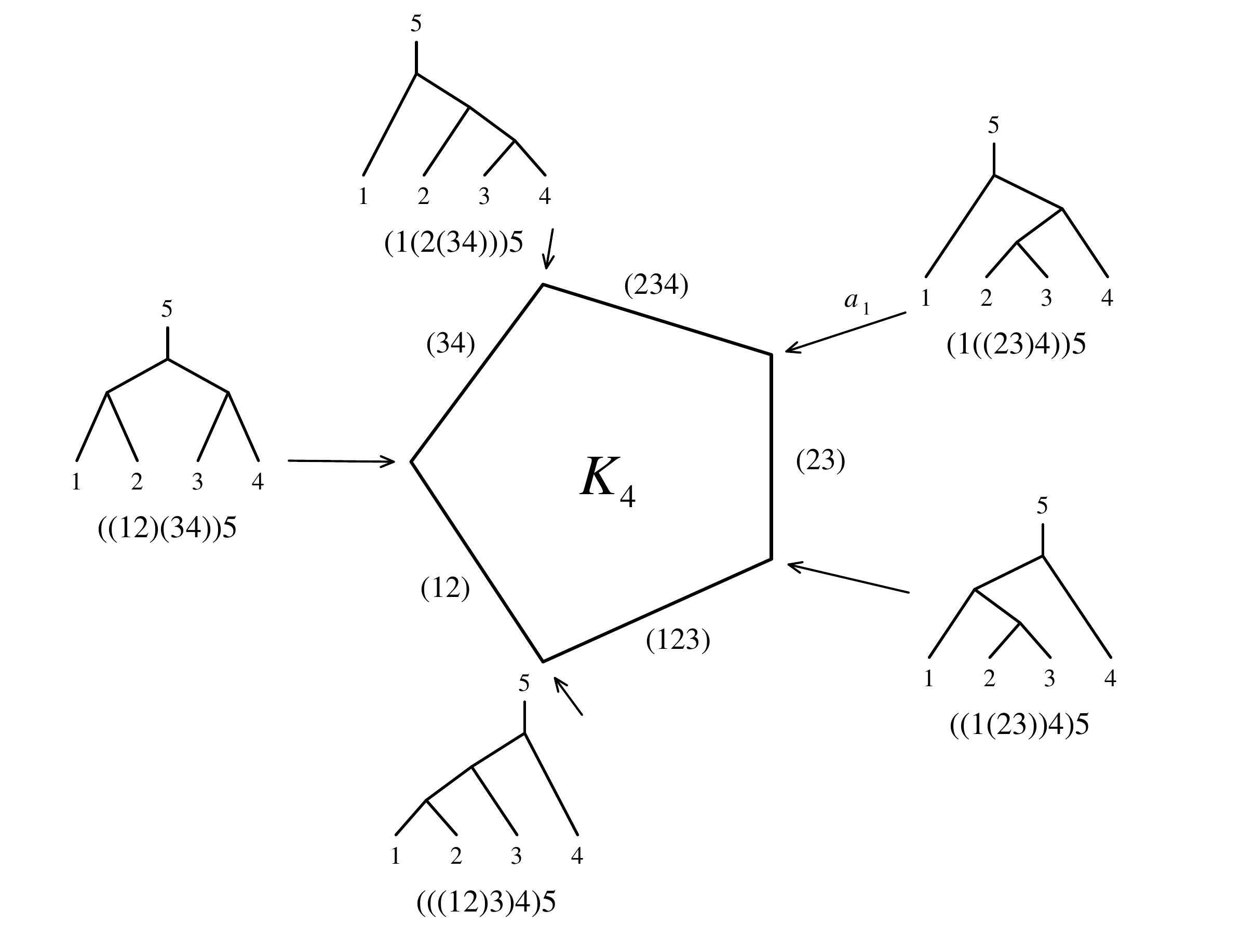}
\vspace{-5pt}
\caption{Fusion trees of \(K_4(\mathbb I_n)\)}
\label{fig:K4}
\end{figure}

The clusters in \(\mathcal N\) specify
\(k\) compatible collision channels. 
In the fusion-tree description, they determine a rooted planar tree whose internal edges encode these channels. Correspondingly, a codimension-\(k\) face is labelled by a partial parenthesization encoding the \(k\) compatible collision channels.  When \(k<n-3\), such a face does not specify a complete asymptotic normalization, but only a partial hierarchy of collision limits. It is therefore represented not by a fully resolved planar binary tree, but by a partially resolved, generally non-binary planar tree. When \(k=n-3\), the face is a vertex and the associated
tree is already binary. Its maximal parenthesization
specifies a complete hierarchy of collisions, or the asymptotic zone determining the corresponding canonical KZ solution. See \fig{fig:K4} as an example for \(n=5\).
A positive-dimensional face, by contrast, specifies only
a partial hierarchy and does not by itself select a unique
vertex-normalized KZ frame.

In the next subsection, we describe the KZ transport between different fusion trees associated with the vertices of \(K_{n-1}(\alpha)\), and explain how braiding relates adjacent associahedra.

\subsection{Fusion-tree transport, braiding, and cables}
\label{sec:2.3}

As discussed in \Sec{sec:2.1}, the Drinfeld associator provides the transport between KZ solutions normalized with respect to different fusion trees. We now formulate the general transport rule between arbitrary fusion trees. Note that, since we are discussing KZ solutions, all fusion trees will be binary fusion trees corresponding to the vertices of \(K_{n-1}\).

{We first record some consequences of the infinitesimal
braid relations that will be used in the transport and
braiding formulas below. For two disjoint clusters \(A,B\), we have 
\begin{equation}
e_{A\cup B} = e_A+e_B+e_{A,B}.
\label{eq:cluster-decomposition}
\end{equation}
where the cross-cluster generator is defined as 
\begin{equation}
e_{A,B}
:=
\sum_{\substack{i\in A\ j\in B}} e_{ij}.
\end{equation}
The internal generator \(e_I\) is central in the Drinfeld--Kohno
subalgebra generated by the labels in \(I\), meaning that it commutes with all generators $e_{ij}$ for $i,j\in I$. Indeed, for \(a,b\in I\),
\begin{equation}
    [e_I,e_{ab}]
=
\sum_{c\in I\setminus\{a,b\}}
[e_{ac}+e_{bc},e_{ab}]
=
0.
\end{equation}
If \(A\cap B=\varnothing\), then
\begin{equation}
[e_A,e_B] = [e_A,e_{A,B}] = [e_B,e_{A,B}]
=
0,
\label{eq:binary-cluster-commutation}
\end{equation}
from
\[
[e_A,e_{A,B}]
=
\sum_{\substack{a<a'\\a,a'\in A}}
\sum_{b\in B}
[e_{aa'},e_{ab}+e_{a'b}]
=
0
\]
by the braid relations \eqn{braid}. }

\begin{figure}[t]
\centering
\includegraphics[width=0.8\textwidth]{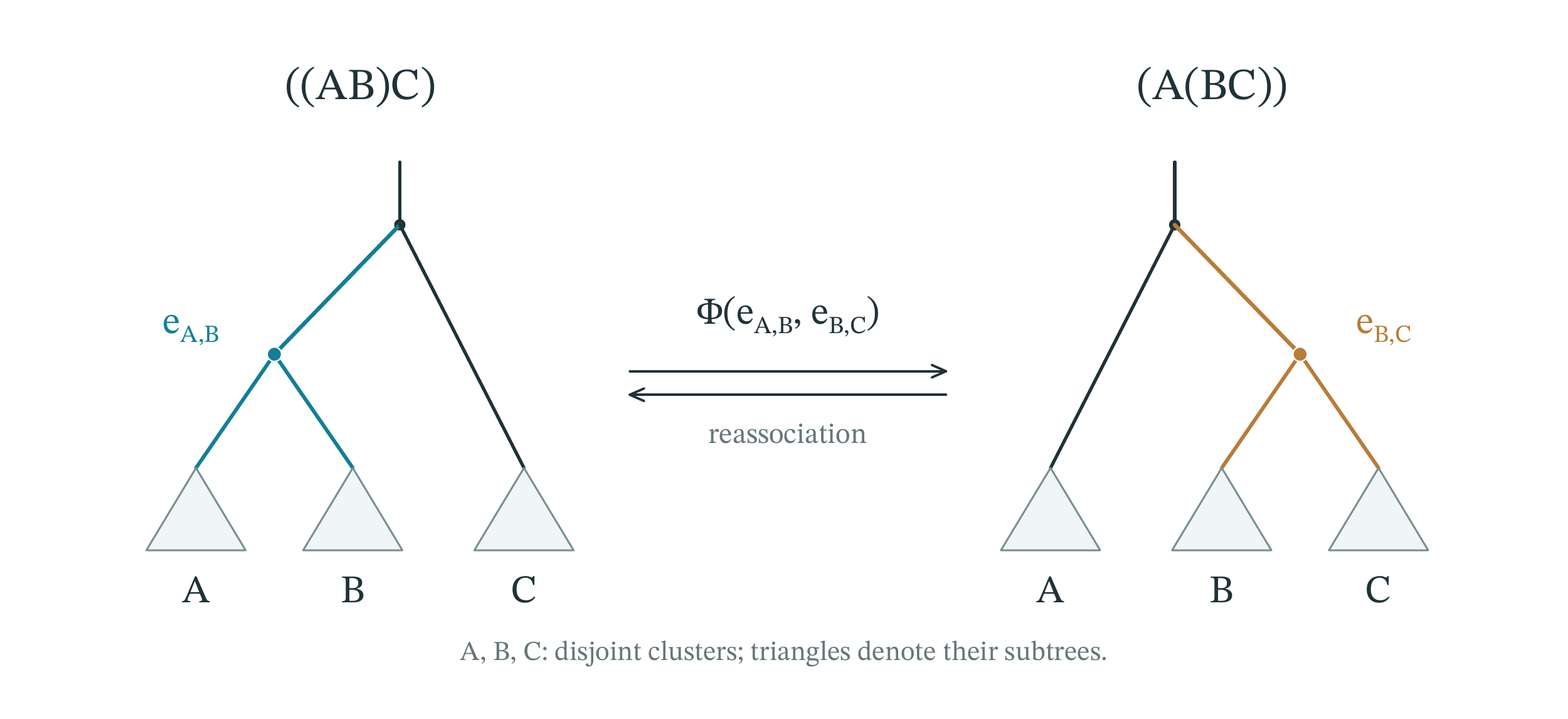}
\vspace{-5pt}
\caption{Fusion tree transport}
\label{fig:fusion_tree_associator}
\end{figure}

For three disjoint clusters \(A\), \(B\), and \(C\),
the elementary reassociation shown in \fig{fig:fusion_tree_associator} 
\begin{equation}
(AB)C\longleftrightarrow A(BC)
\end{equation}
is implemented by the Drinfeld associator \cite{Drinfeld:1991quasi,barnatan1998},
\begin{equation}
G_{(AB)C} = G_{A(BC)}\,\Phi \left(e_{A,B},e_{B,C}\right),
\end{equation}
which satisfies
\begin{equation}
\Phi(A,B)^{-1}=\Phi(B,A).
\end{equation}

Denoting the initial fusion tree by \(T_s\) and the final fusion tree by \(T_f\), we introduce the transport operator \({\cal T}_{s\to f}\) through
\begin{equation}
G_f = G_s\,{\cal T}_{s\to f}.
\end{equation}
By construction, these transport operators satisfy the inverse and composition relations
\begin{equation}
{\cal T}_{s\to f}\,{\cal T}_{f\to s}=1,
\qquad
{\cal T}_{s\to m}\,{\cal T}_{m\to f}
=
{\cal T}_{s\to f},
\end{equation}
for an intermediate fusion tree \(m\). A change of fusion tree within \(K_{n-1}(\alpha)\) can be
realized by different sequences of elementary reassociations.
These sequences give the same transport by the pentagon
identity, together with the commutativity of independent
reassociations. See \app{app:pen-hex} for details.

The reassociations discussed above preserve the ordering $\alpha$. We now consider the transport between adjacent associahedra corresponding to different orderings. Let \(I=(i_1,\ldots,i_k)\) be a state  consecutive block in
the ordering \(\alpha\), with the root label outside \(I\).
The neighbouring chamber across the facet labelled by \(I\)
has ordering
\[
\alpha=LIR,
\qquad
\beta=LI^{\operatorname{mir}}R,
\]
where
\[
I^{\operatorname{mir}}=(i_k,\ldots,i_1)
\]
denotes the reversed ordering of the labels in \(I\).
For example, when \(I=123\), we have \(I^{\operatorname{mir}}=321\),
and the adjacent associahedra $K_4(12345)$
and $K_4(32145)$ share the facet labelled by $123$. The two associahedra share the facet
\[
F_I
=
K_{n-1}(\alpha)\cap D_I
=
K_{n-1}(\beta)\cap D_I.
\]
The transport between adjacent chambers is represented by a \textit{braid} that performs a half-twist of the cluster \(I\),
reversing the order of its labels. We first discuss the exchange of two individual strands,
corresponding to \(|I|=2\), and then describe the general
cluster braid recursively in terms of cables.

\begin{figure}[t]
\centering
\includegraphics[width=0.8\textwidth]
{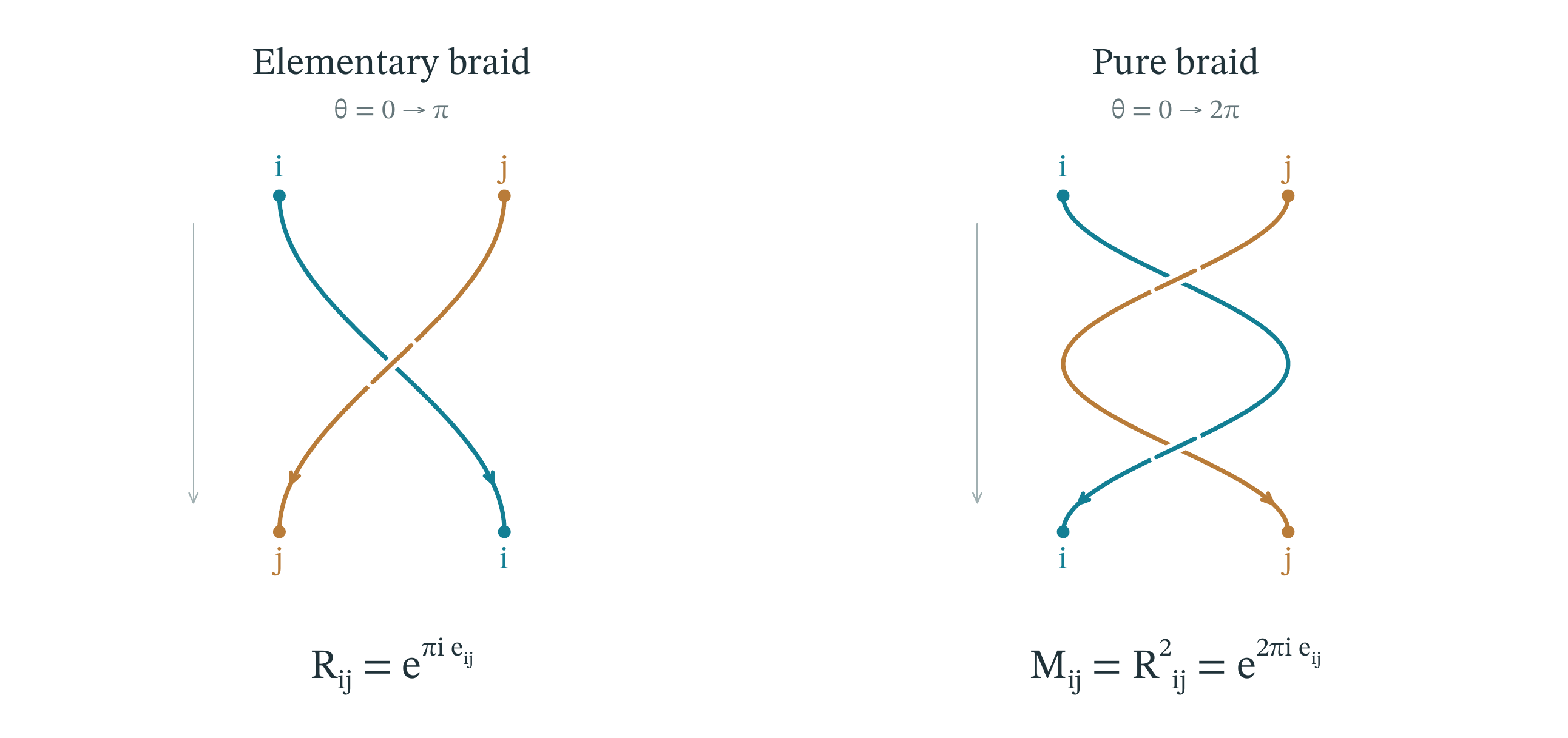}
\vspace{-5pt}
\caption{An elementary braid and the corresponding full monodromy.}
\label{fig:braid_half_full_monodromy}
\end{figure}

A braid of two strands corresponds to an analytic continuation in the configuration space of marked points in which two points move around each other without crossing a collision divisor; see \fig{fig:braid_half_full_monodromy}. An elementary braid exchanges two neighboring marked points by a half-rotation, whereas performing the braid twice gives a pure braid in which the points return to their original ordering after winding once around each other. In the KZ representation, the corresponding local transport factors are therefore
\begin{equation}
B^{\pm}_{ij}=e^{\pm \pi i e_{ij}},
\qquad
M_{ij}^{\pm}=(B_{ij}^{\pm})^2=e^{\pm 2\pi i e_{ij}},
\end{equation}
where \(B^{\pm}_{ij}\) represents the elementary braid, and \(M^{\pm}_{ij}\) the associated monodromy around the singular point \(z_i = z_j\). This exponential form is valid in a fusion frame in which
the leaves \(i\) and \(j\) share the same parent vertex.
Starting from a different frame, we first transport the
KZ solution to such an adapted frame using the appropriate
Drinfeld associators.

{The \textit{hexagon identity} expresses the compatibility
between reassociation and braiding. We adopt the positive
braid convention
\begin{equation}
B_{ij}=e^{\pi i e_{ij}},
\end{equation}
with the exchange of strand labels recorded by the target
fusion tree. In this convention, the hexagon identity reads
\begin{equation}
\label{hexagon}
e^{i\pi(e_{13}+e_{23})}
=
\Phi(e_{23},e_{12})\,e^{i\pi e_{23}}\,
\Phi(e_{13},e_{23})\,e^{i\pi e_{13}}\,
\Phi(e_{12},e_{13}).
\end{equation}
Here \(e_{12}+e_{13}+e_{23}\) is central in the
three-strand Drinfeld--Kohno algebra. Both sides describe the transport
\[
(12)3 \longrightarrow 3(12).
\]
The left-hand side exchanges the cluster \((12)\) with strand
\(3\), preserving its internal ordering, while the right-hand
side resolves this exchange into two elementary braids and
the necessary reassociations.}
Including a positive exchange of the two strands inside
the cluster gives the mirror transport
\[
(12)3 \longrightarrow 3(12) \longrightarrow 3(21).
\]
Its transport factor is
\[
e^{i\pi(e_{13}+e_{23})}\,e^{i\pi e_{12}}
=
e^{i\pi(e_{12}+e_{13}+e_{23})}
=
e^{i\pi e_{123}},
\]
where we used
\([e_{12},e_{13}+e_{23}]=0\).
Thus, a positive half-twist of the three-strand cluster
maps the original fusion tree to its mirror. The pentagon and hexagon identities provide the coherence conditions relating changes of fusion frame and braid transport\footnote{Some explanations of these two identities are given in \app{app:pen-hex}}.

When \(|I|>2\), the exchange involves multiple strands, motivating the introduction of a \textit{cable}, which groups these strands into a single unit. 
Let \(I\subset L_n\) be a cluster, as defined above, and regard
the elements of \(L_n\) as strand labels. Treating the strands
labelled by \(I\) as a single composite strand in external
fusion and braiding operations defines a cable supported on \(I\).
Thus, the cluster specifies which strands belong together,
whereas the cable describes how they are treated collectively
in operations involving other strands.

A cable specifies an ordering
of the strands in \(I\) and an internal fusion tree. For a fixed
ordering, different internal parenthesizations are related by
associator transports; changes of ordering also require braiding.
When several cables are treated as composite inputs, their
reassociation is described by cabled associators.

Before discussing the hexagon identity for cables, we describe the
braid exchanging two cables \cite{kurlinCompressedDrinfeldAssociators2005}
. Let \(A\) and \(B\) be two
disjoint clusters, each equipped with a fixed internal
ordering and fusion tree. In a fusion tree where
they form a local pair, exchanging the two cables while
preserving their internal configurations gives the
transport factor
\[
B_{A,B}^{\pm}=e^{\pm i\pi e_{A,B}},
\qquad
e_{A,B}=\sum_{a\in A}\sum_{b\in B}e_{ab}.
\]
Indeed, in the cable limit, the part of the KZ connection
governing their relative position \(w\) is
\(e_{A,B}\,\di \log w\). A positive or negative half-rotation
of \(w\) therefore produces the stated exponential.
The exchange of labels is recorded by the target fusion
tree.
The hexagon identity expresses the compatibility of this
cabled braid with reassociation: exchanging \(B\) with the
composite cable \((XA)\) gives the same transport as
successively exchanging it with \(A\) and \(X\), with the
appropriate reassociations between these operations. See \app{app:cable-braiding} for the detailed derivation from the strand exchange.

{
For three pairwise disjoint clusters \(X,A,B\), appearing
consecutively in this order and equipped with fixed internal
fusion trees, the cabled hexagon identity reads
\begin{equation}
\label{eq:oriented-cabled-hexagon}
\begin{aligned}
&\Phi(e_{A,B},e_{X,A})\,
e^{\pm i\pi e_{A,B}}\,
\Phi(e_{X,B},e_{A,B})\,
e^{\pm i\pi e_{X,B}}\,
\Phi(e_{X,A},e_{X,B})
\\
&\qquad =
e^{\pm i\pi(e_{X,B}+e_{A,B})}
=
e^{\pm i\pi e_{C,B}},
\qquad C=X\cup A.
\end{aligned}
\end{equation}
This identity describes the exchange
\[
(XA)B \longrightarrow B(XA),
\]
treating \(C=(XA)\) as a single cable.
A further exchange of \(X\) and \(A\) gives
\[
(XA)B \longrightarrow B(XA) \longrightarrow B(AX),
\]
with transport factor
\[
e^{\pm i\pi(e_{X,B}+e_{A,B})}\,
e^{\pm i\pi e_{X,A}}
=
e^{\pm i\pi(e_{X,A}+e_{X,B}+e_{A,B})},
\]
where
\([e_{X,A},e_{X,B}+e_{A,B}]=0\).
This mirrors the three-cable tree while preserving the
internal trees of \(X,A,B\). The full mirror operation
also mirrors these internal trees recursively.}

We now turn to the general mirror operation associated
with a given cluster. Let \(T_I\) be a planar binary fusion tree with leaf set \(I\).
We define the \textit{mirror fusion tree} of \(T_I\) recursively by
\begin{equation}
\operatorname{mir}(i)=i,
\qquad
\operatorname{mir}\bigl((T_A T_B)\bigr)
=
\bigl(\operatorname{mir}(T_B)\operatorname{mir}(T_A)\bigr).
\end{equation}
For example,
\[
\operatorname{mir}((12)3)=3(21),
\qquad
\operatorname{mir}(1(23))=(32)1.
\]
We realize this operation by a recursive braid.
Fix an orientation \(\sigma\in\{+1,-1\}\), corresponding to
positive or negative half-rotations.
For a single leaf, the braid is trivial.
If
\[
T_I=(T_A T_B),
\qquad I=A\cup B,
\]
we first apply the mirror braids to the two subtrees and
then exchange the resulting cables:
\[
(T_A T_B)
\longrightarrow
\bigl(\operatorname{mir}(T_A)\operatorname{mir}(T_B)\bigr)
\longrightarrow
\bigl(\operatorname{mir}(T_B)\operatorname{mir}(T_A)\bigr).
\]
Every exchange is taken with the same orientation \(\sigma\),
without additional windings.
In the corresponding asymptotic fusion frames, the transport
factor of this braid is
\begin{equation}
B^\sigma(I)
=
e^{\sigma\pi i e_I}.
\label{eq:mirror-tree-transport}
\end{equation}
Here the permutation of strand labels is recorded by the
target tree \(\operatorname{mir}(T_I)\).
To prove \eqref{eq:mirror-tree-transport}, we proceed by
induction on the number of leaves.
For a single leaf \(i\), we have \(e_{\{i\}}=0\), and the
transport is the identity.
Suppose the statement holds for \(T_A\) and \(T_B\).
Their internal mirror braids contribute
\(e^{\sigma\pi i e_A}\) and \(e^{\sigma\pi i e_B}\),
respectively, while the exchange of the two cables contributes
\(e^{\sigma\pi i e_{A,B}}\).
Hence
\begin{equation}
B^\sigma(I)
=
e^{\sigma\pi i e_A}
e^{\sigma\pi i e_B}
e^{\sigma\pi i e_{A,B}}.
\end{equation}
{Using \eqref{eq:cluster-decomposition} and
\eqref{eq:binary-cluster-commutation}, we obtain}
\[
B^\sigma(I)
=
e^{\sigma\pi i(e_A+e_B+e_{A,B})}
=
e^{\sigma\pi i e_I},
\]
which completes the induction.
The hexagon identity ensures that the cable exchange used
in this construction agrees with its realization through
elementary braids and reassociations. It should be emphasized that the mirror operation via the facet \(F_I\) is naturally defined only in a fusion tree adapted to the
cluster \(I\). 

Let \(s\) and \(f\) be the source and target
fusion trees of \(G\), let \(I\) in the transport operator be an \(I\)-adapted tree, and let \(I^{\operatorname{mir}}\)
denote tree obtained by mirroring its \(I\)-subtree. The
positive half-continuation is given by
\begin{equation}
G_f
=
G_s\,
\mathcal T_{s\to I}\,
e^{\pi i e_I}\,
\mathcal T_{I^{\operatorname{mir}}\to f}.
\label{eq:half-continuation-adapted-frame}
\end{equation}
To obtain a full positive monodromy, the second half-turn must continue
in the same angular direction and therefore maps \(T_I^{\operatorname{mir}}\) back to
\(T_I\). Closing the path in the original fusion frame gives
\[
\begin{aligned}
    G_s^{\circlearrowleft}
={}&
G_s\,
\mathcal T_{s\to I}\,
R_{I\to I^{\operatorname{mir}}}^{+}\,
\mathcal T_{I^{\operatorname{mir}}\to f}\,
\mathcal T_{f\to I^{\operatorname{mir}}}\,
R_{I^{\operatorname{mir}}\to I}^{+}\,
\mathcal T_{I\to s}
\nonumber\\
={}&
G_s\,
\mathcal T_{s\to I}\,
e^{2\pi i e_I}\,
\mathcal T_{I\to s}.
\end{aligned}
\]
Thus the monodromy in the source frame is the conjugate
\begin{equation}
    G_s M_{I,s}
=
G_s
\mathcal T_{s\to I}\,
e^{2\pi i e_I}\,
\mathcal T_{s\to I}^{-1}.
\label{eq:full-monodromy-source-frame}
\end{equation}

For a given facet \(F_I\), there may be multiple choices of adapted fusion trees. For example, when \(I=\{1,2,3\}\in L_5\), in our left-combo convention, there are two possible adapted fusion trees, distinguished by their internal subtrees:
\[
(12)3 \qquad \text{and} \qquad 1(23).
\]
The monodromy factor we consider is in the form of 
\[
    {\cal T}_{s\to I} e^{2\pi i e_I} {\cal T}_{I\to s}
\]
If we consider another \(I\)-adapted tree \(T'_I\), the monodromy is given by
\[
    {\cal T}_{s\to I'} e^{2\pi i e_I} {\cal T}_{I'\to s}
\]
which can be written as 
\[
 {\cal T}_{s\to I} 
 {\cal T}_{I\to I'}
 e^{2\pi i e_I} 
 {\cal T}_{I' \to I}
 {\cal T}_{I \to s}.
\]
Since the transport operator \(T_{I\to I'}\) preserves \(I\) as a cable, all generators entering the corresponding Drinfeld associators commute with \(e_I\). Consequently, these associators commute with \(e^{2\pi i e_I}\), allowing the transport operators to cancel. This establishes the independence of the monodromy factor from the choice of fusion tree.

To avoid ambiguity in the choice of fusion tree, we adopt the all-left-comb convention for cables, as follows. Fix an oriented chamber ordering \(\alpha\) and a distinguished root label \(n\). We represent the divisor \(D_I=D_{I^c}\) by the consecutive cluster \(I\) not containing \(n\). First, we equip \(I\) with the left-comb subtree \(\operatorname{LC}(I)\). We then collapse \(I\) to a single cable \([I]\), take the left-comb tree of the induced quotient ordering \(\alpha/I\), and substitute \(\operatorname{LC}(I)\) for the leaf \([I]\). This convention selects a unique \(I\)-adapted fusion tree.

\subsection{Single-valued map and dual local system}
\label{svanddual}
The KLT relation, or equivalently the double-copy relation between open-string and closed-string amplitudes, has been recognized to be closely related to the single-valuedness of the worldsheet integrand
\cite{Brown:2018omk,Brown:2019wna,Schlotterer:2018zce,Kakkad:2025klm}. Remarkably, this property persists for the building blocks of AdS string amplitudes: the integrand of the closed-string sphere integral is considered as the single-valued map of the corresponding open-string disk integrand \cite{Alday:2025bjp,Baune:2025hfu,Nocchi:2026yms}. This structure makes it possible to formulate a KLT relation directly at the level of these building blocks, with single-valued MPLs appearing naturally in the closed-string integral. This single-valuedness property is also essential in twisted
de Rham theory, where the pairing between twisted cocycles
and their duals is defined by integrating a globally
single-valued top-degree differential form
\cite{Brown:2019wna,Kakkad:2025klm}.

The single-valued map for single-variable MPLs is well known \cite{Brown:2004ugm,Brown:2013sv}\footnote{The single-variable elliptic MPLs, which are a generalization of MPLs on the torus, also admit a similar single-valued map \cite{Schlotterer:2025qjv,Baune:2025ndr}.}:
\begin{equation}
    \operatorname{sv}(G)(z,\bar{z})
    =
G(z)\mathsf R (\widetilde{G}(\bar z)),
\end{equation}
where the generators appearing in \(\widetilde{G}\) are transformed according to
\begin{equation}
    e_0\rightarrow e_0,\qquad
e_1\rightarrow
\operatorname{sv}(\Phi)(e_1,e_0)e_1\operatorname{sv}(\Phi)(e_0,e_1),
\end{equation}
and \(\mathsf R\) denotes the reversal of words. In \cite{Frost:2025lre}, the single-valued map for multivariable
MPLs is formulated in terms of zeta generators, which takes form
\begin{equation}
\text{sv}(G_{\mathbb{I}_n})(z_i,\bar{z}_i)
=G_{\mathbb{I}_n}(z_i)\,\operatorname{sv}(\mathbb M)\,
\mathsf{R}(G_{\mathbb{I}_n})(\bar{z}_i)\,\operatorname{sv}(\mathbb M)^{-1}
  \label{eq:svG-factorisation}
\end{equation}
where the generating series of single-valued multiple zeta values $\text{sv}(\mathbb{M})$ is defined by
\begin{equation}
  \text{sv}(\mathbb M)
  =
  1
  +
  2\sum_{i_1\in 2\mathbb N+1}\zeta_{i_1}M_{i_1}
  +
  2\sum_{i_1,i_2\in 2\mathbb N+1}
  \zeta_{i_1}\zeta_{i_2}M_{i_1}M_{i_2}
  +\ldots .
  \label{eq:svM-expansion}
\end{equation}
and \(\operatorname{R}\) is
the anti-automorphism of the generators defined by
\begin{equation}
    \mathsf R(e_{i_1j_1}\cdots e_{i_rj_r})
=
e_{i_rj_r}\cdots e_{i_1j_1}.
\end{equation}
Equivalently, this construction can be translated into a transformation acting directly on the KZ generators, in close analogy with the single-variable case, but is much more indirect, being based on the complicated bracket relations between $M_{i}$ and KZ generators given in \cite{Frost:2023stm}. Such a single-valued map allows us to construct the dual local system following the method of \cite{Kakkad:2025klm}, thereby relating the sphere integral of the single-valued integrand to a bilinear combination of disk integrals associated with the multivalued integrand and its dual. 

We denote by \(G^\vee\) the dual antiholomorphic part in the single-valued construction. In our
convention, the single-valued MMPLs are
\begin{equation}
\operatorname{sv}G(z_i,\bar{z}_i)=G(z_i)G^\vee(\bar{z}_i).
\end{equation}
The construction of \(G^\vee\) includes complex conjugation,
a transformation of the generators, and word reversal:
\begin{equation}
    G^\vee (\bar{z}_i)
    =
     \mathsf R\bigl(G(\bar z_i;e_{ij}^\vee)\bigr)
     := \operatorname{sv}(\mathbb M) \mathsf{R}(G_{\mathbb{I}_n})(\bar{z}_i;e_{ij}) \operatorname{sv}(\mathbb M)^{-1}
     .
\end{equation}
Here \(e_{ij}\mapsto e_{ij}^\vee\) denotes the  generator transformation from the zeta generators. Writing
\begin{equation}
    \widetilde G=G(\bar z_i;e_{ij}^\vee),
\qquad
\widetilde\Omega
=
\sum_{i<j}e_{ij}^\vee\,d\log(\bar z_i-\bar z_j),
\end{equation}
we have the differential equation
\[\di \widetilde G=\widetilde\Omega\,\widetilde G.\]
Since \(\mathsf R\) reverses multiplication,
\[
\di G^\vee
=
\mathsf R(\widetilde\Omega\,\widetilde G)
=
\mathsf R(\widetilde G)\,
\mathsf R(\widetilde\Omega),
\]
\(G^\vee\) satisfies the right differential equation
\begin{equation}
\label{eqn:duallocalsystem}
    \di G^\vee=G^\vee\,\Omega^\vee,
    \qquad
    \Omega^\vee
    :=
    \mathsf R(\widetilde\Omega).
\end{equation}

The single-valued map construction in \cite{Frost:2023stm, Frost:2025lre} is formulated in a chosen
fibration basis, which in our fusion-tree language means the identical reference fusion
tree for both factors \(G(z)\) and \(G^\vee(\bar{z})\).
We normalize our sMMPLs as 
\begin{equation}
    \operatorname{sv}G_{\mathbb{I}_n} = G_{\mathbb{I}_n}
G^\vee_{\mathbb{I}_n}.
\end{equation}
Now let us consider the sMMPLs constructed from the different chambers of the real slice of the moduli space, corresponding to distinct color orderings. 
However, if we consider another chamber \(\alpha\), which is not the identity permutation, the canonical solution's single-valued map will be 
\begin{equation}
    \operatorname{sv}G_{\alpha} = G_{\alpha}G^\vee_{\alpha}
\end{equation}
which changes the normalization. 
They are connected by the single-valued transport operator 
\begin{equation}
    \operatorname{sv}G_{\mathbb{I}_n} =
    \operatorname{sv}G_{\alpha} \operatorname{sv}{\cal T}_{\alpha\to \mathbb{I}_n}
\end{equation}
{where 
\(\operatorname{sv}{\cal T}_{\alpha \to {\mathbb I}_n}\) is regarded as a generalization of Deligne associators \cite{Brown:2013sv,Alday:2025bjp}.}
For our application, we want to consider the integrand \(\operatorname{sv} G_{\mathbb{I}_n}\) and of course fix the normalization on the whole Riemann sphere. The chamber loadings of \(K_{n-1}\) compatible with this choice will be specified in \Sec{basis}.

\subsection{Review of AdS KLT relations}
\label{2.5}
In this subsection we will review the KLT relation of general AdS building blocks derived in \cite{Nocchi:2026yms}. 
After fixing \(z_1=0\), \(z_{n-1}=1\), and \(z_n=\infty\), the general AdS string amplitude building blocks are defined as follows:
\begin{align}
&\begin{aligned}\label{eq:Open-string BB in AdS}
    Z&^{(\text{AdS})}(\tau|\rho):= \\[1ex]  & \int_{\gamma(\tau)}\hspace{5pt} \prod_{i=2}^{n-2}d z_i\hspace{4pt} \, \text{KN}_\tau(z) \, \text{PT}_\rho(z)\, G_{(\cdots((\tau_1\tau_2)\tau_3)\cdots)\tau_n}(z_2,\ldots,z_{n-2}) 
\end{aligned}\\[4ex]
&\begin{aligned}
\widetilde Z^{(\mathrm{AdS})}
(\beta|\tau )
:=
\mathcal{U}_\beta\,
\mathsf{R}\bigl[
Z^{(\mathrm{AdS})}
(\beta| \tau)
\bigr]\,
\operatorname{sv}(\mathbb M)^{-1}.
\label{eq:right-moving-Z-def}
\end{aligned}\\[4ex]
&\begin{aligned}\label{eq:Closed-string BB in AdS}
    J&^{(\text{AdS})}(\tau|\rho):=\\[1.5ex] 
    &  \,\int_{\mathbb{C}^{n-3}} \,\prod_{i=2}^{n-2} \, d^2z_i\, \prod_{1\leq i<j\leq n-1} \hspace{-5pt}|z_{ij}|^{2s_{ij}} \,\text{PT}_\rho(z) \, \text{PT}_\tau(\bar{z})\, \text{sv}(G_{\mathbb{I}})(z_2,\ldots,z_{n-2}) \,. 
\end{aligned}
\end{align}
with
\begin{equation}
\mathcal U_\beta
:=
\mathcal T_{\beta\to\mathbb I}\,
\operatorname{sv}(\mathbb M)\,
\mathsf{R} [\overline{\mathcal T_{\beta\to\mathbb I}}]\,,
\label{eq:U-beta-def}
\end{equation}
and here the Parke--Taylor factors are understood to include
the corresponding gauge-fixing Jacobians. {The right-moving building block involves the antiholomorphic
factor matched to the fixed single-valued normalization
\(\operatorname{sv}G_{\mathbb I_n}\) according to \Sec{svanddual}.
Writing
\(
G_{\mathbb I_n}=G_\beta\mathcal T_{\beta\to\mathbb I_n},
\)
we define
\begin{equation}
\begin{aligned}
\widetilde G_\beta(\bar z_i)
&:=
G_\beta(z_i)^{-1}
\operatorname{sv}G_{\mathbb I_n}(z_i,\bar z_i)
\\
&=
\mathcal U_\beta\,
\mathsf{R} \bigl[\overline{G_\beta(z_i)}\bigr]\,
\operatorname{sv}(\mathbb M)^{-1},
\end{aligned}
\end{equation}
Restricting the antiholomorphic factor to the real chamber
with the standard branch convention, we obtain
\begin{equation}
\begin{aligned}
\widetilde Z^{(\mathrm{AdS})}(\beta|\tau)
&:=
\int_{\gamma(\beta)}
\prod_{i=2}^{n-2} d z_i\,
\mathrm{KN}_\beta(z)\,
\widetilde G_\beta(z)\,
\mathrm{PT}_\tau(z)
\\
&=
\mathcal{U}_\beta\,
\mathsf{R}\bigl[
Z^{(\mathrm{AdS})}
(\beta|\tau)
\bigr]\,
\operatorname{sv}(\mathbb M)^{-1}.
\end{aligned}
\end{equation}
}

Then the KLT relation for these building blocks reads
\begin{align}\label{eq:AdS KLT}
J^{\rm(AdS)}(\tau|\rho)=\sum_{\alpha,\beta\in S_{n-3}}Z^{\rm(AdS)}(\alpha\,1\,n-1\,n|\rho)\,S_{\rm (AdS)}[\alpha|\beta]\,\tilde{Z}^{\rm(AdS)}(1\,\beta\,n-1\,n|\tau)\,.
\end{align}
with
\be
\begin{aligned}
S_{(\text{AdS})}[\alpha|\beta]
&\equiv
S[\alpha(2)\,\alpha(3)\ldots\alpha(n-2)\,1\,n{-}1\,n
\,|\,
1\,\beta(2)\,\beta(3)\ldots\beta(n-2)\,n{-}1\,n]
\\
&=
\left(\frac{i}{2}\right)^{n-3}
F^{(\alpha)}K_{\beta(n-2)}^{(\alpha)}
K_{\beta(n-3)}^{(\alpha)}
\ldots
K_{\beta(2)}^{(\alpha)} \,,
\end{aligned}
\fe
where the factors are given by
\be
\begin{aligned}
K_{\beta(k)}^{(\alpha)}
=&
\bigg[
\sum_{\sigma}
\Theta_\alpha(\sigma\circ\beta)
\bigg(
\theta_{\alpha}\bigl(\beta(k),\sigma_{\beta(k-1)}\bigr)
\\
&
+
\sum_{j=2}^{k-1}
\prod_{\ell=j}^{\overrightarrow{k-1}}
C_{\sigma_{\beta(\ell)} \leftarrow \beta(k)}
^{(\sigma_{\beta(\ell+1)},\ldots,\sigma_{\beta(k-1)},1)}
\,
\theta_\alpha\bigl(\sigma_{\beta(j)},\beta(k)\bigr)
\theta_\alpha\bigl(\beta(k),\sigma_{\beta(j-1)}\bigr)
\bigg)
\bigg]
e^{i\pi E_{1\beta(k)}}\\
&\times\prod_{j=2}^{\overleftarrow{k-1}}\Phi(e_{1\sigma_{\beta(j)}}+\sum_{\ell=j+1}^{k-1}e_{\sigma_{\beta(j)}\sigma_{\beta(\ell)}},e_{1\beta(k)}+\sum_{\ell=j+1}^{k-1}e_{\sigma_{\beta(\ell)} \beta(k)})\Biggr]
-\text{c.c.}
\end{aligned}
\fe
Here,  
\begin{align}\label{Rcross}
    C^{(R)}_{j \leftarrow k}=\Phi\left(e_{jk},\sum_{r\in R}e_{jr}\right)e^{i\pi E_{jk}}\Phi\left(\sum_{r\in R}e_{kr},e_{jk}\right)
\end{align}
and \(\theta_\alpha(i,j)=1\) if \(i\) lies to the right of \(j\) in the left ordering \(\alpha=(\alpha(2)\,\alpha(3)\, \ldots \alpha(n-2))\) of the KLT kernel, and
\(\theta_\alpha(i,j)=0\) otherwise. The last factor $\theta_\alpha\bigl(\beta(k),\sigma_{\beta(j-1)}\bigr)$ is understood to be $1$ for $j=2$ and
\[
\Theta_\alpha(\sigma\circ\beta)
:=
\prod_{r=2}^{k-2}
\theta_\alpha\bigl(\sigma_{\beta(r+1)},\sigma_{\beta(r)}\bigr).
\]
In addition, the overall factor $F^{(\alpha)}$ reads\footnote{The factor can be canceled if we do the reflection and cyclic permutation on $Z(\alpha\,1\,n{-}1\,n|\rho)$ to change it to $Z(1\,\alpha^{T}\,n\,n{-}1|\rho)$.}
\be
F^{(\alpha)}=\prod_{j=3}^{\overleftarrow{n-2}}\Phi(e_{1\alpha(j)}+\sum_{\ell=j+1}^{n-2}e_{\alpha(j)\alpha(\ell)},\sum_{\ell=2}^{j-1}e_{\alpha(\ell)\alpha(j)})
\fe

To compare the KLT relation with the twisted period relation,
we use differential-form measures. Writing \(m=n-3\), the
complex integration measure is related to the wedge-product
measure by
\begin{equation}
    \prod_{i=2}^{n-2}d^2z_i
=
\left(\frac{i}{2}\right)^m
\bigwedge_{i=2}^{n-2}(dz_i\wedge d\bar z_i),
\qquad d^2z_i=dx_i\,dy_i.
\end{equation}
Placing all holomorphic differentials before the
antiholomorphic ones introduces the additional orientation
sign \((-1)^{m(m-1)/2}\).
In what follows, we absorb the scalar measure-conversion
factor into the normalization of the sphere
building blocks and the corresponding KLT kernel.

\section{The noncommutative generalization of twisted de Rham theory}
\label{derham}
In this section, we review the mathematical preliminaries of twisted de Rham theory \cite{Aomoto:2011ggg,Mizera:2017cqs,Mizera:2019gea} and its generalization to the setting of a noncommutative coefficient ring \cite{Kakkad:2025klm}. The twisted de Rham theory has its roots in the study of multivalued
hypergeometric integrals and their Gauss--Manin connections
\cite{10.3792/pjaa.58.97,Aomoto:1982configurations}.
For the general theory of regular singular connections and the
hypergeometric setting, see
\cite{deligne1970equations,yoshida1997hypergeometric,Aomoto:2011ggg}.

\subsection{Mathematical preliminaries}

In this subsection, we briefly review the basic framework of twisted de Rham theory, in which open-string integrals are interpreted as \textit{periods}, namely, as pairings between twisted cycles and twisted cocycles. The notion of a twist arises from the multivaluedness of the functions appearing in the integrand: their branch data are incorporated into the twisted geometric structure, so that the corresponding differential forms are naturally regarded as twisted cocycles.

Let us consider an integral of the form
\begin{equation}
\label{stringy}
    \int_C u(z)\,\omega(z),
\end{equation}
where \(u(z)\), the multivalued function on the \(m\)-dimensional complex manifold, satisfies the differential equation
\begin{equation}
\label{diff}
    \mathrm{d}u(z)=\tau\,u(z),
    \qquad
    \tau=\sum_{i=1}^{k}\alpha_i\,\di \log f_i(z), \quad \alpha_i \in\mathbb{C}\backslash \mathbb{Z}
\end{equation}
While \(u(z)\) is in general multivalued, the one-form \(\tau=\di \log u\) is single-valued. The singular locus associated with the functions \(f_i(z)\) defines divisors
\begin{equation}
    D_i:=\{z\in\mathbb{C}^m\mid f_i(z)=0\},
\end{equation}
and we denote their union by
\begin{equation}
    D:=\bigcup_{i=1}^{k}D_i.
\end{equation}
We then work on the complement of these divisors,
\begin{equation}
    X:=\mathbb{C}^m\setminus D.
\end{equation}
The differential form \(\omega(z)\) in \eqn{stringy} is taken to be a single-valued holomorphic top form on \(X\), and \(C\) is a geometric \(m\)-chain on \(X\).

For the multivalued function \(u(z)\) along \(C\), we choose a branch \(u_C\). Locally, \(u_C\) can be regarded as a single-valued holomorphic function on a simply connected open subset of \(X\) containing \(C\). The single-valuedness of the chosen branch \(u_C\) allows us to apply Stokes' theorem,
\begin{equation}
    \int_{\partial C} u_{\partial C}\,\omega
    =
    \int_C \di (u_C\omega)
    =
    \int_C u_C\,(\di+\tau\wedge)\omega,
\end{equation}
where \(u_{\partial C}\) denotes the branch induced on the boundary \(\partial C\). It is therefore natural to regard the integration region \(C\) as being \textit{loaded} with the branch choice \(u_C\), and to denote the corresponding twisted chain by
\begin{equation}
    C\otimes u_C.
\end{equation}
In terms of this loaded chain, Stokes' theorem takes the form
\begin{equation}
    \int_{\partial C\otimes u_{\partial C}} \omega
    =
    \int_{C\otimes u_C} \nabla_\tau \omega,
\end{equation}
where
\begin{equation}
    \nabla_\tau := \di +\tau\wedge,
    \qquad
    \tau:=\di \log u,
\end{equation}
is the twisted de Rham differential, with \(\tau\) encoding the twist. Here and throughout, we use the notation
\begin{equation}
    \int_{C\otimes u_C} \omega := \int u_C \omega.
\end{equation}

To keep track of the branch data globally, we cover the manifold \(X\) by simply connected open sets \(\{U_i\}\), and choose a branch \(u_i\) of \(u\) on each \(U_i\). Since the differential equation
\(
    \mathrm{d}u=\tau\,u
\)
determines its local solutions only up to multiplication by a nonzero constant, the space of local solutions is one-dimensional. On the overlap \(U_i\cap U_j\), two branch choices are therefore related by a constant \textit{monodromy factor},
\begin{equation}
    u_i(z)=M_{ij}\,u_j(z),
    \qquad z\in U_i\cap U_j.
\end{equation}
These monodromy factors play the role of transition functions and satisfy
\begin{equation}
    M_{ik}=M_{ij}M_{jk},
    \qquad
    M_{ij}=M_{ji}^{-1}.
\end{equation}
Thus, the collection of local branches \({u_i}\), together with the transition functions \({M_{ij}}\), defines a rank-one \textit{local system} \(\mathcal{L}_\tau\) on \(X\). The branch data of the multivalued function can then be regarded as local coefficients attached to the simplices of the triangulation of the manifold.

More precisely, given an ordinary \(p\)-chain
\begin{equation}
 C=\sum_a c_a \sigma_a
\in C_p(X,\mathbb{C}), \quad c_a \in \mathbb{C}
\end{equation}
we associate to each simplex \(\sigma_a\) a local branch \(u_{\sigma_a}\) of \(u\). The corresponding \textit{twisted chain} is then defined by
\begin{equation}
\mathscr C
:=
\sum_a c_a
\underbrace{
\sigma_a\otimes u_{\sigma_a}
}_{\text{twisted simplex}}
\in
C_p(X,\mathcal{L}_\tau), \quad c_a \in \mathbb{C}.
\label{eq:twistedchain}
\end{equation}
Here, \(u_{\sigma_a}\) denotes the local coefficient, or equivalently the branch of the twisting function, carried by the simplex \(\sigma_a\). Correspondingly, the twisted boundary operator is defined on a generic $p$-simplex $\sigma=\braket{0,\!1,\!2,\!\cdots\!,p}$ spanned by $p\;\!{+}1$ points:
\begin{equation}\!\!
\partial_\tau \big(\braket{0,\!1,\!2,\!\cdots\!,p} \otimes {u}_{\braket{0,1,2,\cdots\;\!\!,p}} \big)
:= \sum_{j=0}^p (-1)^j \braket{0,\!1,\!2,\! \cdots\!,j\;\!{-}1,j\;\!{+}1,\!\cdots\!,p} \otimes {u}_{\braket{0,1,2,\cdots\;\!\!, j-1,j+1,\cdots\;\!\!,p}} ,
\label{eq:TwistedBoundarySimplex}
\end{equation}
where the branch on each codimension-one face is obtained by restricting, or equivalently transporting, the branch \(u_\sigma\) to that face.
The twisted boundary operator is nilpotent, \(\partial_\tau^2=0\).

The twisted boundary operator $\partial_\tau$
defines a chain complex
\begin{equation}
    C_m(X,{\cal L}_\tau)\xrightarrow{\partial_\tau}
    C_{m-1}(X,{\cal L}_\tau).
\end{equation} A \textit{twisted cycle} is a twisted chain of the form
\begin{equation}
    \mathscr C
    = \sum_a c_a \sigma_a\otimes u_{\sigma_a}
\end{equation}
that is closed under the twisted boundary operator, i.e.,
\begin{equation}
    \partial_\tau \mathscr C=0.
\end{equation}
The corresponding \textit{twisted homology group} is defined as
\begin{equation}
H_p(X,\mathcal L_\tau)
:=
\frac{
\ker\!\left(
\partial_\tau:
C_p(X,\mathcal L_\tau)
\rightarrow
C_{p-1}(X,\mathcal L_\tau)
\right)
}{
\operatorname{im}\!\left(
\partial_\tau:
C_{p+1}(X,\mathcal L_\tau)
\rightarrow
C_p(X,\mathcal L_\tau)
\right)
}.
\label{thomo}
\end{equation}

In terms of the twisted boundary operator, Stokes' theorem takes the form
\begin{equation}
    \int_{\partial_\tau \mathscr C}\omega
    =
    \int_{\mathscr C}\nabla_\tau\omega.
\end{equation}
Now, we turn to the corresponding homology group. Since the connection is flat,
\begin{equation}
    \nabla_\tau^2=0,
\end{equation}
the twisted differential defines the holomorphic twisted de Rham complex
\begin{equation}
 {\nabla_\tau}
    \Omega^{p}(X)
    \xrightarrow{\nabla_\tau}
    \Omega^{p+1}(X)
\end{equation}
A differential form \(\omega\in\Omega^{p}(X)\) is called a \textit{twisted cocycle} if
\begin{equation}
    \nabla_\tau\omega=0.
\end{equation}
The corresponding \textit{twisted cohomology group} is defined by
\begin{equation}
H^p(X,\nabla_\tau)
:=
\frac{
\ker\!\left(
\nabla_\tau:\Omega^{p}(X)\rightarrow\Omega^{p+1}(X)
\right)
}{
\operatorname{im}\!\left(
\nabla_\tau:\Omega^{p-1}(X)\rightarrow\Omega^{p}(X)
\right)
}.
\label{eq:TwistedCohomologyGroup}
\end{equation}
For the string integrals considered in this paper, we focus on holomorphic top-degree forms
\[
\omega\in\Omega^{m,0}(X),
\]
which represent classes in the twisted cohomology group \(H^m(X,\nabla_\tau)\). Since \(X\) has complex dimension \(m\), such forms are automatically \(\nabla_\tau\)-closed, and the nontrivial information is therefore encoded in their equivalence classes modulo \(\nabla_\tau\)-exact forms. These cohomology classes naturally pair with twisted cycles in \(H_m(X,\mathcal L_\tau)\).

\paragraph{Dual local system.}
A dual local system is defined by the differential equation
\begin{equation}
\di u^\vee = \tau^\vee u^\vee .
\end{equation}
Two common choices are the inverse dual and the complex-conjugate dual \cite{Mizera:2017cqs,Hanamura_Yoshida_1999}. For the inverse dual,
\begin{equation}
u^\vee=u^{-1},
\qquad
\tau^\vee=-\tau,
\end{equation}
whereas for the complex-conjugate dual,
\begin{equation}
u^\vee=\overline{u},
\qquad
\tau^\vee=\overline{\tau}.
\end{equation}
In the inverse case, if the original local system has monodromy \(M\), the dual local system carries the inverse monodromy \(M^{-1}\). Consequently, the product
\begin{equation}
\operatorname{sv}(u) = uu^\vee
\end{equation}
is single-valued. For the complex-conjugate choice, the dual monodromy is instead \(\overline{M}\); the product \(u\,\overline{u}\) is single-valued when the monodromy is unitary, so that \(\overline{M}=M^{-1}\).

The dual local system similarly gives rise to
the dual twisted homology $H_m(X,\mathcal L_{\tau^\vee})$
and twisted cohomology $H^m(X,\nabla_{\tau^\vee})$,
following the same construction described above. In particular, we have the following isomorphisms:
\[
\begin{aligned}
H^m(X,\nabla_\tau) \cong H_m(X,\mathcal L_{\tau^\vee}), \quad
H^m(X,\nabla_{\tau^\vee}) \cong H_m(X,\mathcal L_\tau).
\end{aligned}
\]

\subsection{Twisted de Rham theory over a noncommutative ring}

In their study of the four-point AdS double copy, the authors of \cite{Kakkad:2025klm} extend twisted de Rham theory from the complex numbers to the completed free associative algebra
\(
{\cal R}=\widehat{\mathbb C\langle e_0,e_1\rangle},
\)the algebra of noncommutative formal power series in \(e_0\) and \(e_1\). Here we extend this framework to higher multiplicities by considering the local system defined by the full KZ equation, with local coefficients valued in the completed universal enveloping algebra of the spherical Drinfeld--Kohno Lie algebra \cite{Brown:2009qja} 
\begin{equation}
    \mathcal R:=\widehat{U(\mathfrak t_n^{\mathrm{sph}})}.
\end{equation}
The completion is taken with respect to the degree in the generators \(e_{ij}\), allowing formal power series of arbitrary degree.

The Koba--Nielsen factor itself can be regarded as a solution of a KZ-type equation in which the generators are replaced by commuting kinematic invariants. It is therefore convenient to introduce the combined generators
\begin{equation}
    E_{ij}=s_{ij}\mathbf 1+e_{ij},
\end{equation}
which simultaneously encode the Koba--Nielsen factor and the multivariable MPLs. In the massless case, the kinematic invariants satisfy
\begin{equation}
    \sum_{j\neq i}s_{ij}=0,
\qquad i=1,\ldots,n,
\end{equation}
while the spherical Drinfeld--Kohno generators obey
\[
\sum_{j\neq i}e_{ij}=0, \qquad i=1,\ldots,n.
\]Consequently,
\begin{equation}
    \sum_{j\neq i}E_{ij}=0,
\qquad i=1,\ldots,n.
\end{equation}

Including the contribution from Mandelstam variables, the full multivalued integrand, given by the product of the Koba--Nielsen factor and the MMPL generating function, satisfies the differential equation
\begin{equation}
\label{noncomtwist}
    \di {\cal I}=\tau\,{\cal I},
\end{equation}
where
\begin{equation}
    \tau
    =
    \sum_{i<j} E_{ij}\,d\log(z_i-z_j),
\end{equation}
and \(\cal I\) is \(\cal R\)-valued. This equation provides the natural noncommutative analogue of the twist in the flat spacetime
\begin{equation}
\di u=\tau\,u,
\qquad
\tau=\di \log u.
\end{equation}
Indeed, since the ordering of factors is essential in the noncommutative case, the analogue of the logarithmic derivative \cite{Kakkad:2025klm} is 
\begin{equation}
\tau= d\mathcal I\,\mathcal I^{-1},
\end{equation}
rather than an ordinary expression \(\di \log\mathcal I\) which is not well defined.

Correspondingly, the solution space of \eqn{noncomtwist} is no longer a one-dimensional vector space over \(\mathbb C\). Instead, it forms a rank-one \(\cal{R}\)-right module. This module-valued structure provides the noncommutative counterpart of the rank-one local system appearing in the flat-space twisted de Rham theory. On an overlap of the open sets of the manifold \(U_i\cap U_j\), two local solutions differ by a \(z\)-independent right multiplication by an invertible element of \(\mathcal R\), namely the monodromy factor
\begin{equation}
    \mathcal I_i
    =
    \mathcal I_j M_{ij},
    \qquad
    M_{ij}\in\mathcal R^\times, 
\end{equation}
Since the transition functions act from the right, their composition law on a triple overlap \(U_i\cap U_j\cap U_k\) takes the form
\begin{equation}
    M_{ik}=M_{jk}M_{ij}.
\end{equation}
Thus, the local solution space defines a rank-one right \(\mathcal R\)-module local system.

A twisted \(p\)-chain is a finite sum
\begin{equation}
    \mathscr C=\sum_a \sigma_a \otimes s_{\sigma_a},
\end{equation}
where \(s_\sigma\) is the local coefficient. After choosing an invertible local solution
\(\mathcal I_\sigma\), we may write
\(s_\sigma=\mathcal I_\sigma c_\sigma\), with a constant
\(c_\sigma\in\mathcal R\). Thus,
\begin{equation}
    \mathscr C
=
\sum_a(\sigma_a\otimes\mathcal I_{\sigma_a})c_a.
\end{equation}
The closedness condition for a twisted cycle is
\begin{equation}
    \partial_\tau\mathscr C=0.
\end{equation}
The twisted boundary operator has the same simplicial form as in the commutative case. 
Denoting by \(\sigma_{a,j}\) the \(j\)-th codimension-one face of \(\sigma_a\), we therefore have
\begin{equation}
\partial_\tau
\Big[
(\sigma_a\otimes\mathcal I_{\sigma_a})c_a
\Big] = 
\sum_{j=0}^{p}(-1)^j
( \sigma_{a,j}\otimes \mathcal I_{\sigma_{a_j}} ) c_a .
\label{eq:NCtwistedboundary}
\end{equation}

Let us now consider a constant right multiplication by \(R\in\mathcal R\),
\begin{equation}
\mathscr C\cdot R = \sum_a (\sigma_a\otimes\mathcal I_{\sigma_a})C_aR.
\end{equation}
Applying the twisted boundary operator gives
\begin{equation}
\begin{aligned}
\partial_\tau(\mathscr C\cdot R)
&= \sum_{a,j}(-1)^j\left(
\sigma_{a,j}\otimes
\mathcal I_{\sigma_a}
\right)C_a R\\
&=
\left[
\sum_{a,j}(-1)^j
\left(
\sigma_{a,j}\otimes
\mathcal I_{\sigma_a}
\right)C_a
\right]R
\
&=
(\partial_\tau\mathscr C)\cdot R .
\end{aligned}
\end{equation}
Hence, \(\partial_\tau\) is right \(\mathcal R\)-linear. In particular,
\begin{equation}
\partial_\tau\mathscr C=0
\qquad\Longrightarrow\qquad
\partial_\tau(\mathscr C\cdot R)=0,
\end{equation}
so right multiplication preserves the twisted-cycle condition. The obstruction to a left \(\mathcal R\)-module structure appears when the boundary contributions of different simplices are glued along a common face. Consider two simplices \(\sigma_a\) and \(\sigma_b\) sharing a codimension-one face
\begin{equation}
F=\sigma_a\cap\sigma_b.
\end{equation}
The restrictions of their local branches to \(F\) need not coincide, but are related by a monodromy factor \(M_{ab}\in\mathcal R^\times\),
\begin{equation}
\left.\mathcal I_{\sigma_a}\right|_F
=
\left.\mathcal I_{\sigma_b}\right|_F
M_{ab}.
\end{equation}
Assuming opposite induced orientations on the common face, cancellation of the corresponding boundary contributions requires
\begin{equation}
\left.\mathcal I_{\sigma_a}\right|_F c_a
=
\left.\mathcal I_{\sigma_b}\right|_F c_b,
\end{equation}
or equivalently,
\begin{equation}
c_b=M_{ab}c_a.
\end{equation}
Under a common right multiplication,
\begin{equation}
c_a\longmapsto c_aR,
\qquad
c_b\longmapsto c_bR,
\end{equation}
the gluing condition is preserved, since
\begin{equation}
M_{ab}(C_aR)
=
(M_{ab}C_a)R
=
C_bR.
\end{equation}
By contrast, under left multiplication of the coefficients,
\begin{equation}
C_a\longmapsto RC_a,
\qquad
C_b\longmapsto RC_b,
\end{equation}
the new gluing condition would require
\begin{equation}
M_{ab}RC_a
=
RC_b
=
RM_{ab}C_a.
\end{equation}
Since \(\mathcal R\) is noncommutative, in general
\begin{equation}
M_{ab}R\neq RM_{ab},
\end{equation}
and the cancellation of the common boundary face is therefore not preserved. Thus, left multiplication does not in general preserve the twisted-cycle condition.

Consequently, both \(\ker\partial_\tau\) and \(\operatorname{im}\partial_\tau\) are right \(\mathcal R\)-modules, and the twisted homology group
\begin{equation}
H_p(X,\mathcal L_\tau)
:=
\frac{
\ker\!\left(
\partial_\tau:
C_p(X,\mathcal L_\tau)
\rightarrow
C_{p-1}(X,\mathcal L_\tau)
\right)
}{
\operatorname{im}\!\left(
\partial_\tau:
C_{p+1}(X,\mathcal L_\tau)
\rightarrow
C_p(X,\mathcal L_\tau)
\right)
}.
\end{equation}
naturally carries the structure of a right \(\mathcal R\)-module. For a twisted simplex, we define the period pairing between twisted cycles and cocycles by
\begin{equation}
    \int_{\mathscr C}\omega = \int_{C\otimes\mathcal I_C}\omega
    :=
    \int_C \omega\,\mathcal I_C ,
\end{equation}
where the ordering of elements in the integrand is essential. The corresponding Stokes' theorem remains valid as an \(\mathcal R\)-valued identity,
\begin{equation}
    \int_{\partial_\tau\mathscr C}\omega
    =
    \int_{\mathscr C}\nabla_\tau\omega.
\end{equation}
In contrast to the flat-space case, the differential form \(\omega\) does not in general commute with the noncommutative twist \(\tau\). For an \(\mathcal R\)-valued \(p\)-form \(\omega\), using
\(
    \di \mathcal I=\tau\,\mathcal I,
\)
we obtain
\begin{equation}
\begin{aligned}
    \di (\omega\,\mathcal I)
    &=
    \di \omega\,\mathcal I
    +
    (-1)^p\omega\wedge d\mathcal I
    \\
    &=
    \left[
        \di \omega
        +
        (-1)^p\omega\wedge\tau
    \right]\mathcal I .
\end{aligned}
\end{equation}
This naturally defines the noncommutative twisted differential
\begin{equation}
    \nabla_\tau\omega
    :=
    \di \omega
    +
    (-1)^p\omega\wedge\tau .
\label{eq:TwistedDifferentialNC1}
\end{equation}
In the commutative limit, graded commutativity gives
\begin{equation}
    (-1)^p\omega\wedge\tau
    =
    \tau\wedge\omega,
\end{equation}
and hence this definition reduces to the usual flat-space twisted differential
\(\nabla_\tau=\di+\tau\wedge\). 
The flatness condition
\begin{equation}
    \di\tau-\tau\wedge\tau=0
\end{equation}
implies
\begin{equation}
    \nabla_\tau^2=0.
\end{equation}
A twisted \(p\)-cocycle is therefore an \(\mathcal R\)-valued \(p\)-form satisfying
\begin{equation}
    \nabla_\tau\omega=0.
\end{equation}
The space of solutions to the closedness condition, and hence the associated twisted cohomology, naturally carries the structure of a left \(\mathcal R\)-module, rather than that of a vector space over \(\mathbb C\) as in the commutative case.
The twisted cohomology group 
\begin{equation}
H^p(X,\nabla_\tau)
:=
\frac{
\ker\!\left(
\nabla_\tau:\Omega^{p}(X)\rightarrow\Omega^{p+1}(X)
\right)
}{
\operatorname{im}\!\left(
\nabla_\tau:\Omega^{p-1}(X)\rightarrow\Omega^{p}(X)
\right)
}.
\end{equation}
carries the structure of a left \({\cal R}\)-module.

\paragraph{Dual local system.}
Following \sec{svanddual}, we fix the reference ordering
$\mathbb I_n=(1,\ldots,n)$ of the single-valued integrand and its left-comb normalization.
Including the Koba--Nielsen factor, we define the fixed
single-valued integrand
\begin{equation}
\label{eq:fixed-sv-integrand}
    W(z,\bar z)
    :=\operatorname{sv}\mathcal I_{\mathbb I_n}
    =\operatorname{sv}\mathrm{KN}\,
      \operatorname{sv}G_{\mathbb I_n}
    =\mathcal I_{\mathbb I_n}(z)
      \mathcal I_{\mathbb I_n}^{\vee}(\bar z),
\end{equation}
where
$\operatorname{sv}\mathrm{KN}
=\prod_{i<j}|z_i-z_j|^{2s_{ij}}$.
The antiholomorphic connection associated with this reference
normalization is
\begin{equation}
    \tau^\vee
    =\sum_{i<j}s_{ij}\,d\log(\bar z_i-\bar z_j)
     +\Omega^\vee,
\end{equation}
with $\Omega^\vee$ defined in \eqn{eqn:duallocalsystem}.
The reference factors satisfy
\begin{equation}
    \partial\mathcal I_{\mathbb I_n}
    =\tau\mathcal I_{\mathbb I_n},
    \qquad
    \bar\partial\mathcal I_{\mathbb I_n}^{\vee}
    =\mathcal I_{\mathbb I_n}^{\vee}\tau^\vee,
\end{equation}
and hence
\begin{equation}
    \partial W=\tau W,
    \qquad
    \bar\partial W=W\tau^\vee.
\end{equation}
For any invertible local holomorphic solution $\mathcal I$
of \eqn{noncomtwist}, we define its dual function on the chamber \(\alpha\) by
\begin{equation}
\label{dual}
    \mathcal I^\vee_{\alpha}(\bar z)
    :=\mathcal I^{-1}_{\alpha}(z)W(z,\bar z).
\end{equation}
The same $W$ is used for all chambers. Thus the chosen
holomorphic loading is retained, while its dual is determined
by the requirement $\mathcal I\mathcal I^\vee=W$.
In particular, the dual factor is matched to the fixed reference
normalization, rather than obtained from a separately normalized
single-valued completion in each chamber.

Indeed, \eqn{dual} gives
\begin{align}
    \partial\mathcal I^\vee
    &=-\mathcal I^{-1}\tau W
      +\mathcal I^{-1}\partial W=0,
    \\
    \bar\partial\mathcal I^\vee
    &=\mathcal I^{-1}\bar\partial W
      =\mathcal I^\vee\tau^\vee.
\end{align}
Consequently, all dual loadings are antiholomorphic solutions
of the same right differential equation
\begin{equation}
\label{duallocalsystem}
    \di \mathcal I^\vee=\mathcal I^\vee\tau^\vee.
\end{equation}
The dual loading itself is generally multivalued; its product
with the corresponding holomorphic loading is single-valued. On an overlap where
$\mathcal I_i=\mathcal I_jM_{ij}$, with
$M_{ij}\in\mathcal R^\times$ constant, the dual loadings obey
\begin{equation}
    \mathcal I_i^\vee
    =M_{ij}^{-1}\mathcal I_j^\vee.
\end{equation}
Thus a change of holomorphic normalization by constant right
multiplication induces the inverse left multiplication on the
dual factor, leaving $W$ and $\tau^\vee$ unchanged.
These dual loadings therefore define a common rank-one left
$\mathcal R$-module local system $\mathcal L_{\tau^\vee}$.
The chamber loadings used in this paper will be specified
in \sec{basis}.

The corresponding logarithmic derivative is given by \cite{Kakkad:2025klm}
\begin{equation}
    \tau^\vee=(\mathcal I^\vee)^{-1}\di \mathcal I^\vee.
\end{equation}
For an $\mathcal R$-valued differential form $\omega$, we have
\begin{equation}
    \di (\mathcal I^\vee\omega)
    =\mathcal I^\vee
      \bigl(\di \omega+\tau^\vee\wedge\omega\bigr).
\end{equation}
This defines the dual twisted differential
\begin{equation}
    \nabla_{\tau^\vee}\omega
    :=d\omega+\tau^\vee\wedge\omega.
\end{equation}
Accordingly, the dual twisted homology
$H_p(X,\mathcal L_{\tau^\vee})$ carries a left
$\mathcal R$-module structure, while the dual twisted
cohomology $H^p(X,\nabla_{\tau^\vee})$ carries a right
$\mathcal R$-module structure.

\paragraph{Duality network.}
There exists a nondegenerate twisted period pairing between twisted homology and cohomology \cite{Mizera:2017cqs,Kakkad:2025klm,Brown:2019wna},
\[
H_m(X,\mathcal L_\tau)\times H^m(X,\nabla_\tau)
\longrightarrow \mathcal R,
\]
together with its counterpart for the dual local system,
\[
H_m(X,\mathcal L_{\tau^\vee})\times H^m(X,\nabla_{\tau^\vee})
\longrightarrow \mathcal R.
\]
In addition, the twisted homology groups admit an intersection pairing \cite{mana.19941660122}
\[
H_m(X,\mathcal L_\tau)\times H_m(X,\mathcal L_{\tau^\vee})
\longrightarrow \mathcal R,
\]
while the corresponding intersection pairing between twisted cohomology groups is given by \cite{ChoMatsumoto:1995periods}
\[
H^m(X,\nabla_\tau)\times H^m(X,\nabla_{\tau^\vee})
\longrightarrow \mathcal R.
\]

In the noncommutative setting, the twisted pairing becomes an \(\mathcal R\)-valued pairing between a left \(\mathcal R\)-module and a right \(\mathcal R\)-module \cite{Kakkad:2025klm}, defined by
\begin{equation}
    \langle\omega,\mathscr C\rangle
=\int_{\mathscr C}\omega
=\int_C\omega\,\mathcal I_C,
\qquad
\omega\in H^m(X,\nabla_\tau),\quad
\mathscr C\in H_m(X,\mathcal L_\tau).
\end{equation}
The \(\mathcal R\)-bilinearity of the pairing is expressed as
\begin{equation}
    \langle a\omega,\mathscr C b\rangle
=a\langle\omega,\mathscr C\rangle b,
\qquad a,b\in\mathcal R.
\end{equation}
All four pairings introduced above satisfy analogous bilinearity properties with respect to their corresponding left and right \(\mathcal R\)-module structures. We use the same bracket notation for the four pairings involving
twisted homology, twisted cohomology, and their duals. Let
\[
\begin{aligned}
\omega &\in H^m(X,\nabla_\tau),
&\qquad
\mathscr C &\in H_m(X,\mathcal L_\tau),\\
\omega^\vee &\in H^m(X,\nabla_{\tau^\vee}),
&\qquad
\mathscr C^\vee &\in H_m(X,\mathcal L_{\tau^\vee}).
\end{aligned}
\]
The period pairings and the cocycle pairing are defined by
\[
\begin{aligned}
\langle \omega,\mathscr C\rangle
&:= \int_{\mathscr C}\omega,\\
\langle \mathscr C^\vee,\omega^\vee\rangle
&:= \int_{\mathscr C^\vee}\omega^\vee,\\
\langle \omega,\omega^\vee\rangle
&:= \int_X \omega\,\mathcal I
   \wedge \mathcal I^\vee\,\omega^\vee.
\end{aligned}
\]
Throughout, the left and right entries of each bracket belong
to the left and right $\mathcal R$-modules, respectively. For compact twisted cycles intersecting transversely, the
intersection pairing is
\begin{equation}
\label{eq:cyclespairing}
\langle \mathscr C^\vee,\mathscr C\rangle
:=
\sum_{p\in C\cap\bar C}
\epsilon_p\,
\left.
\mathcal I^\vee_{\bar C}
\bigl(\mathcal I_{\bar C}\mathcal I^\vee_{\bar C}\bigr)^{-1}
\mathcal I_C
\right|_p .
\end{equation}
Here $C$ and $\bar C$ denote the geometric supports of
$\mathscr C$ and $\mathscr C^\vee$, respectively, and
$\epsilon_p=\pm1$ is the local oriented intersection sign.
In one complex dimension, our sign convention is
\begin{equation}
\label{eq:IntersectionSigns}
\vcenter{\hbox{\includegraphics[width=0.12\textwidth]{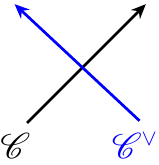}}}
\quad\Longrightarrow\quad +1,
\qquad\qquad
\vcenter{\hbox{\includegraphics[width=0.12\textwidth]{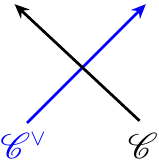}}}
\quad\Longrightarrow\quad -1.
\end{equation}
For non-compact twisted cycle pairing, one must introduce the regularized cycle \cite{Mizera:2017cqs,mana.19941660122,Mimachi:2002gi}, as we will show in the string amplitude case.
\subsection{Bases and chambers}
\label{basis}

{Explicit bases of twisted cycles and cocycles for
genus-zero configuration-space integrals were studied in
\cite{Britto:2021prf}.} 
In our setting up, the twisted homology and cohomology are \(\mathcal R\)-modules of
rank \((n-3)!\). This rank is preserved under the formal
KZ deformation of the usual Koba--Nielsen system,
whose (co)homology is concentrated in degree \(n-3\)
\cite{CohenDimcaOrlik2003}. 
We now briefly review the rank of twisted (co)homology group following \cite{Mizera:2019gea}. 
Let $X=\mathcal M_{0,n}$ and $d=n-3$. Under generic
nonresonant kinematics, which is translated into the generic condition that the kinematic variables \(s_{ij}\) are not integers, the cohomology associated with the Koba--Nielsen factor is concentrated in degree $d$
\cite{amoto1975,CohenDimcaOrlik2003,Mizera:2019gea}. The twisted Euler characteristic coincides with the
ordinary Euler characteristic $\chi(X)$.
Hence
\begin{equation}
    \chi(X)
=\sum_{k=0}^{2d}(-1)^k
 \dim H^k(X,\nabla_{\mathrm{KN}})
=(-1)^d\dim H^d(X,\nabla_{\mathrm{KN}}).
\end{equation}
Evaluating the Poincar\'e polynomial
$P_X(t)=\prod_{j=2}^{n-2}(1+jt)$ at $t=-1$ gives the Euler characteristic 
\[
\chi(X)=(-1)^d d!,
\]
and we get
\begin{equation}
    \qquad
\dim H^d(X,\nabla_{\mathrm{KN}})=(n-3)!.
\end{equation}
The corresponding twisted homology has the same
dimension by duality. Twisted
cohomology remains concentrated in degree $d$ when it takes matrix values \cite{CohenDimcaOrlik2003}, yielding twisted
(co)homology modules of rank $(n-3)!$ over $\mathcal R$ in our KZ setting.

For the application to string amplitudes, bases are chosen for the
twisted cycles and cocycles. Fixing the \(\text{SL}(2,\mathbb R)\)
redundancy by setting \(z_1=0\), \(z_{n-1}=1\), and \(z_n=\infty\),
we define the Koba--Nielsen factor on the real chamber of ordering
\(\beta\) as
\begin{equation}
\mathrm{KN}_\beta(z)
:={}
\prod_{i=2}^{n-2}
|z_i|^{s_{1i}}
|1-z_i|^{s_{i,n-1}}
\times
\prod_{2\leq i<j\leq n-2}
|z_j-z_i|^{s_{ij}}.
\end{equation}
These real-chamber values determine the holomorphic branch of
\(\mathrm{KN}_\beta\).
{In the flat-space case, the interiors of the associahedron, loaded with the Koba--Nielsen factors, 
\begin{equation}
    \tilde{\mathbf C}(\alpha)
=K^\circ_{n-1}(\alpha)\otimes\mathrm{KN}_\alpha,
\end{equation}
where
\begin{equation}
    K^\circ_{n-1}(\alpha)
=
\left\{
0<z_{\alpha(2)}<z_{\alpha(3)}<\cdots
<z_{\alpha(n-2)}<1
\right\}
\subset\mathcal M_{0,n}(\mathbb R),
\end{equation}
defines the loaded chambers on the blown-up moduli space.}
Its closure in
\(\overline{\mathcal M}_{0,n}(\mathbb R)\)
is the associahedron \(K_{n-1}(\alpha)\). 
The set of permutations
\begin{equation}
    \{\tilde{\mathbf{C}}(\beta)\,|\,\beta \in (1,{\cal P}(2,3,\dots n-2),n-1,n)\}
\end{equation}
forms a basis of the twisted homology group \cite{Mizera:2017cqs,Casali:2019ihm}.

\begin{figure}[t]
\centering
\includegraphics[width=0.6\textwidth]{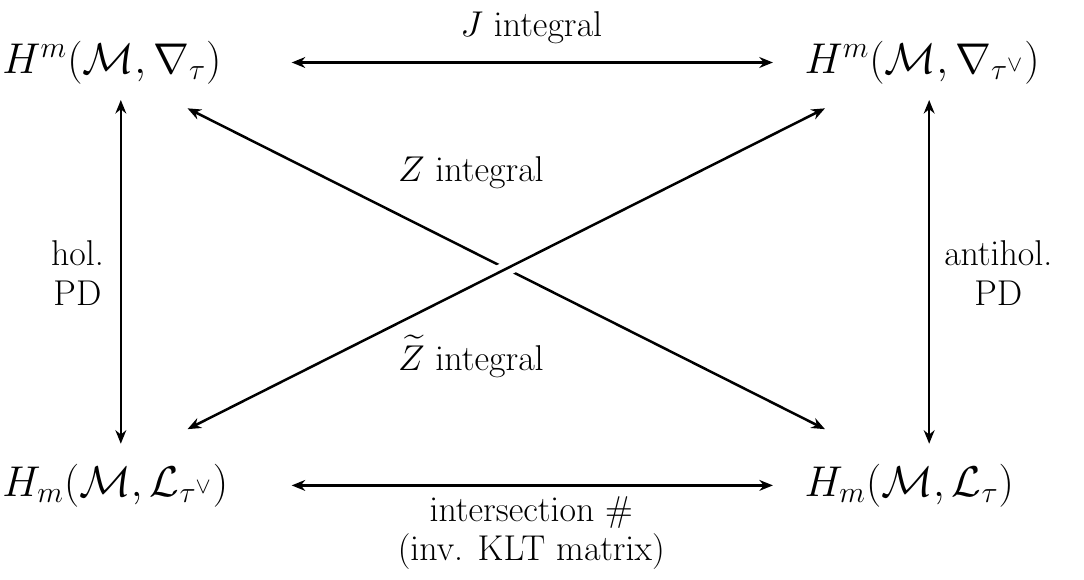}
\vspace{-5pt}
\caption{All relevant pairings of twisted de Rham theory as applied to string amplitudes.}
\label{fig:Dualities}
\end{figure}

We now specify the chamber loadings used in the period
and intersection matrices for the KZ setup. For each chamber
$K^\circ_{n-1}(\beta)$, let $G_\beta$ be the canonical KZ
solution normalized at the prescribed left-comb fusion
tree for the ordering $\beta$. On the non-dual side,
we use the \emph{standard loading}
\begin{equation}
    {\cal I}(\beta) =\mathcal I_\beta(z)
    := \mathrm{KN}_\beta(z)\,G_\beta(z),
\end{equation}
On the dual side, the loading is determined by the
construction in \eqref{dual}. With the same fixed
single-valued integrand
$W=\operatorname{sv}\mathcal I_{\mathbb I_n}$
in every chamber, we set
\begin{equation}
    {\cal I}^\vee (\beta) = \mathcal I_\beta^\vee(\bar z)
    := \mathcal I_\beta(z)^{-1}W(z,\bar z),
    \qquad
    {\cal I}(\beta) {\cal I}^\vee(\beta)= \mathcal I_\beta(z)\mathcal I_\beta^\vee(\bar z)
    =W(z,\bar z).
\end{equation}
Thus the dual loading is fixed by the standard
holomorphic loading and the common single-valued
normalization; no independent chamber normalization
is imposed on the dual side. The corresponding locally finite twisted cycles are
\begin{equation}
\begin{aligned}
    \mathcal C(\beta)
    &:=K^o_{n-1}(\beta)\otimes\mathcal I(\beta),\\
    \mathcal C^\vee(\beta)
    &:=K^o_{n-1}(\beta)\otimes\mathcal I^\vee(\beta).
\end{aligned}
\end{equation}
Their compact representatives, when required for
intersection pairings, are obtained by regularization
as described in \sec{regularization}.
The set forms the basis of the right \(\cal R\)-module 
\begin{equation}
    \{{\cal C}(\beta)\,|\,\beta \in (1,{\cal P}(2,3,\dots n-2),n-1,n)\}.
\end{equation}
For the cohomology group \(H^m(X,\nabla_{\tau})\), we still choose the Parke--Taylor form in the flat-space case as a basis \cite{Mizera:2017cqs}
\begin{equation}
    \mathrm{PT}(\beta) = \mathrm{PT}_\beta(z)
    :=
    \frac{
        dz_2\wedge dz_3\wedge\cdots\wedge dz_{n-2}
    }{
        (0-z_{\beta(2)})
        (z_{\beta(2)}-z_{\beta(3)})
        \cdots
        (z_{\beta(n-2)}-1)
    },
\end{equation}
and its dual as a basis of \(H^m(X,\nabla_{\tau^\vee})\)
\begin{equation}
    \mathrm{PT}^\vee(\beta) = \mathrm{PT}_\beta(\bar{z})
    :=
    \frac{
        d\bar{z}_2\wedge d\bar{z}_3\wedge\cdots\wedge d\bar{z}_{n-2}
    }{
        (0-\bar{z}_{\beta(2)})
        (\bar{z}_{\beta(2)}-\bar{z}_{\beta(3)})
        \cdots
        (\bar{z}_{\beta(n-2)}-1)
    }.
\end{equation}
With the above basis, the \((n-3)!\times (n-3)!\) period matrix is given by
\begin{equation}
    P_{\alpha}{}^{\beta}:=
    \<\mathrm{PT}(\alpha),{\cal C}(\beta)\> = 
    \int_{{K}^o_{n-1}(\beta)} {\mathrm{PT}(\alpha ){\cal I}(\beta)}
\end{equation}
which is the $Z$ integral with MMPLs  insertions, the building block for AdS open-string amplitudes \(Z^{\rm (AdS)}(\beta|\alpha)\) in \cite{Nocchi:2026yms}. 
The dual period matrix is given by 
\begin{equation}
    \tilde{P}_{\alpha}{}^\beta 
    := \<{\cal C}^\vee(\alpha),{\mathrm{PT}}^\vee(\beta)\>
    =
    \int_{{K}^o_{n-1}(\alpha)}
    {\cal I}^\vee(\alpha) \mathrm{PT}^\vee(\beta)
\end{equation}
which is the corresponding dual $Z$ integral \({\tilde{Z}}^{\rm (AdS)}(\alpha|\beta)\) in \cite{Nocchi:2026yms}. The pairing matrix between twisted cocycles gives 
\begin{equation}
    F_{\alpha}{}^\beta 
    := \<\mathrm{PT}(\alpha),\mathrm{PT}^\vee(\beta)\> = \int_X {\mathrm{PT}(\alpha) {\operatorname{sv}{\cal I}_n}\wedge\mathrm{PT}^\vee(\beta)} 
\end{equation}
which is the $J$-integral with single-valued MPL insertions, the building block for AdS closed-string amplitudes \(J^{(\mathrm{AdS})}(\beta|\alpha)\) in \cite{Nocchi:2026yms}. The pairing between twisted cycles gives the definition of the intersection number matrix 
\begin{equation}
    I_{\alpha}{}^\beta =
    \<{{\cal C}}^\vee(\alpha),  \mathrm{reg}\,{\cal C}(\beta)\> 
    = \sum_{p_i=\tilde{K}_{n-1}(\alpha)\cap {K_{n-1}}(\beta)} 
    \pm {\cal I}^\vee_{\alpha
    }({\cal I I^\vee})^{-1} {\cal I}_{\beta} R_{p_i}.
\end{equation}
Please note that the pairing of twisted cycles requires at least one of them to be compact, while our bases for string amplitudes are non-compact. Therefore we use the regularized cycle \(\mathrm{reg}\, \cal C(\beta)\) in the pairing and \(R_{p_i}\) is the factor needed for the regularization.
We will talk about this problem in detail in \sec{regularization}. Here, $\widetilde{K}_{n-1}(\alpha)$ denotes a deformation
of the unregularized associahedron $K_{n-1}(\alpha)$ \cite{Mizera:2017cqs}.  Figure~\ref{fig:Dualities} provides an overview of the
various pairings appearing in string amplitudes,
adapted from \cite{Kakkad:2025klm}.

In the flat-space setting, the compatibility of period
and intersection pairings is expressed by twisted Riemann period
relations \cite{ChoMatsumoto:1995periods}.
In \cite{Kakkad:2025klm}, it was shown that Poincaré duality and the
bilinearity of the pairing between modules imply the following matrix identity, namely the twisted period relation:
\begin{equation}
    \mathbf{F} = \mathbf{P}\mathbf{I}^{-1}\tilde{\mathbf{P}},
\end{equation}
which gives the KLT relations of AdS building blocks. Although the explicit intersection-number calculation in
\cite{Kakkad:2025klm} is restricted to four points, the twisted
period relation is derived in a general setting.
We refer the reader to that work for the derivation and
further details.

The intersection number matrix is geometrically the intersection of different chambers of the moduli space \(\overline{\cal M}_{0,n}\). For \(\overline{\cal M}_{0,n}\), the number of chambers is \((n-1)!/2\). The chambers labelled by cyclically equivalent permutations introduce an ambiguity in the KZ setting. 
The adjoint chambers share the divisor, or the face of associahedra, where the intersection happens. 

For the simplicity of the result of the intersection number, we make such a convention for the integrand loaded on the chamber. First, we choose the identity permutation as the reference chamber \(K_{n-1}(\mathbb{I}_n)\) with the left-comb tree integrand. The adjoint chamber, whose intersection with the reference chamber is not zero, is obtained from the mirror transformation for the \(I\) subtree corresponding to the divisor \(D_I\) with the reversal and the cyclic transformation. For example, normal to the boundary \((123)\) of the associahedron \(K_4(\mathbb{I}_5)\), the mirror subtree gives the \[{\cal I}_{((3(21))4)5}\]
which is not the left-comb tree for the ordering \((32145)\). The chambers intersecting at lower-codimension faces can be constructed recursively, like the chamber intersecting \(K_4 (\mathbb{I}_5)\) at the vertices \((12)\cap(123)\). 
It can be obtained from the chamber 21345 via the boundary \((213)\) 
\[(((21)3)4)5 \to ((3(12))4)5\]
or from the chamber 32145 via the boundary \((12)\)
\[((3(21))4)5 \to ((3(12))4)5.\]
The cyclic and reflection symmetries of \(\mathcal I\) \cite{Nocchi:2026yms} ensure that this choice of loading is compatible with our previously chosen basis with only constant factor difference. The mirror operation specifies the adjacent ordering
and an intermediate fusion tree. The resulting loading
is then transported to the canonical left-comb
normalization for that ordering, in accordance with
the standard-loading convention above.

{
The two entries of the intersection pairing need not use
the same chamber basis. For the explicit calculations below, let \(\sigma \in S_{n-3}\)
we choose orderings
\[
\begin{aligned}
\alpha_\sigma
&=(1,\sigma(2),\ldots,\sigma(n-2),n-1,n),\\
\beta_\sigma
&=(\sigma(n-2),\ldots,\sigma(2),1,n-1,n),
\end{aligned}
\]
and use the dual cycles labelled by \(\alpha_\sigma\)
and the non-dual cycles labelled by \(\beta_\sigma\).
These collections provide bases on the respective sides. 
The orderings \(\beta_\sigma\) are obtained by reversing
the consecutive block
\((1,\sigma(2),\ldots,\sigma(n-2))\), while keeping
\(n-1\) and \(n\) fixed.
All chamber loadings retain the prescribed left-comb
normalization.}

In the next section, we show how to compute the intersection number in the KZ framework which is identified as the inverse KLT kernel of AdS building blocks.

\section{Inverse KLT kernel and intersection numbers}
\label{inter}
\subsection{Regularization of twisted cycles}
\label{regularization}
\begin{figure}[t]
\centering
\includegraphics[width=0.8\textwidth]{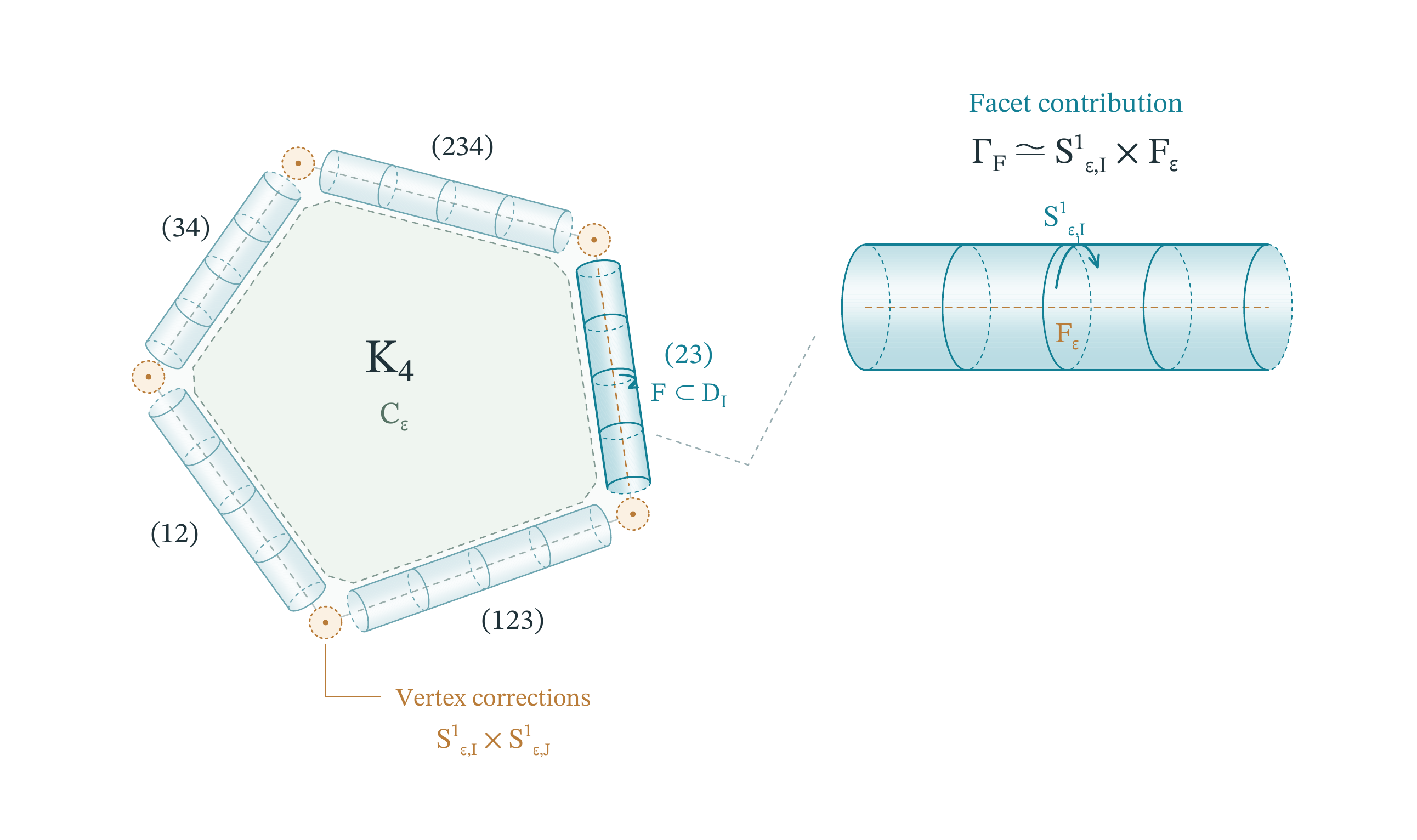}
\vspace{-5pt}
\caption{Regularization of \(K_4(\mathbb I_n)\)}
\label{fig:K4reg}
\end{figure}

For the pairing between twisted cycles in \eqref{eq:cyclespairing} to be well defined, at least one of the two twisted cycles must be compact, or more generally have compact support \cite{mana.19941660122,Aomoto:2011ggg}. However, the manifold $X$ relevant to string amplitudes is
generally non-compact. We therefore consider the locally finite
twisted homology group $H_m^{\mathrm{lf}}(X,\mathcal L_\tau)$,
defined using a locally finite open cover of $X$
\cite{Mizera:2017cqs}.
The loaded chamber cycles chosen above form a basis of this group. We introduce the
regularization map to ordinary twisted homology
\begin{equation}
    \text{Reg}: H^{\text{lf}}_m(X,{\cal L}_\tau)\to H_m(X,{\cal L}_\tau).
\end{equation}

The regularization proceeds by cancelling the twisted boundary
of a truncated associahedron. Let
\(d=\dim K_{n-1}=n-3\).
After removing small neighbourhoods of its boundary, we attach
a tube along each truncated facet by taking a small circle
in the complex normal direction to the corresponding divisor.
Each circle is one-dimensional, while the resulting tube has dimension \(d\). With suitable orientations and inverse
monodromy factors, the twisted boundaries of these tubes
cancel the facet contributions from the truncated associahedron
\cite{Mizera:2017cqs,Kakkad:2025klm}. An illustration for $n=5$ is shown in \fig{fig:K4reg}.

The tubes give rise to additional boundary terms near codimension-two faces, which are cancelled by further corrections involving the two normal circles arising from the intersecting codimension-one faces. Generally, a codimension-\(k\) face \(F\) requires
a correction locally modelled on
\[
    F_\varepsilon\times(S^1)^k,
\]
where \(F_\varepsilon\) denotes a truncated copy of \(F\).
All such corrections have the same dimension as the associahedron itself,
\[
    \dim F_\varepsilon+k=(d-k)+k=d.
\]
At each codimension, the corresponding corrections cancel the residual twisted boundary terms arising from the regularization at the preceding codimension. Following \cite{Mizera:2017cqs}, we assume generic kinematics
such that $s_I\notin\mathbb Z$ for every boundary channel $I$. With the contribution of the Koba--Nielsen factor, the relevant
monodromy differences \(M_I-1\) are always invertible. Consequently, with the inverse of the monodromy factor, the resulting
chain is both compact and closed, and therefore defines a compact twisted \(m\)-cycle on \(X\) for computing intersection numbers.

We first introduce the regularization contribution associated
with a single facet
\(F_I=K_{n-1} \cap D_I\).
Let \(\mathscr C= F_{I} \otimes\mathcal I_s\), where \(s\) denotes
the reference fusion tree of the chamber. With our transport convention
\(
\mathcal{I}_{{I}}
=
\mathcal I_s\,\mathcal T_{s\to I},
\)
the monodromy around \(F_I\), expressed in the reference
frame, is
\begin{equation}
M_{I,s}
=
\mathcal T_{s\to I}\,
e^{2\pi i E_I}\,
\mathcal T_{s\to I}^{-1}.
\end{equation}
Here \(E_I=s_I\mathbf1+e_I\) is the full residue, including
the Koba--Nielsen contribution. Since \(e^{2\pi iE_I}-\mathbf1\) is invertible,
the corresponding regularization factor needed is given by
\begin{equation}
R_{I,s}
:=
(M_{I,s}-\mathbf1)^{-1}
=
\mathcal T_{s\to I}\,
\bigl(e^{2\pi iE_I}-\mathbf1\bigr)^{-1}
\mathcal T_{s\to I}^{-1}.
\end{equation}

Let \(K_{\varepsilon}\) be the truncated chamber, namely the interior part of \(K_{n-1}\) and let
\(F_{I,\varepsilon}\) denote its boundary piece
corresponding to \(F_I\), equipped with the induced
boundary orientation. We attach the normal tube
\[
\Gamma_I
\simeq
S^1_{\varepsilon,I}\times F_{I,\varepsilon},
\]
with the circle factor placed first in the product
orientation. The circle is positively oriented around
\(F_I\) and is regarded as a based loop along which the
loading is analytically continued. Writing \(\mathcal I_s^{\Gamma_I}\) for this continued
loading, the facet correction of \(F_{I}\) is
\begin{equation}
\mathscr C_I^{\mathrm{corr}}
=
-\bigl(\Gamma_I\otimes\mathcal I_s^{\Gamma_I}\bigr)
R_{I,s}.
\end{equation}
Transporting the loading once around the normal circle,
thereby crossing the branch cut, transforms
$\mathcal I_s$ into $\mathcal I_s M_{I,s}$.
This monodromy manifests itself in the twisted boundary
as a factor of $M_{I,s}-\mathbf 1$,
which is cancelled by the regularization factor $R_{I,s}$.
Consequently,
\[
\partial_\tau\mathscr C_I^{\mathrm{corr}}
=
-F_{I,\varepsilon}\otimes\mathcal I_s
+
\text{terms supported near }
\partial F_{I,\varepsilon}.
\]
The first term cancels the corresponding boundary
contribution of the truncated associahedron. The remaining
terms will be cancelled by corrections associated
with higher-codimension faces.

\paragraph{Higher dimensional faces.} Now let us consider the regularization of the codimension-\(k\) face \(F\). Let 
\begin{equation}
    {\cal N} = \{I_1,\dots,I_k\}
\end{equation}
a planar nested set, 
and clusters in \({\cal N}\) are pairwise compatible, which determines the codimension-\(k\) face \(F_{\cal N}(\alpha)\). 
\begin{equation}
F_{\mathcal N}(\alpha)
=
\bigcap_{I\in\mathcal N}F_I(\alpha)
=
K_{n-1}(\alpha)\cap\bigcap_{I\in\mathcal N}D_I.
\end{equation}

As the intersection of facets, \(F_{\cal N}\) inherits the tube and the regularization factor of the facet. 
We are going to show the coherence of the product of regularization factors. 
Without loss of generality, let us first consider  the intersection between \(F_{I_1}\) and \(F_{I_2}\). The compatibility between clusters \(I_1\) and \(I_2\) allows three possibilities, \(I_1\subset I_2\), \(I_2\subset I_1\) and \(I_1\cap I_2 =\varnothing\).

Now let us first consider the disjoint case \(I_1\cap I_2 = \varnothing \). We have the commutation relation for the cluster generators
\[[E_{I_1},E_{I_2}] = 0.\]
The regularization factor for facet \(F_{I_1}\) is given by
\[R_{I_1,s} = 
{\cal T}_{ s\to I_1} 
\left(e^{2\pi i E_{I_1}}-1\right)^{-1} {\cal T}_{I_1\to s}\]
And for  \( F_{I_2} \), it is given by 
\[R_{I_2,s}=
{\cal T}_{ s \to I_2}
\left(e^{2\pi i E_{I_2}}-1\right)^{-1} {\cal T}_{I_2 \to s}
\]
The regularization factors associated with the two faces
$F_{I_1}$ and $F_{I_2}$ can be multiplied in either order.
The first ordering gives
\begin{equation}
\begin{aligned}
R_{I_1,s} R_{I_2,s}
={}&
    {\cal T}_{s \to I_1}
    \left(e^{2\pi i E_{I_1}}-\mathbf{1}\right)^{-1}
    {\cal T}_{I_1 \to I_2}
    \left(e^{2\pi i E_{I_2}}-\mathbf{1}\right)^{-1}
    {\cal T}_{I_2 \to s},
\end{aligned}
\label{eq:order1}
\end{equation}
while reversing the order gives
\begin{equation}
\begin{aligned}
R_{I_2,s} R_{I_1,s}
={}&
    {\cal T}_{s \to I_2}
    \left(e^{2\pi i E_{I_2}}-\mathbf{1}\right)^{-1}
    {\cal T}_{I_2 \to I_1}
    \left(e^{2\pi i E_{I_1}}-\mathbf{1}\right)^{-1}
    {\cal T}_{I_1 \to s}.
\end{aligned}
\label{eq:order2}
\end{equation}
Here the intermediate transport operators are defined
by composition through the reference configuration $s$,
\begin{equation}
    {\cal T}_{I_1 \to I_2}
    = {\cal T}_{I_1 \to s}\,
      {\cal T}_{s \to I_2},
    \qquad
    {\cal T}_{I_2 \to I_1}
    = {\cal T}_{I_2 \to s}\,
      {\cal T}_{s \to I_1}.
\end{equation}
To show the coherence of the two products, let us first suppose that \(I_1\) lies to the left of
\[
I_2=(b_1,\ldots,b_r),
\]
and let
\[
x_1,\ldots,x_m
\]
be the labels lying between \(I_1\) and \(I_2\). We denote by \(A\) the left-comb prefix ending at the cable \([I_1]\). Thus \(I_1\) occurs as an intact subtree of \(A\). If no labels occur to the left of \(I_1\), then \(A=[I_1]\). Define
\[
P
:=
\operatorname{LC}(A,x_1,\ldots,x_m)
=
(((A x_1)x_2)\cdots x_m).
\]Hence the relevant part of the ordering is
\[
(A,x_1,\ldots,x_m,b_1,\ldots,b_r),
\]while all labels lying to the right of \(I_2\) will be suppressed, since the following transformations act trivially on them. The relevant part of the \({I_1}\)-adapted left-comb tree is
\[
\left.T_{I_1}\right|_{\mathrm{rel}}
=
(((P b_1)b_2)\cdots b_r).
\]In this tree, \(I_2\) does not yet occur as a cable. Indeed, the intermediate subtrees are
\[
P b_1,\qquad
(P b_1)b_2,\qquad\ldots,
\]rather than
\[
b_1b_2,\qquad
(b_1b_2)b_3,\qquad\ldots.
\]
Therefore, the transformation to an \({I_2}\)-adapted tree cannot in general be represented by a single reassociation
\[
(AB)C\longrightarrow A(BC)
\]with \(B\cup C=I_2\). Instead, the left-comb subtree of \(I_2\) must be constructed one label at a time.

Let us define
\[
L_1:=b_1,
\qquad
L_k:=(L_{k-1}b_k),
\qquad 2\leq k\leq r.
\]Thus
\[
L_r
=
(((b_1b_2)b_3)\cdots b_r)
=
\operatorname{LC}(I_2).
\]
For \(k=1,\ldots,r\), define the intermediate tree
\[
T^{(k)}
=
(((P L_k)b_{k+1})\cdots b_r).
\]
For every \(k=2,\ldots,r\), the transformation
\[
T^{(k-1)}\longrightarrow
T^{(k)}
\]is the elementary reassociation
\[
((P L_{k-1})b_k)
\longrightarrow
P(L_{k-1}b_k)
=
P L_k.
\]The corresponding cabled associator is
\[
\widehat\Phi_k
:=
\Phi_{P,L_{k-1},b_k}
=
\Phi\left(
e_{L_{k-1},b_k},e_{P,L_{k-1}}
\right).
\]
Since successive transports multiply on the right, the
complete cable-preserving transformation is
\[
\begin{aligned}
\mathcal T_{I_1\to I'_2}
&=
\widehat\Phi_2
\widehat\Phi_3
\cdots
\widehat\Phi_r
\\
&=
\Phi_{P,b_1,b_2}
\Phi_{P,L_2,b_3}
\cdots
\Phi_{P,L_{r-1},b_r}.
\end{aligned}
\]
At the \(k\)-th step, the three blocks in the elementary reassociation are
\[
A=P,\qquad B=L_{k-1},\qquad C=b_k.
\]Consequently,
\[
B\cup C=L_k,
\]which equals the whole cluster \(I_2\) only at the final step \(k=r\).
We now prove that this transport operator commutes with \(E_{I_1}\). For each elementary factor, set
\[
X_k:=e_{P,L_{k-1}},
\qquad
Y_k:=e_{L_{k-1},b_k}.
\]Since \(Y_k\) is entirely inside \(I_2\), we have
\[
[E_{I_1},Y_k]=0.
\]Moreover, because \(I_1\subseteq P\),
\[
X_k
=
e_{I_1,L_{k-1}}
+
e_{P\setminus I_1,L_{k-1}}.
\]The second term has support disjoint from \(I_1\). For the first term, the infinitesimal braid relations give
\[
\begin{aligned}
[e_{I_1},e_{I_1,L_{k-1}}]
&=
\sum_{\ell\in L_{k-1}}
\sum_{\substack{i<j\\ i,j\in I_1}}
[e_{ij},e_{i\ell}+e_{j\ell}]\\
&=0.
\end{aligned}
\]
It follows that
\[
[E_{I_1},X_k]
=
[E_{I_1},Y_k]
=
0,
\]and therefore
\[
[E_{I_1},\widehat\Phi_k]=0.
\]Since every factor in the ordered product has this property,
\[{
\left[
E_{I_1},
\mathcal T_{{I_1}\to  {I'_2}}
\right]=0.
}
\]If \(I_2\) lies to the left of \(I_1\), then the cable \([I_1]\) belongs to the unchanged part of the tree, and the same commutation relation follows directly from disjoint support.

In our all cabled left-comb fusion tree, \({I_2'}\) may not be the same fusion tree for \(I_2\) if there are other labels on the left side of \(I_2\). Let us decompose the transport operator as
\[{{\cal T}_{{I_1}\to {I_2}}} = \mathcal T_{{I_1}\to {I_2'}} 
\mathcal T_{{I_2'}\to {I_2}}.\]
The transport operator that keeps \(I_2\) as a cable commutes with the cluster generators
\[[\mathcal T_{{I'_2} \to  {I_2}},
E_{I_2} ]  =0. \]
We can rewrite \eqn{eq:order2} as 
\[
\begin{aligned}
    &{\cal T}_{ s\to I_2} 
    \left(e^{2\pi i E_{I_2}}-1\right)^{-1}
{\cal T}_{ I_2 \to I_1} 
\left(e^{2\pi i E_{I_1}}-1\right)^{-1} 
{\cal T}_{I_1\to s}\\
&= 
{\cal T}_{ s\to I_2} 
\mathcal T_{{I_2} \to {I'_2}} 
\left(e^{2\pi i E_{I_2}}-1\right)^{-1}
\left(e^{2\pi i E_{I_1}}-1\right)^{-1} 
{\cal T}_{{I_2'}\to  {I_1}} 
{\cal T}_{I_1\to s}\\
&={\cal T}_{s\to {I'_2}}
\left(e^{2\pi i E_{I_2}}-1\right)^{-1}
\left(e^{2\pi i E_{I_1}}-1\right)^{-1}
{\cal T}_{{I'_2}\to {s}}
\end{aligned}
\]
and \eqn{eq:order1} as
\[\begin{aligned}
    &{\cal T}_{s\to I_1} 
\left(e^{2\pi i E_{I_1}}-1\right)^{-1}
{\cal T}_{ {I_1}\to {I_2}} 
\left(e^{2\pi i E_{I_2}}-1\right)^{-1} 
{\cal T}_{ {I_2}\to s}\\
&={\cal T}_{ s \to I_2'}
\left(e^{2\pi i E_{I_1}}-1\right)^{-1}
\left(e^{2\pi i E_{I_2}}-1\right)^{-1}
{\cal T}_{{I'_2} \to s}.
\end{aligned}
\]
Note here the tree denoted above by \(T_{I'_2}\) is obtained by
forming the \(I_2\)-subtree while retaining the
\(I_1\)-cable. It therefore contains both clusters,
and need not coincide with the independently chosen
single-facet tree \(T_{I_2}\).

This motivates the notion of an
\({\mathcal N}\)-adapted fusion tree, where
\(\mathcal N=\{I_1,I_2\}\): a  fusion tree is
adapted to \({\mathcal N}\) if it contains both
\(I_1\) and \(I_2\) as cables.
In the disjoint case, our convention selects such a
tree by taking the left-comb subtree of each cluster
and the left comb of the quotient ordering in which
both clusters are contracted to cables.
We denote the resulting tree by
\(T_{\mathcal N}\).
Thus \(T_{I'_2}=T_{\mathcal N}\) in the preceding
calculation.

For each cluster \(I\) and compatible nested set
\(\mathcal N\), let \(T_I\) and \(T_{\mathcal N}\)
denote the adapted fusion trees selected by the
conventions above.
We use the corresponding cluster labels as
shorthand in transport subscripts; for example,
\[
\mathcal T_{I\to\mathcal N}
:=
\mathcal T_{T_I\to T_{\mathcal N}}.
\]
In particular, \(T_I=T_{\{I\}}\).

We next consider the nested case
\(I_1\subset I_2\).
The definition of an \({\mathcal N}\)-adapted tree
is unchanged, but the left-comb convention must now
be applied recursively. First, we equip \(I_1\) with its left-comb subtree
and contract it to a cable \([I_1]\).
Inside \(I_2\), this gives the ordered quotient blocks
\[
I_2/I_1
:=(I_2\setminus I_1)\cup\{[I_1]\},
\]
with their order inherited from the chamber.
We take the left comb of these blocks and substitute
the \(I_1\)-subtree back into \([I_1]\).
The resulting \(I_2\)-subtree is then treated as a
single cable when forming the left comb of the
remaining outer blocks.
This defines our preferred tree
\(\tau_{\mathcal N}\), which retains both clusters. In particular, the left-comb convention inside
\(I_2\) is imposed on the quotient blocks, rather
than directly on its original labels.
The resulting tree therefore need not coincide
with the single-facet tree \(T_{I_2}\).

Now, let us consider the second case of the compatibility condition. Without loss of generality, we set \(I_1 \subset I_2\) and the commutation relation holds:  
\[[E_{I_1},E_{I_2}] = 0.\]
Using the common adapted tree, we now compare the
products in \eqref{eq:order1} and \eqref{eq:order2}.
We factor the intermediate transport as
\[
\mathcal T_{I_1\to I_2}
=
\mathcal T_{I_1\to {\mathcal N}}
\mathcal T_{{\mathcal N}\to I_2}.
\]
The transport between the \({I_2}\)- and \({\cal N}\)-adapted fusion trees keeps the cluster \(I_2\) as a cable, which commutes with the cabled generators:
\[[E_{I_2},{\cal T}_{{\cal N}\to I_2}]=0\]
The transport between the \(I_1\)- and \({\cal N}\)-adapted fusion trees keeps \(I_1\) as a cable, which can be viewed as 
\[(AB)C \to A(BC)\]
where \(B=I_1\), and \(A\),\(B\)  and \(C\) are pairwise disjoint.  Using the commutation relation
\[[E_{A,B},E_B]=0,\qquad [E_{B,C},E_{B}]=0,\]
we have
\[[E_{I_1},{\cal T}_{I_1 \to {\cal N}}]=0.\]
Changing the order of product in \eqn{eq:order2} gives
\[ 
\begin{aligned}
    {\cal T}_{{s}\to I_2} {\cal T}_{{I_2}\to {{\cal N}}} 
    \left(e^{2\pi i E_{I_2}}-1\right)^{-1}
 \left(e^{2\pi i E_{I_1}}-1\right)^{-1} 
 {\cal T}_{{{\cal N}}\to I_1} 
 {\cal T}_{I_1\to s}\\
 ={\cal T}_{s \to {\cal N}} \left(e^{2\pi i E_{I_2}}-1\right)^{-1}
 \left(e^{2\pi i E_{I_1}}-1\right)^{-1} 
  {\cal T}_{{\cal N}\to s}.
\end{aligned}
\]
Similarly, \eqn{eq:order1} is rewritten as 
\[
\begin{aligned}
    {\cal T}_{s \to {I_1}}
    {\cal T}_{I_1 \to {\cal N}}
    \left(e^{2\pi i E_{I_1}}-1\right)^{-1}
 \left(e^{2\pi i E_{I_2}}-1\right)^{-1}
 {\cal T}_{ {\cal N} \to {I_2}} 
 {\cal T}_{I_2 \to s}\\
 ={\cal T}_{s \to {\cal N}} \left(e^{2\pi i E_{I_2}}-1\right)^{-1}
 \left(e^{2\pi i E_{I_1}}-1\right)^{-1} 
  {\cal T}_{{\cal N}\to s}.
\end{aligned}
\]
Thus, the product of the regularization factors is independent of their ordering, yielding
\[{\cal T}_{s\to {\cal N}} 
\left(e^{2\pi i E_{I_1}}-1\right)^{-1}
\left(e^{2\pi i E_{I_2}}-1\right)^{-1}
{\cal T}_{{\cal N} \to s}.
\]

Extending the above analysis to a collection
$\mathcal N$ of $k$ mutually compatible clusters,
we find that the product of the corresponding
regularization factors takes the form
\begin{equation}
    R_{{\cal N},s}
    ={\cal T}_{s\to {\cal N}} \prod_{I \in {\cal N}} \left(e^{2\pi i E_{I}}-1\right)^{-1} {\cal T}_{{\cal N}\to s},
\end{equation}
where \({\cal N}\) in \(\cal T\) stands for the left-comb tree which keeps all cables for \(I\in{\cal N}\). Now, we have obtained the factors needed for all faces of \(K_{n-1}\), the cycle after regularization can be written as: Combining the contributions from all faces, we obtain
a compact representative of the original locally finite
twisted cycle. Let \(\mathfrak N_\alpha\) denote the set of all planar
nested sets associated with \(K_{n-1}(\alpha)\), including
the empty set. For
\(\mathcal N=\{I_1,\ldots,I_k\}\), choose a fusion tree
adapted to \(\cal N\), and define
\begin{equation}
R_{\mathcal N,s}
:=
\mathcal T_{s\to {\mathcal N}}
\left[
\prod_{I\in\mathcal N}
\bigl(e^{2\pi iE_I}-\mathbf1\bigr)^{-1}
\right]
\mathcal T_{{\mathcal N}\to s}.
\label{eq:face-regularization-factor}
\end{equation}
The factors inside the brackets commute because the
clusters in \(\mathcal N\) are pairwise compatible.

Choose compatible truncations and normal tubes. The
contribution associated with \(F_{\mathcal N}\) is
supported on
\[
\Gamma_{\mathcal N}
\simeq
S^1_{\varepsilon,I_1}\times\cdots\times
S^1_{\varepsilon,I_k}\times F_{\mathcal N,\varepsilon},
\]
where each circle winds positively around the
corresponding divisor. Each circle is treated as a
based loop, and the loading is analytically continued
from the truncated chamber along these loops. We orient \(F_{\mathcal N,\varepsilon}\) by placing
the ordered outward normals to the corresponding
boundary hypersurfaces before its tangent orientation.
The circle factors in \(\Gamma_{\mathcal N}\) are
ordered in the same way and placed before the face
factor. With these conventions, the regularized cycle is
\begin{equation}
{
\operatorname{Reg}_{\varepsilon}\mathscr C
=
\sum_{\mathcal N\in\mathfrak N_\alpha}
(-1)^{|\mathcal N|}
\bigl(
\Gamma_{\mathcal N}
\otimes
\mathcal I_s^{\Gamma_{\mathcal N}}
\bigr)
R_{\mathcal N,s}.
}
\label{eq:regularized-chamber}
\end{equation}
For the empty nested set, we set
\begin{equation}
\Gamma_{\varnothing}=K_\varepsilon,
\qquad
R_{\varnothing,s}=\mathbf1,
\qquad
\mathcal I_s^{\Gamma_{\varnothing}}
=\mathcal I_s|_{K_\varepsilon}.
\end{equation}
Thus the empty-set contribution is the truncated chamber,
while the terms with \(|\mathcal N|=k\) are the corrections
associated with codimension-\(k\) faces. Every term has real dimension
\[
\dim\Gamma_{\mathcal N}
=
k+\dim F_{\mathcal N}
=
n-3.
\]
The twisted boundary terms cancel between contributions
from incident faces, giving
\begin{equation}
    \partial_\tau
\operatorname{Reg}_{\varepsilon}\mathscr C=0.
\end{equation}
Since all relevant monodromy factors are
invertible, this defines a class in
\(H_{m}(X,\mathcal L_\tau)\).
Its homology class is independent of the sufficiently
small truncation parameters.

After the regularization, we use the notation for the pairing of twisted cycles 
\begin{equation}
    \<{\cal C}(\alpha),{\cal C}(\beta)\>:=
    \<{\cal C}^\vee (\alpha),\operatorname{Reg} {\cal C}(\beta)\>.
\end{equation}

\paragraph{Four-point example.}
At four points, the chamber \(K_3(\mathbb I_4)\) is the interval $(0,1)$,
oriented from left to right. Its truncation
$K_\varepsilon=[\varepsilon,1-\varepsilon]$ has boundary
\[
\partial K_\varepsilon=[p_1]-[p_0],
\qquad
p_0=\varepsilon,\qquad p_1=1-\varepsilon.
\]
Thus the two boundary faces, with their induced orientations,
are
\[
F_{0,\varepsilon}=-[p_0],
\qquad
F_{1,\varepsilon}=+[p_1].
\]
Let $S^1_{\varepsilon,0}$ and $S^1_{\varepsilon,1}$ be
positively oriented small circles around $0$ and $1$,
based at $p_0$ and $p_1$, respectively. The corresponding
tubes therefore satisfy, as oriented chains,
\[
\Gamma_0=-S^1_{\varepsilon,0},
\qquad
\Gamma_1=+S^1_{\varepsilon,1}.
\]

Denote the full monodromies in the chamber reference frame by
$ M_0$ and $ M_1$, and set
\[
R_0=(M_0-\mathbf1)^{-1},
\qquad
R_1=(M_1-\mathbf1)^{-1}.
\]
Writing $\mathcal I_j$ for the loading analytically
continued along $S^1_{\varepsilon,j}$, the facet correction
$-(\Gamma_j\otimes\mathcal I^{\Gamma_j})R_j$ gives
\begin{equation}
\begin{aligned}
\operatorname{Reg}_\varepsilon\mathscr C
={}&
\bigl(S^1_{\varepsilon,0}\otimes\mathcal I_0\bigr)R_0
+\bigl(K_\varepsilon\otimes\mathcal I\bigr)
\\
&-
\bigl(S^1_{\varepsilon,1}\otimes\mathcal I_1\bigr)R_1.
\end{aligned}
\label{eq:four-point-regularization-check}
\end{equation}
In particular, the left and right endpoint corrections
carry coefficients $+R_0$ and $-R_1$, respectively. This reproduces the result in \cite{Kakkad:2025klm}.

Indeed, the positively oriented based loops obey
\[
\partial_\tau
\bigl(S^1_{\varepsilon,j}\otimes\mathcal I_j\bigr)
=
\bigl(p_j\otimes\mathcal I(p_j)\bigr)
( M_j-\mathbf1).
\]
Hence
\[
\begin{aligned}
\partial_\tau\operatorname{Reg}_\varepsilon\mathscr C
={}&
\bigl(p_0\otimes\mathcal I(p_0)\bigr)
\bigl[( M_0-\mathbf1)R_0-\mathbf1\bigr]
\\
&+
\bigl(p_1\otimes\mathcal I(p_1)\bigr)
\bigl[\mathbf1-(M_1-\mathbf1)R_1\bigr]
=0.
\end{aligned}
\]
All regularization coefficients multiply on the right,
as required by the right $\mathcal R$-module structure.

\subsection{Intersection numbers}
In the flat-space case, the self-intersection number of the associahedron \(K_{n-1}(\mathbb{I}_n)\) from \cite{mana.19941680111,Mizera:2017cqs} is given by 
\begin{equation}
    \<\textbf{C}(\mathbb{I}_n),\textbf{C}(\mathbb{I}_n)\> = (-1)^{n-1} \sum_k^{n-3}\sum_{F=H_1\cap\cdots\cap H_k} \prod_{i=1}^k (e^{2\pi i s_{H_i}}-1)^{-1}
\label{eqn:flatselfinter}
\end{equation}
which is a sum over contributions from barycenters of all the codimension-\(k\) faces of the associahedron. Obviously, the number comes from the regularization factor of the twisted cycle. 

In the KZ setting, as we showed in the last subsection, since the monodromy factor is local in the fusion tree, the contribution to the intersection number naturally carries information about the transport of different fusion trees:
\begin{equation}
\label{eq:selfintersection}
    H_{K_{n-1}}^{\mathrm{self}}(\mathbb{I}_n) 
    = \<{\cal C}(\mathbb{I}_n),{\cal C}(\mathbb{I}_n)\>=
        (-1)^{n-1}
\sum_{{F_{\cal N}}\leq K_{n-1}(\mathbb I_n)}
\mathcal T_{s \to {\cal N}}
\left[
\prod_{I \in {\cal N}}
(e^{2\pi iE_I}-\mathbf1)^{-1}
\right]
\mathcal T_{{\cal N}\to s}.
\end{equation}
where \(s\) is the left-comb tree of \(\mathbb I_n\).

Let us take the four-point example \cite{Alday:2025cxr,Kakkad:2025klm} where \(n=4\) and \(\mathbb{I}_4 = 1234\), where we have the codimension-one contribution from the vertices of the interval. 
Starting from the left-comb tree $G_{1234}$,
the vertex at $z=0$ corresponds to the original
fusion tree $T_{{\cal N}_1}=s$.
Therefore, the first contribution involves no
Drinfeld associators and is simply given by
\[
    \frac{1}{e^{2\pi i E_0}-1}.
\]
The vertex at \(z=1\) corresponds to the fusion tree \((1(23))4\), hence, a transport operator is needed: 
\[{\cal T}_{s \to {\cal N}_2} = \Phi(e_{1},e_0)\]
and the local intersection number  is given by 
\[\Phi(e_1,e_0) \frac{1}{e^{2\pi i E_1}-1} \Phi(e_0,e_1)\]
With the last intersection point  located at the middle of the whole associahedron \(K_3\), which gives the contribution 1, the self-intersection number is 
\begin{equation}
    \<{\cal C}(\mathbb{I}_4),{\cal C}(\mathbb{I}_4)\> = -\frac{1}{e^{2\pi i E_0}-1}-1-\Phi(e_1,e_0) \frac{1}{e^{2\pi i E_1}-1} \Phi(e_0,e_1)
\end{equation}
where the minus sign is from the combination of the relative orientation and the sign of regularization factors. See \cite{Kakkad:2025klm} for further details. In the flat-space case, Mizera shows that the calculation of the intersection number follows the graphic rule of the bi-adjoint scalar amplitudes, we provide the version with transport operators in 
\app{app:graphic-rule}, which shows that, in general, the intersection number can be written in the form of cotangent propagators.

Now we move to the intersection between different chambers.
The intersection number between different chambers carries information about the relation between the analytic branch and the local branch, which is shown in the formula,
\[I_{\alpha}{}^\beta =
\<{{\cal C}}(\alpha),{\cal C}(\beta)\>
= 
\sum_{p_i = \tilde{K}_{n-1}(\alpha)\cap 
K_{n-1}(\beta)} 
\pm {\cal I}^\vee_{\alpha}({\cal I I^\vee})^{-1} 
{\cal I}_{\beta}R_{p_i}\]
Since the dual MMPLs are constructed to obtain globally
single-valued MMPLs, we can always choose a single chamber for the single-valued MMPLs in the middle.
To put it explicitly, the dual multi-valued function \eqn{dual} loaded on chamber \(\alpha\) is 
\[{\cal I}_{\alpha}^\vee 
= {\cal I}_{\alpha}^{-1}
\operatorname{sv} {\cal I}_{\mathbb{I}_n}
\]
We choose the loading of \(K_{n-1}(\alpha)\) as the local reference frame. Its relative continuation factor is therefore the identity, while the loading of \(K_{n-1}(\beta)\), analytically continued to the same local branch, contributes the transition matrix.
\begin{equation}
\label{eq:intersectionloading}
    \sum_{\tilde{K}_{n-1}(\alpha)\cap K_{n-1}(\beta)} 
    \pm \, {\cal I}^\vee_{\alpha}({\cal I}_{\alpha} {\cal I}^\vee_{\alpha})^{-1} {\cal I}_{\beta} 
    =\sum_{\tilde{K}_{n-1}(\alpha)\cap K_{n-1}(\beta)} 
    \pm \,
    {\cal I}_{\alpha}^{-1} {\cal I}_{\beta}
\end{equation} 
For example, normal to the facet \(F_I\), the difference between the chamber's own loading and the one 
from the adjacent chamber is the analytical continuation factor 
\(e^{i\pi E_{I}}.\)
With the regularization factor and the transport operator, it gives the common contribution of the intersection number normal to the face \(F_I\) \cite{Mizera:2017cqs}
\[\frac{1}{2i}{\cal T}_{s\to I}\frac{1}{\sin{\pi E_I}}{\cal T}_{I\to s}.\]

Suppose that \(K_{n-1}(\alpha)\) and \(K_{n-1}(\beta)\)
intersect along a nonempty face of codimension \(k\).
Let
\[
\mathcal N=\{I_1,\ldots,I_k\}
\]
be the compatible set of clusters labelling the
\(k\) facets of \(K_{n-1}(\alpha)\) that contain this
intersection. Denoting these facets by \(F_I\),
we write the common face as
\begin{equation}
F_{\mathcal N}
:=
K_{n-1}(\alpha)\cap K_{n-1}(\beta)
=
\bigcap_{I\in\mathcal N}F_I.
\end{equation}
The face \(F_{\mathcal N}\) belongs to both
associahedra, and the clusters in \(\mathcal N\)
specify the collision channels common to the two
chambers. Since the facets are mutually compatible,
the corresponding clusters in $\mathcal N$
are pairwise either nested or disjoint.
By the cabled Drinfeld coherence relations
discussed in \Sec{sec:2.3}, the corresponding
operators commute:
\[
    [E_I,E_J]=0,\qquad I,J\in\mathcal N.
\]

To separate the normal and tangential contributions for the intersection by the face \(F_{\cal N}\),
we first consider an individual intersection point.
Let \(p_i\) be an intersection point associated with
a regularization term supported near a subface
\(F_{\mathcal N_i}\subseteq F_{\mathcal N}\).
The corresponding nested sets satisfy
\[
\mathcal N\subseteq\mathcal N_i,
\qquad
\mathcal J_i:=\mathcal N_i\setminus\mathcal N.
\]
The channels in \(\mathcal N\) are normal to the
common face \(F_{\mathcal N}\), whereas those in
\(\mathcal J_i\) describe its further boundary
degenerations. 

Choose an \(\mathcal N\)-adapted tree
\(T_{\mathcal N}^{\alpha}\) on the \(\alpha\)-side.
The chosen normal continuation determines the
corresponding tree on the \(\beta\)-side:
\[
T_{\mathcal N}^{\beta}
:=
\operatorname{Mirr}_{\mathcal N}
\bigl(T_{\mathcal N}^{\alpha}\bigr),
\]
where \(\operatorname{Mirr}_{\mathcal N}\) denotes
the subtree mirror operations prescribed by the
normal continuation along the channels
\(I\in\mathcal N\).
Thus the superscript \(\beta\) indicates the chamber
containing the resulting tree; it does not specify
an independently chosen left-comb representative.

For the local regularization term, we choose an
\(\mathcal N_i\)-adapted tree
\(T_{\mathcal N_i}^{\beta}\).
Since \(\mathcal N\subseteq\mathcal N_i\), this tree
also retains every cluster as a cable in \(\mathcal N\). Let \(B_{\mathcal N}\) denote the normal continuation
between the two \({\mathcal N}\)-adapted frames.
With the positive half-turn convention used above,
\[
B_{\mathcal N}
=
\prod_{I\in\mathcal N}e^{\pi iE_I}.
\]
The factors commute because the clusters are compatible. The remaining continuation takes place in the
tangential problem on \(F_{\mathcal N}\).
We write it as the reference fusion-tree transport
from \(T_{\mathcal N}^{\beta}\) to
\(T_{{\cal N}_i}^\beta\), followed by an additional monodromy
\(M_i\), expressed in the \(T_{{\cal N}_i}^\beta\)-frame.
Thus, using \eqref{eq:intersectionloading},
the local loading difference is
\[
\begin{aligned}
L_i
&:=
\mathcal I_\alpha(p_i)^{-1}\mathcal I_\beta(p_i)
\\
&=
\mathcal T_{\alpha\to\mathcal N^\alpha}
B_{\mathcal N}
\mathcal T_{\mathcal N^\beta\to\mathcal N_i^\beta}
M_i
\mathcal T_{\mathcal N_i^\beta\to\beta}.
\end{aligned}
\]
Here the loadings are continued to the local branches
used at \(p_i\).
A deformation of the \(\alpha\)-side  
\(K_{n-1}(\alpha)\) can be chosen, with its loading transported along
the deformation, that preserves its twisted homology
class and introduces no additional tangential winding. This is the choice in the self-intersection number formula \eqn{eqn:flatselfinter}. We use such a  deformation underlying the scalar self-intersection
calculation \cite{mana.19941680111,Mizera:2017cqs}, for which
the local contributions are indexed by all nonempty faces of
\(F_{\mathcal N}\), including \(F_{\mathcal N}\) itself.
This geometric construction is independent of the coefficient
local system, and we apply it to our KZ loadings, giving \(M_i=1\).

The regularization factor for this term, expressed
in the \(\beta\)-frame, is
\[
R_{\mathcal N_i,\beta}
=
\mathcal T_{\beta\to\mathcal N_i^\beta}
R_{\mathcal N_i}
\mathcal T_{\mathcal N_i^\beta\to\beta},
\qquad
R_{\mathcal N_i}
:=
\prod_{I\in\mathcal N_i}
\bigl(e^{2\pi iE_I}-\mathbf1\bigr)^{-1}.
\]
Consequently, the contribution of \(p_i\) is
\begin{equation}
\label{eq:localintersection}
\begin{aligned}
\operatorname{Cont}_{p_i}
&=
\epsilon_i L_iR_{\mathcal N_i,\beta}
\\
&=
\epsilon_i
\mathcal T_{\alpha\to\mathcal N^\alpha}
B_{\mathcal N}
\mathcal T_{\mathcal N^\beta\to\mathcal N_i^\beta}
M_i R_{\mathcal N_i}
\mathcal T_{\mathcal N_i^\beta\to\beta}.
\end{aligned}
\end{equation}
Here \(\epsilon_i\) includes the sign of the
regularization term and the local oriented
intersection sign. Finally, compatibility allows us to split the
regularization factor into its normal and tangential
parts:
\[
R_{\mathcal N_i}
=
\underbrace{
\prod_{I\in\mathcal N}
\bigl(e^{2\pi iE_I}-\mathbf1\bigr)^{-1}
}_{R_{\mathcal N}}
\underbrace{
\prod_{J\in\mathcal J_i}
\bigl(e^{2\pi iE_J}-\mathbf1\bigr)^{-1}
}_{R_{\mathcal J_i}}.
\]
By the argument in \Sec{regularization}, the transport
from \(T_{\mathcal N}^{\beta}\) to
\(T_{\mathcal N_i}^{\beta}\) preserves every cluster
in \(\mathcal N\). Hence
\[
\bigl[
\mathcal T_{\mathcal N^\beta\to\mathcal N_i^\beta},
E_I
\bigr]=0,
\qquad I\in\mathcal N.
\]
The same commutation property holds for the tangential
monodromy \(M_i\).
Combining each normal braid factor with its
regularization factor gives
\[
e^{\pi iE_I}
\bigl(e^{2\pi iE_I}-\mathbf1\bigr)^{-1}
=
\frac{1}{2i}(\sin\pi E_I)^{-1}.
\]
We can therefore rewrite \eqref{eq:localintersection}
as
\begin{equation}
\label{eq:recursivelocal}
\begin{aligned}
\operatorname{Cont}_{p_i}
={}&
\frac{\epsilon_i}{(2i)^k}
\mathcal T_{\alpha\to\mathcal N^\alpha}
\left[
\prod_{I\in\mathcal N}(\sin\pi E_I)^{-1}
\right]
\mathcal T_{\mathcal N^\beta\to\mathcal N_i^\beta}
\\
&\qquad\times
\left[
M_i
\prod_{J\in\mathcal J_i}
\bigl(e^{2\pi iE_J}-\mathbf1\bigr)^{-1}
\right]
\mathcal T_{\mathcal N_i^\beta\to\beta},
\qquad k=|\mathcal N|.
\end{aligned}
\end{equation}
Using the deformation of the \(\alpha\)-side described above, we take \(M_i=\mathbf1\).
Summing the local contributions gives
\begin{equation}
\label{eq:recursive}
\begin{aligned}
\langle\mathcal C(\alpha),\mathcal C(\beta)\rangle
={}&
\frac{1}{(2i)^k}
\mathcal T_{\alpha\to\mathcal N^\alpha}
\left[
\prod_{I\in\mathcal N}(\sin\pi E_I)^{-1}
\right]
\\
&\times
\sum_{F_{\mathcal N_i}\leq F_{\mathcal N}}
\epsilon_i\,
\mathcal T_{\mathcal N^\beta\to\mathcal N_i^\beta}
\left[
\prod_{J\in\mathcal J_i}
\bigl(e^{2\pi iE_J}-\mathbf1\bigr)^{-1}
\right]
\mathcal T_{\mathcal N_i^\beta\to\beta}.
\end{aligned}
\end{equation}
With the compatible orientation convention for the
tangential pairing, this takes the form
\begin{equation}
\begin{aligned}
\boxed{\langle\mathcal C(\alpha),\mathcal C(\beta)\rangle
=
(-1)^{\omega(\alpha|\beta)+1}\left(\frac{i}{2}\right)^k
\mathcal T_{\alpha\to\mathcal N^\alpha}
\left[
\prod_{I\in\mathcal N}(\sin\pi E_I)^{-1}
\right]
H_{F_{\mathcal N}}^{\mathrm{self}}
\bigl(T_{\mathcal N}^{\beta}\bigr)
\mathcal T_{\mathcal N^\beta\to\beta},\quad  k=|\mathcal N|.}
\end{aligned}
\end{equation}
where \(\omega(\alpha|\beta)\) is the relative winding number of the two permutations described and proved to be the exponent of the overall sign of intersection numbers  in \cite{Mizera:2016jhj,Mizera:2017cqs}.
The self-intersection of the face \(F_{\cal N}\) is introduced by analogy with \eqn{eq:selfintersection} where \(\epsilon_i =1\):
\begin{equation}
\begin{aligned}
H_{F_{\mathcal N}}^{\mathrm{self}}
\bigl(T_{\mathcal N}^{\beta}\bigr)
:={}&
(-1)^{d_{F_{\mathcal N}}}
\sum_{F_{\mathcal N_i}\leq F_{\mathcal N}}
\mathcal T_{\mathcal N^\beta\to\mathcal N_i^\beta}
\left[
\prod_{J\in\mathcal J_i}
\bigl(e^{2\pi iE_J}-\mathbf1\bigr)^{-1}
\right]
\mathcal T_{\mathcal N_i^\beta\to\mathcal N^\beta},
\\
d_{F_{\mathcal N}}
={}&n-3-k,
\qquad
\mathcal J_i=\mathcal N_i\setminus\mathcal N.
\end{aligned}
\end{equation}
The tangential self-intersection
\(H_{F_{\mathcal N}}^{\mathrm{self}}
(T_{\mathcal N}^{\beta})\)
is expressed in the reference frame
\(T_{\mathcal N}^{\beta}\), using the same reference
orientation in both entries of the tangential pairing.

\subsection{Five-point example}
\label{sec:five-point}

We illustrate the intersection number calculation at five points.
Let us choose the intersection matrix to have rows labelled by
\(\alpha_1=12345,\alpha_2=13245\) and columns labelled by
\(\beta_1=32145,\beta_2=23145\):
\begin{equation}
    \mathbf I_5=(I_{ab})_{a,b=1,2},
\qquad
I_{ab}:=
\langle\mathcal C(\alpha_a),\mathcal C(\beta_b)\rangle.
\end{equation}
We use the chamber reference loadings fixed above and retain
the order of all noncommuting factors. Write
\[
R_I:=\bigl(e^{2\pi iE_I}-\mathbf1\bigr)^{-1}.
\]
Here labels such as \(12\) and \(123\) denote clusters.
The residue \(E_{123}\) commutes with \(e_{12},e_{13},e_{23}\)
and hence with all associators and regularization factors below.

\paragraph{A common facet.}
First take \(\alpha=\alpha_1=12345\) and
\(\beta=\beta_1=32145\). Their common face is
\[
F_{\mathcal N}
=K_4(\alpha)\cap K_4(\beta)=F_{123},
\qquad
\mathcal N=\{123\}.
\]
The chamber reference fusion trees and the fusion trees after the mirror transformation via the face \(F_{\cal N}\) are
\[
\begin{aligned}
T_\alpha=T_{\mathcal N}^{\alpha}
&=(((12)3)4)5,
&T_\beta&=(((32)1)4)5,
\\
T_{\mathcal N}^{\beta}
&=\operatorname{Mirr}_{\mathcal N}
  (T_{\mathcal N}^{\alpha})
=((3(21))4)5.
\end{aligned}
\]
In particular, the mirror tree \(T_{\mathcal N}^{\beta}\)
is different from the chamber reference tree \(T_\beta\).
The relevant transport operator and the cabled braid operator are given by
\[
\mathcal T_{\alpha\to\mathcal N^\alpha}=\mathbf1,
\qquad
\mathcal T_{\mathcal N^\beta\to\beta}
=\Phi(e_{23},e_{12}),
\qquad
B_{\mathcal N}=e^{\pi iE_{123}}.
\]

\begin{figure}[b]
\centering
\includegraphics[width=0.6\textwidth]{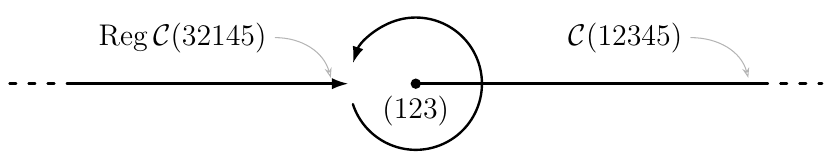}
\vspace{-5pt}
\caption{Normal contours at the common facet \(F_{123}\)}
\label{fig:inside}
\end{figure}

We deform the \(\alpha\)-side simplex, with its loading transported
along the deformation, while keeping the regularized \(\beta\)-cycle
fixed. Choose the deformation without additional monodromy factor,
so that \(M_i=\mathbf1\).
There are three local intersection points, near the relative interior of
\(F_{\mathcal N}\) and its two endpoints:
\[
\begin{array}{c|c|c}
\text{point}&\text{associated face}&\text{nested set}
\\ \hline
p_1&F_{123}&\mathcal N_1=\{123\}
\\
p_2&F_{12}\cap F_{123}&\mathcal N_2=\{12,123\}
\\
p_3&F_{23}\cap F_{123}&\mathcal N_3=\{23,123\}.
\end{array}
\]
The contours are illustrated in \figs{fig:inside}{fig:normal}.
All local contributions below include the regularization and
intersection orientation signs fixed in the preceding subsection.

\begin{figure}[t]
\centering
\includegraphics[width=0.6\textwidth]{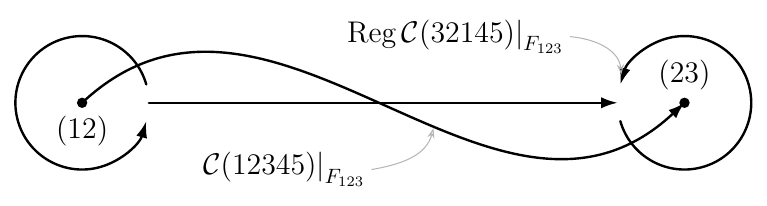}
\vspace{-5pt}
\caption{The three local contributions associated with
\(F_{123}\), \(F_{12}\cap F_{123}\), and
\(F_{23}\cap F_{123}\).}
\label{fig:normal}
\end{figure}

At \(p_1\), 
the loading difference and the regularization factor in this frame are
\[
L_1=B_{\mathcal N}\Phi(e_{23},e_{12}),
\qquad
R_{\mathcal N_1,\beta}=R_{123}.
\]
The contribution is therefore
\[
\operatorname{Cont}_{p_1}
=B_{\mathcal N}\Phi(e_{23},e_{12})R_{123}
=B_{\mathcal N}R_{123}\Phi(e_{23},e_{12}).
\]

At \(p_2\), the \({\cal N}_2\)-adapted fusion tree is
\[
T_{\mathcal N_2}^{\beta}
=((3(21))4)5=T_{\mathcal N}^{\beta},
\qquad
\mathcal T_{\mathcal N_2^\beta\to\beta}
=\Phi(e_{23},e_{12}).
\]
Consequently,
\[
L_2=B_{\mathcal N}\Phi(e_{23},e_{12}),
\qquad
R_{\mathcal N_2,\beta}
=
\Phi(e_{12},e_{23})
R_{123}R_{12}
\Phi(e_{23},e_{12}),
\]
and the adjacent transport factors cancel in the local pairing:
\[
\begin{aligned}
\operatorname{Cont}_{p_2}
&=
B_{\mathcal N}\Phi(e_{23},e_{12})
\bigl[
\Phi(e_{12},e_{23})
R_{123}R_{12}
\Phi(e_{23},e_{12})
\bigr]
\\
&=
B_{\mathcal N}R_{123}R_{12}\Phi(e_{23},e_{12}).
\end{aligned}
\]

At \(p_3\), the \({\cal N}_3\)-adapted regularization tree is already \(T_\beta\):
\[
T_{\mathcal N_3}^{\beta}=(((32)1)4)5,
\qquad
\mathcal T_{\mathcal N^\beta\to\mathcal N_3^\beta}
=\Phi(e_{23},e_{12}),
\qquad
\mathcal T_{\mathcal N_3^\beta\to\beta}=\mathbf1.
\]
Thus
\[
L_3=B_{\mathcal N}\Phi(e_{23},e_{12}),
\qquad
R_{\mathcal N_3,\beta}=R_{123}R_{23},
\]
and
\[
\operatorname{Cont}_{p_3}
=
B_{\mathcal N}R_{123}\Phi(e_{23},e_{12})R_{23}.
\]
Only the channel \(123\) contributes to the normal half-continuation;
the additional channels \(12\) and \(23\) enter through the
regularization at the endpoints of the common face.

We now consider the sign of the intersection points. The \(\alpha\)-chamber
points towards \(F_{123}\) in the normal direction and from
\((12)\) to \((23)\) tangentially.
The normal geometric intersection sign is \(+1\), while the
tangential signs at \((p_1,p_2,p_3)\) are \((-1,-1,+1)\).
The normal regularization contributes a minus sign, which combines
with the tangential regularization signs \((+1,+1,-1)\) to give
\((-1,-1,+1)\).
Thus the combined signs are
\[
\epsilon_1=\epsilon_2=\epsilon_3=+1,
\]
as already incorporated in the local contributions below.

Summing the three contributions gives
\begin{equation}
\label{eq:five-point-I11}
\begin{aligned}
I_{11}
&=
B_{\mathcal N}R_{123}
\bigl[
\Phi(e_{23},e_{12})
+R_{12}\Phi(e_{23},e_{12})
+\Phi(e_{23},e_{12})R_{23}
\bigr]
\\
&=
-\frac14(\sin\pi E_{123})^{-1}
\bigl[
\cot(\pi E_{12})\Phi(e_{23},e_{12})
+\Phi(e_{23},e_{12})\cot(\pi E_{23})
\bigr].
\end{aligned}
\end{equation}
Here we used
\[
B_{\mathcal N}R_{123}
=
\frac{1}{2i}(\sin\pi E_{123})^{-1},
\qquad
R_I=-\frac12\mathbf1-\frac{i}{2}\cot(\pi E_I).
\]
In particular, the two constant terms
\(-\Phi(e_{23},e_{12})/2\) from the endpoint contributions
cancel the interior term \(\Phi(e_{23},e_{12})\). The same result follows from the face recursion.
The common face is an interval, whose tangential self-intersection
in the mirror frame is
\begin{equation}
    \begin{aligned}
H_{F_{\mathcal N}}^{\mathrm{self}}
(T_{\mathcal N}^{\beta})
&=
-\bigl[
\mathbf1+R_{12}
+\Phi(e_{23},e_{12})R_{23}\Phi(e_{12},e_{23})
\bigr]
\\
&=
\frac{i}{2}
\bigl[
\cot(\pi E_{12})
+\Phi(e_{23},e_{12})
 \cot(\pi E_{23})
 \Phi(e_{12},e_{23})
\bigr].
\end{aligned}
\end{equation}
For the reversal of the three-label cluster \(123\), the winding number \cite{Mizera:2016jhj} \[\omega(\alpha|\beta) = \omega(12345|32145)=3\]  and the sign is given
\[
(-1)^{\omega(\alpha|\beta)+1}=1.\]
Hence
\[
I_{11}
=
\frac{i}{2}(\sin\pi E_{123})^{-1}
H_{F_{\mathcal N}}^{\mathrm{self}}
(T_{\mathcal N}^{\beta})
\Phi(e_{23},e_{12}),
\]
which reproduces \eqref{eq:five-point-I11} with the same signs.

\paragraph{A common vertex.}
Next consider \(\alpha_2=13245\) and \(\beta_1=32145\).
Their common face is the vertex
\[
F_{\mathcal N_{\mathrm v}}
=
K_4(\alpha_2)\cap K_4(\beta_1)
=
F_{23}\cap F_{123},
\qquad
\mathcal N_{\mathrm v}=\{23,123\}.
\]
The source reference tree and the adapted trees are
\[
\begin{aligned}
T_{\alpha_2}
&=(((13)2)4)5,
&T_{\mathcal N_{\mathrm v}}^{\alpha_2}
&=((1(32))4)5,
\\
T_{\mathcal N_{\mathrm v}}^{\beta_1}
&=
\operatorname{Mirr}_{\mathcal N_{\mathrm v}}
(T_{\mathcal N_{\mathrm v}}^{\alpha_2})
=
(((32)1)4)5
=
T_{\beta_1}.
\end{aligned}
\]
Thus
\[
\mathcal T_{\alpha_2\to\mathcal N_{\mathrm v}^{\alpha_2}}
=
\Phi(e_{23},e_{13}),
\qquad
\mathcal T_{\mathcal N_{\mathrm v}^{\beta_1}\to\beta_1}
=
\mathbf1.
\]
Both channels now contribute normal half-continuations,
\[
B_{\mathcal N_{\mathrm v}}
=
e^{\pi iE_{123}}e^{\pi iE_{23}}.
\]
The relative sign of the cycles at the vertex is \(-1\). 
Therefore
\begin{equation}
\label{eq:five-point-I21}
\begin{aligned}
I_{21}
&=
-\frac{1}{(2i)^2}
\Phi(e_{23},e_{13})
(\sin\pi E_{123})^{-1}
(\sin\pi E_{23})^{-1}
\\
&=
\frac14
\Phi(e_{23},e_{13})
(\sin\pi E_{123})^{-1}
(\sin\pi E_{23})^{-1}.
\end{aligned}
\end{equation}
Exchanging the labels \(2\leftrightarrow3\) in
\eqref{eq:five-point-I21} and \eqref{eq:five-point-I11}
gives the remaining entries:
\begin{equation}
\label{eq:five-point-remaining}
\begin{aligned}
I_{12}
&=
\frac14
\Phi(e_{23},e_{12})
(\sin\pi E_{123})^{-1}
(\sin\pi E_{23})^{-1},
\\
I_{22}
&=
-\frac14(\sin\pi E_{123})^{-1}
\bigl[
\cot(\pi E_{13})\Phi(e_{23},e_{13})
+\Phi(e_{23},e_{13})\cot(\pi E_{23})
\bigr].
\end{aligned}
\end{equation}
Consequently, the full intersection matrix is
\begin{equation}
\label{eq:five-point-matrix}
\mathbf I_5
=
\begin{pmatrix}
I_{11}&I_{12}\\
I_{21}&I_{22}
\end{pmatrix},
\end{equation}
with entries given in \eqref{eq:five-point-I11},
\eqref{eq:five-point-I21}, and \eqref{eq:five-point-remaining}.
In the normalization of the twisted period relation, the corresponding
kernel is \(\mathbf I_5^{-1}\), with rows labelled by
\((\beta_1,\beta_2)\) and columns by \((\alpha_1,\alpha_2)\).
Setting the KZ generators to zero sends all associators to \(1\)
and recovers the flat-space five-point intersection matrix, including
the overall factor \((i/2)^2=-1/4\) relative to the trigonometric
bi-adjoint matrix \cite{Mizera:2017cqs}.

Our five-point intersection matrix is in exact agreement with the inverse of the KLT kernel obtained in \cite{Nocchi:2026yms}, after accounting for the different basis orderings and normalization conventions. This provides a nontrivial consistency check of our intersection-theoretic construction and establishes its agreement with the independently derived five-point AdS KLT relations.

\subsection{Field-theory limit and bi-adjoint building blocks}
In flat space, the field-theory limit of the inverse KLT kernel gives the bi-adjoint scalar amplitudes. As an analog, we also take the field-theory limit here and try to define what the bi-adjoint building block in AdS spacetime is. 

Let us assume that the KZ generators $e_{ij}$ are proportional to the string length parameter $\alpha'$. Hence all the Drinfeld associators will degenerate to 1, and the only $e_{ij}$ left will come from the trigonometric functions. For example
\[
m(1234|2134)\propto\frac{1}{E_{12}}
\]
Actually, the results are obtained by replacing $s_{ij}$ with $E_{ij}$ in the bi-adjoint scalar amplitudes in flat spacetime, and hence they admit a Berends-Giele (BG) recursion \cite{Berends:1987me} after introducing $\phi(P|Q)$, which is the analog of bi-adjoint scalar BG currents \cite{Mafra:2016ltu}:
\[
E_P\phi(P|Q)=\sum_{P=XY,Q=RS}\phi(X|R)\phi(Y|S)-\phi(Y|R)\phi(X|S)
\]
and the final building block is
\[
m(Pn|Qn)=\lim_{E_P\to0}E_P\phi(P|Q)\phi(n|n)
\]
where the single-particle state $\phi(i|j)$ is defined to be $\delta_{ij}$. Note that for  every monomial in the AdS bi-adjoint building block, the factors commute because the KZ generators satisfy the braid relations. These bi-adjoint building blocks have the potential to reveal the relation between the AdS string amplitudes and field amplitudes in the language of worldsheet bootstrap.

\section{Conclusion}

We have studied the inverse KLT kernel of AdS string integrals through regularized intersections of associahedral
cycles with noncommutative KZ loadings, extending the scalar
approach of \cite{Mizera:2017cqs} to MMPL insertions.
The holomorphic and dual antiholomorphic factors define KZ-type
local systems, and the twisted period relation expresses
single-valued sphere integrals as bilinears of disk periods
with the inverse intersection matrix as kernel.

The loading data organize the local boundary contributions to
the inverse kernel: channel residues determine normal monodromies,
while fusion trees specify asymptotic frames related by Drinfeld
associators. Endpoint transports track the dependence on chamber
loadings, and the common-face structure determines which channels
can occur together. The five-point example illustrates how the
tangential endpoint contributions and the interior contribution
of a shared facet combine after the appropriate fusion-tree
transports are included. Under the field-theory limit $\alpha'\to0$, the leading contributions come from
cubic trees. The resulting operator-valued tree sums suggest
bi-adjoint building blocks whose propagators retain the
noncommutative data of the polylogarithmic generating series.
The interpretation of their coefficients in a particular
AdS theory remains a separate physical question.

An immediate direction is to apply our framework
to explicit AdS curvature corrections of flat-space string amplitudes.
At the level of building blocks, the twisted period relation
expresses sphere integrals through a bilinear combination of
disk periods, with the kernel determined by the inverse
intersection matrix. Expanding this relation at fixed
polylogarithmic weight yields identities that can be used
to organize and check the integral basis entering a
worldsheet bootstrap ansatz. To turn these identities into
constraints on an amplitude, one must also specify the
external-state dependence and the appropriate coefficient
extraction like the monodromy relation \cite{Alday:2025cxr}.
An open question is whether physical AdS string amplitudes
inherit an analogous twisted-period or intersection-theoretic
structure. This requires understanding how the structure of
the generating series behaves under the truncation and
coefficient extraction appropriate to a given curvature
correction.

A second direction is to see how the bi-adjoint building blocks connect to
color--kinematics duality and double copy in AdS spacetime \cite{Albayrak:2020fyp,Alday:2021odx,Zhou:2021gnu}. In the flat space,
these relations \cite{Bern:2008qj,Bern:2010ue} have been explored
in theories with fundamental matter
\cite{Johansson:2014zca,Johansson:2015oia},
spontaneously broken gauge and gravity theories
\cite{Chiodaroli:2015rdg},
and amplitudes involving massive scalar and spinning particles
\cite{Johansson:2019dnu,Bautista:2019evw,Plefka:2019wyg};
see \cite{Bern:2019prr} for a review. String-theoretic perspectives on color--kinematics duality
and the double copy have also been developed through worldsheet
structures and their field-theory limits, both at tree level
\cite{HenryTye:2010tcy,Mafra:2011kj}
and at one loop \cite{Ochirov:2013xba,Casali:2020knc}.
The bi-adjoint building blocks are expected to be the AdS curvature correction to the flat-space bi-adjoint scalar amplitudes; hence it will be interesting to see whether it can bridge the amplitude relations in flat and AdS spacetime.

\begin{acknowledgments}
We thank Hiren Kakkad and Alexander Ochirov for enlightening
conversations, Alexander Ochirov for his helpful comments on
the draft, and Oliver Schlotterer for organizing the discussions.
YT is supported by the National Natural Science Foundation of China (NSFC) (Grant No.~124B2094). SZ is supported by the Science and Technology Commission of the Shanghai
Municipality via STCSM grant No. 24ZR1450600.

{\bf Statement on use of AI:} AI was used for assistance with Python and \LaTeX{} code for figure preparation and numerical verification. All outputs were reviewed by the authors, who take full responsibility for the content of this manuscript.

\end{acknowledgments}

\appendix

\section{Pentagon and hexagon identities}
\label{app:pen-hex}
Drinfeld associators obey the following pentagon identity \cite{Drinfeld:1991quasi,barnatan1998}
\be
\Phi(e_{ab},e_{bc}+e_{bd})\Phi(e_{ac}+e_{bc},e_{cd})=\Phi(e_{bc},e_{cd})\Phi(e_{ab}+e_{ac},e_{bd}+e_{cd})\Phi(e_{ab},e_{bc})
\fe
and hexagon identity:
\be
e^{\pm i\pi e_{ab}}\Phi(e_{cd},e_{ab})e^{\pm i\pi e_{cd}}\Phi(e_{ef},e_{cd})e^{\pm i\pi e_{ef}}\Phi(e_{ab},e_{ef})=1
\fe
if $e_{ab}+e_{cd}+e_{ef}=0$. 

Applying the cabling homomorphism to the Drinfeld pentagon
identity gives its cabled version
\cite{Drinfeld:1991quasi,barnatan1998}. For four pairwise disjoint clusters $A,B,C,D$ appearing
consecutively in this order, with their internal fusion
trees fixed, the identity reads
\begin{equation}
\label{pentagon}
\begin{aligned}
&\Phi(e_{A,B},e_{B,C}+e_{B,D})
 \Phi(e_{A,C}+e_{B,C},e_{C,D})
\\
&\qquad =
 \Phi(e_{B,C},e_{C,D})
 \Phi(e_{A,B}+e_{A,C},e_{B,D}+e_{C,D})
 \Phi(e_{A,B},e_{B,C}).
\end{aligned}
\end{equation}
The identity states that the two-edge
and three-edge paths connecting the same initial and final
trees give the same transition matrix.

For the hexagon identity, there are some useful corollaries. Let us set
\be
&e_{ab}\to e_{ab}-\frac{1}{3}e_{abc}\\
&e_{cd}\to e_{bc}-\frac{1}{3}e_{abc}\\
&e_{ef}\to e_{ac}-\frac{1}{3}e_{abc}
\fe
where
\be
e_{abc}=e_{ab}+e_{bc}+e_{ac}
\fe
Then we have
\be
e^{\pm i\pi e_{ab}}\Phi(e_{bc},e_{ab})e^{\pm i\pi e_{bc}}\Phi(e_{ac},e_{bc})e^{\pm i\pi e_{ac}}\Phi(e_{ab},e_{ac})=e^{\pm i\pi e_{abc}}
\fe
since $e_{abc}$ commutes with all variables in the equation. There are some corollaries. If we take the inverse of both sides:
\be
\Phi(e_{ac},e_{ab})e^{\pm i\pi e_{ac}}\Phi(e_{bc},e_{ac})e^{\pm i\pi e_{bc}}\Phi(e_{ab},e_{bc})e^{\pm i\pi e_{ab}}=e^{\pm i\pi e_{abc}}
\fe
Since $e_{ab}$ commutes with $e_{bc}+e_{ac}$, we have
\be
\Phi(e_{bc},e_{ab})e^{\pm i\pi e_{bc}}\Phi(e_{ac},e_{bc})e^{\pm i\pi e_{ac}}\Phi(e_{ab},e_{ac})=e^{\pm i\pi (e_{bc}+e_{ac})}
\fe
and
\be
\Phi(e_{ac},e_{ab})e^{\pm i\pi e_{ac}}\Phi(e_{bc},e_{ac})e^{\pm i\pi e_{bc}}\Phi(e_{ab},e_{bc})=e^{\pm i\pi (e_{bc}+e_{ac})}
\fe

The pentagon identity is easy to show via the fusion tree. We now consider the left-comb tree
\begin{equation}
G_{12345}
\equiv
G_{(((12)3)4)5}
\end{equation}
as the reference fusion frame. We now move this counterclockwise around the pentagon, they may be ordered as
\begin{equation}
(((12)3)4)5\to
((1(23))4)5\to
(1((23)4))5\to
(1(2(34)))5\to
((12)(34))5\to
(((12)3)4)5.
\end{equation}

Starting from the left-comb tree, we choose \(G_{(1(2(34)))5}\) as the target fusion tree. Obviously, there are two inequivalent paths along its edges: one consisting of two elementary reassociations and the other of three. The first path is the counterclockwise order, which gives the associator right multiplication:
\[
G_{(1(2(34)))5}
=
G_{12345}\,
\Phi(e_{23},e_{12})\,
\Phi(e_{24}+e_{34},e_{12}+e_{13})\,
\Phi(e_{34},e_{23}).
\]
The second one is the anticlockwise order and its associator right multiplication is given by
\[
G_{(1(2(34)))5}
=
G_{12345}\,
\Phi(e_{34},e_{13}+e_{23})\,
\Phi(e_{23}+e_{24},e_{12}).
\]
The equality between the two possible sequences of reassociations connecting the same initial and final fusion trees is precisely encoded by the pentagon identity.

The hexagon identity can be regarded as a consequence of the flatness of the KZ connection. The flatness requires that the analytic continuation along different paths should give the same result. Let us consider the analytic continuation starting from \(G_{12345}\) to \(G_{((32)1)4)5}\), consisting of associators and braids.
The first path is 
\[(12)3 \xrightarrow{e^{i\pi e_{12}}} (21)3 \xrightarrow{\Phi(e_{13},e_{12})}
    2(13) \xrightarrow{e^{i\pi e_{13}}}
    2(31) \xrightarrow{\Phi(e_{23},e_{13})}
    (23)1 \xrightarrow{e^{i\pi e_{23}}}
    (32)1 \xrightarrow{\Phi(e_{12},e_{23})}
    3(21)\]
which gives the right multiplication factor
\[
R_1 = e^{i\pi e_{12}} \Phi(e_{13},e_{12}) e^{i\pi e_{13}}\Phi(e_{23},e_{13})e^{i\pi e_{23}} \Phi(e_{12},e_{23}).
\]
The second path is 
\[(12)3 \xrightarrow{\Phi(e_{23},e_{12})} 
1(23) \xrightarrow{e^{i\pi e_{23}}}
    1(32) \xrightarrow{\Phi(e_{13},e_{23})}
    (13)2 \xrightarrow{e^{i\pi e_{13}}}
    (31)2 \xrightarrow{\Phi(e_{12},e_{13})}
    3(12) \xrightarrow{e^{i\pi e_{12}}}
    3(21)\]
which gives the right multiplication factor
\[R_2=\Phi(e_{23},e_{12})e^{i\pi e_{23}}\Phi(e_{13},e_{23})e^{i\pi e_{13}}\Phi(e_{12},e_{13})e^{i\pi e_{12}}.\]

The hexagon identities express the compatibility of associator
transport with braiding. We compare two paths from
\(G_{(((12)3)4)5}\) to \(G_{((3(21))4)5}\), whose underlying
braid words are
\[
\sigma_1\sigma_2\sigma_1
\qquad\text{and}\qquad
\sigma_2\sigma_1\sigma_2,
\]
respectively \cite{etingofLecturesRepresentationTheory1998}.
These words represent the same braid, and the corresponding
paths are homotopic with fixed tangential endpoints.
The flatness of the KZ connection therefore implies that
their right multiplication factors agree:
\begin{equation}
R_1=R_2.
\end{equation}
This equality is the braid (Yang--Baxter) relation expressed
in fusion-tree frames, with associators accounting for
changes of parenthesization. Algebraically, it follows
from the coherence identities of the Drinfeld associator,
in particular the hexagon identities.

\section{Cable braiding from elementary strand exchanges}
\label{app:cable-braiding}

In this appendix, we derive the braid factor for two cables
from elementary strand exchanges and the hexagon identities.
Let $T_A$ and $T_B$ be planar binary fusion trees supported
on two nonempty disjoint sets of labels $A$ and $B$.
We consider the standard positive cable braid
\begin{equation}
(T_A T_B)\longrightarrow(T_B T_A),
\label{eq:app-cable-braid}
\end{equation}
which exchanges the two cables while preserving their
internal trees. We show that its connection factor is
\begin{equation}
R(A,B)
=
\exp\bigl(i\pi e_{A,B}\bigr),
\qquad
e_{A,B}
:=
\sum_{\substack{a\in A\\ b\in B}}e_{ab}.
\label{eq:app-cable-result}
\end{equation}
The exchange of labels is recorded by the target tree in
\eqref{eq:app-cable-braid}. If the braid is instead represented
as an operator on a fixed tensor ordering, the corresponding
block permutation must also be included.

The argument takes place in the completed universal
enveloping algebra of the local Drinfeld--Kohno Lie algebra.
All exponentials and associators are understood as formal
series.

\subsection{Recursive decomposition of the cable braid}

We first derive two recursive formulas without assuming
\eqref{eq:app-cable-result} for composite cables.
In the following discussion,
{$a$, $b$, and $c$ denote individual strand labels.}

\paragraph{Splitting the left cable.}
Suppose the left cable has tree $(ab)$. The exchange
\begin{equation}
(ab)c\longrightarrow c(ab)
\end{equation}
can be performed through the sequence
\begin{equation}
\begin{aligned}
((ab)c)
&\longrightarrow a(bc)
\longrightarrow a(cb)
\\
&\longrightarrow (ac)b
\longrightarrow (ca)b
\longrightarrow c(ab).
\end{aligned}
\label{eq:app-left-braid-path}
\end{equation}
The five connection factors, listed in chronological order,
are
{
\begin{equation}
\Phi(e_{bc},e_{ab}),
\qquad
R(b,c),
\qquad
\Phi(e_{ac},e_{bc}),
\qquad
R(a,c),
\qquad
\Phi(e_{ab},e_{ac}).
\label{eq:app-left-path-factors}
\end{equation}
}
{
Using $R(b,c)=e^{i\pi e_{bc}}$ and
$R(a,c)=e^{i\pi e_{ac}}$, the hexagon identity gives
\begin{align}
R(ab,c)
&=
\Phi(e_{bc},e_{ab})e^{i\pi e_{bc}}
\Phi(e_{ac},e_{bc})e^{i\pi e_{ac}}
\Phi(e_{ab},e_{ac})
\nonumber\\
&=
e^{i\pi(e_{bc}+e_{ac})}
=
e^{i\pi e_{a\cup b,c}}.
\label{eq:app-left-recursion-result}
\end{align}
}

\paragraph{Splitting the right cable.}
Similarly, the exchange
\begin{equation}
a(bc)\longrightarrow (bc)a
\end{equation}
can be performed through
\begin{equation}
\begin{aligned}
a(bc)
&\longrightarrow ((ab)c)
\longrightarrow (ba)c
\\
&\longrightarrow b(ac)
\longrightarrow b(ca)
\longrightarrow (bc)a.
\end{aligned}
\label{eq:app-right-braid-path}
\end{equation}
{
Multiplying the corresponding right factors in chronological
order and applying the hexagon identity yields
\begin{align}
R(a,bc)
&=
\Phi(e_{ab},e_{bc})e^{i\pi e_{ab}}
\Phi(e_{ac},e_{ab})e^{i\pi e_{ac}}
\Phi(e_{bc},e_{ac})
\nonumber\\
&=
e^{i\pi(e_{ab}+e_{ac})}
=
e^{i\pi e_{a,b\cup c}}.
\label{eq:app-right-recursion-result}
\end{align}
}

\subsection{Proof for arbitrary cables}
{
The same path decompositions apply when the individual
strands are replaced by disjoint cables with fixed internal
trees. Indeed, for three such cables $X,Y,Z$, the generators
\begin{equation}
x=e_{X,Y},
\qquad
y=e_{Y,Z},
\qquad
z=e_{X,Z}
\end{equation}
satisfy the infinitesimal braid relations
\begin{equation}
[x,y+z]=[y,z+x]=[z,x+y]=0.
\end{equation}
Consequently, the hexagon identities remain valid after
cabling. In the recursive path decompositions, the
exchanges of smaller cables are retained as $R$-factors
until the induction hypothesis is applied.
}

We now prove \eqref{eq:app-cable-result} by induction.
For $|A|=|B|=1$, both cables consist of a single strand,
and the claim is trivial.

Let us first assume $|B|=1$ but $|A|=N>1$, and suppose
the claim holds for all $|A|<N$.
Then we can split its prescribed tree at the root,
\begin{equation}
T_A=(T_XT_Y),
\qquad A=X\cup Y.
\end{equation}
{
Since $|X|,|Y|<N$, the induction hypothesis applies to
both $(X,B)$ and $(Y,B)$. The left-cable decomposition,
followed by the hexagon identity, therefore gives
\begin{align}
R(A,B)
&=
\Phi(e_{Y,B},e_{X,Y})\,
R(Y,B)\,
\Phi(e_{X,B},e_{Y,B})
\nonumber\\
&\qquad\times
R(X,B)\,
\Phi(e_{X,Y},e_{X,B})
\nonumber\\
&=
e^{i\pi(e_{X,B}+e_{Y,B})}
=
e^{i\pi e_{A,B}}.
\end{align}
}
{
Now let $|A|$ be arbitrary and $|B|=N>1$.
Having established the case $|B|=1$, we assume that the
claim holds for every $A$ and all $|B|<N$, and split
the right tree:
}
\begin{equation}
T_B=(T_YT_Z),
\qquad B=Y\cup Z.
\end{equation}
{
Applying the induction hypothesis to $(A,Y)$ and $(A,Z)$
in the right-cable decomposition, followed by the hexagon
identity, gives
\begin{align}
R(A,B)
&=
\Phi(e_{A,Y},e_{Y,Z})\,
R(A,Y)\,
\Phi(e_{A,Z},e_{A,Y})
\nonumber\\
&\qquad\times
R(A,Z)\,
\Phi(e_{Y,Z},e_{A,Z})
\nonumber\\
&=
e^{i\pi(e_{A,Y}+e_{A,Z})}
=
e^{i\pi e_{A,B}}.
\end{align}
}
This completes the induction.

For an explicit four-strand example, consider
\begin{equation}
(12)(34)\longrightarrow(34)(12).
\end{equation}
Write
\begin{equation}
x=e_{12},
\qquad
y=e_{23}+e_{24},
\qquad
z=e_{13}+e_{14}.
\end{equation}
Here $R(12,34)$ denotes the exchange of the parenthesized
cables $(12)$ and $(34)$.
Splitting the left cable gives
{
\begin{equation}
R(12,34)
=
\Phi(y,x)\,
R(2,34)\,
\Phi(z,y)\,
R(1,34)\,
\Phi(x,z).
\label{eq:app-four-strand-outer}
\end{equation}
}
Each remaining exchange is reduced to individual strands
by splitting the right cable:
\begin{align}
R(1,34)
&=
\Phi(e_{13},e_{34})e^{i\pi e_{13}}
\nonumber\\
&\quad\times
\Phi(e_{14},e_{13})e^{i\pi e_{14}}
\Phi(e_{34},e_{14}),
\label{eq:app-four-strand-first}
\\
R(2,34)
&=
\Phi(e_{23},e_{34})e^{i\pi e_{23}}
\nonumber\\
&\quad\times
\Phi(e_{24},e_{23})e^{i\pi e_{24}}
\Phi(e_{34},e_{24}).
\label{eq:app-four-strand-second}
\end{align}
Substituting these two expressions into
\eqref{eq:app-four-strand-outer} displays the four
elementary strand exchanges, together with all nine
associator factors in this decomposition.

Two applications of the right hexagon give
\begin{equation}
R(1,34)=e^{i\pi z},
\qquad
R(2,34)=e^{i\pi y}.
\end{equation}
A final application of the left hexagon then yields
{
\begin{align}
R(12,34)
&=
\Phi(y,x)e^{i\pi y}
\Phi(z,y)e^{i\pi z}\Phi(x,z)
\nonumber\\
&=
e^{i\pi(y+z)}
\nonumber\\
&=
\exp\bigl(
i\pi(e_{13}+e_{14}+e_{23}+e_{24})
\bigr).
\label{eq:app-four-strand-result}
\end{align}
}
\section{Graphic rules for intersection numbers}
\label{app:graphic-rule}
The dual polygon rules reorganize the regularized self-intersection 
of a loaded associahedron in \eqn{eq:selfintersection} into a sum of diagrams that correspond to Feynman diagrams of bi-adjoint scalar amplitudes \cite{Mizera:2016jhj}. The dual polygon
encodes the channels.
It does not measure crossings of curves in the polygon.

Fix a cyclic ordering \(\alpha\) of \(n\) labels, and a reference fusion tree for the ordering \(\alpha\) (left-comb tree in our convention). The corresponding
associahedron \(K_{n-1}(\alpha)\) has dimension \(d=n-3\). And we introduce an auxiliary convex \(n\)-gon \(P\).
We identify a diagonal of \(P\) with the stable planar split it
represents, and write \(I\) for its cluster representative not
containing the chosen root \(n\). A set \({\cal N}\) of mutually non-crossing
diagonals labels a codimension-$|{\cal N}|$ face $F_{\cal N}$ of \(K_{n-1}(\alpha)\) and the empty set labels the whole associahedron itself. The requirement of non-crossing diagonals of \(P\) corresponds to the compatibility of facets of \(K_{n-1}\)

For each \(\cal N\), choose a fusion tree completion $\tau_N$ containing all its
channels, namely keeping all cluster \(I\in{\cal N}\) as cables. In this adapted frame the residues $E_I$, $I\in {\cal N}$,
commute. Set the regularization factor 
\begin{equation}
R_I:=\bigl(e^{2\pi iE_I}-\mathbf1\bigr)^{-1}.
\label{eq:app-self-R}
\end{equation}
We work in the coefficient algebra in which the required monodromy
differences are invertible. Using the chamber orientation convention,
the regularization gives
\begin{equation}
{
H_{K_{n-1}}^{\mathrm{self}}(\alpha)
=(-1)^d\sum_{{\cal N}\ \mathrm{non\text{-}crossing}}
\mathcal T_{\alpha \to\tau_{\cal N}}
\left[\prod_{I\in {\cal N}}R_I\right]
\mathcal T_{\tau_{\cal N}\to \alpha}.
}
\end{equation}
For \({\cal N}=\varnothing\), we define \(R_{\varnothing} =\mathbf1\) before multiplication
by $(-1)^d$. In particular, $H_{K_2}^{\mathrm{self}}=\mathbf1$.
The sign agrees with $(-1)^{r-1}$ in the flat-space face-sum formula.
We include it in $H^{\mathrm{self}}$ throughout this appendix.

Introduce $c_I:=\cot(\pi E_I)$. The identity
\begin{equation}
R_I=-\frac12\bigl(\mathbf1+i c_I\bigr)
\label{eq:app-self-cot-identity}
\end{equation}
and commutativity of the compatible residues imply
\begin{equation}
\prod_{I\in {\cal N}}R_I
=\left(-\frac12\right)^{|{\cal N}|}
\sum_{S\subseteq {\cal N}}i^{|S|}\prod_{J\in S}c_J.
\label{eq:app-self-expand}
\end{equation}
The subset $S$ records the channels whose cotangent factors are
retained. The remaining diagonals in ${\cal N}\setminus S$ contribute
only to a numerical coefficient.

To collect all terms with the same \(S\), their transport operators must first
be compared. Choose a fusion tree \(\tau_S\) adapted to \(S\).
For ${\cal N}\supseteq S$, the trees $\tau_S$ and $\tau_{\cal N}$ can be connected
by reassociations preserving every cable in \(S\). The resulting
transport commutes with each retained residue \(E_I\), \(I\in S\).
Composition of transports therefore gives
\begin{equation}
\begin{aligned}
&\mathcal T_{\alpha\to\tau_{\cal N}}
\left[\prod_{I\in S}c_I\right]
\mathcal T_{\tau_{\cal N}\to \alpha}
\\
&\qquad=
\mathcal T_{\alpha\to\tau_S}
\left[\prod_{I\in S}c_I\right]
\mathcal T_{\tau_S\to \alpha}
=:\widehat c_S^{(\alpha)}.
\end{aligned}
\label{eq:app-retained-transport}
\end{equation}
Thus $\widehat c_S^{(\alpha)}$ is independent of the fusion tree
within the cable-preserving class of transports, and
$\widehat c_{\varnothing}^{(\alpha
)}=\mathbf1$. Equation~\eqref{eq:selfintersection}
becomes
\begin{equation}
H_{K_{n-1}}^{\mathrm{self}}(\alpha)
=(-1)^d\sum_{S\ \mathrm{non\text{-}crossing}}
i^{|S|}A_{n-1}(S)\,\widehat c_S^{(\alpha)},
\qquad
A_{n-1}(S):=\sum_{\substack{{\cal N}\supseteq S
\\{\cal N}\ \mathrm{non\text{-}crossing}}}
\left(-\frac12\right)^{|\cal N|}.
\label{eq:app-self-collected}
\end{equation}

Let $\operatorname{Cells}(S)$ denote the subpolygons obtained by
cutting the polygon \(P\) along \(S\). Every refinement ${\cal N}\supseteq S$ has a unique
decomposition
\[
{\cal N}=S\cup\bigcup_{Q\in\operatorname{Cells}(S)}{\cal N}_Q,
\]
where \({\cal N}_Q\) is a possibly empty non-crossing dissection inside \(Q\).
Consequently,
\begin{equation}
A_{n-1}(S)=\left(-\frac12\right)^{|S|}
\prod_{Q\in\operatorname{Cells}(S)}q_{|Q|},
\qquad
q_\ell:=\sum_{{\cal N}_Q\ \mathrm{in\ an}\ \ell\mathrm{\text{-}gon}}
\left(-\frac12\right)^{|{\cal N}_Q|}.
\label{eq:app-refinement-factorization}
\end{equation}
Here $|Q|$ is the number of sides of $Q$, and the sum defining
$q_\ell$ includes the empty dissection.

The coefficients are known as the \textit{Catalan numbers} \cite{Mizera:2017cqs}:
\begin{equation}
q_\ell=
\begin{cases}
\displaystyle
\left(\frac{i}{2}\right)^{\ell-3}
\operatorname{Cat}_{(\ell-3)/2},&\ell\ \text{odd},\\[2mm]
0,&\ell\ \text{even}.
\end{cases}
\label{eq:app-q-catalan}
\end{equation}
In particular, $q_3=1$, $q_4=0$, $q_5=-1/4$, and $q_7=1/8$.
Even-sided cells disappear because their refinement contributions
cancel, rather than because such cells are forbidden in the original
face sum.

Let $\mathscr O(P)$ be the dissections \(S\) for which every cell has
an odd number of sides. There are \(|S|+1\) cells and
\[
\sum_{Q\in\operatorname{Cells}(S)}(|Q|-3)=d-|S|.
\]
\begin{figure}[t]
\centering
\includegraphics[width=0.8\textwidth]{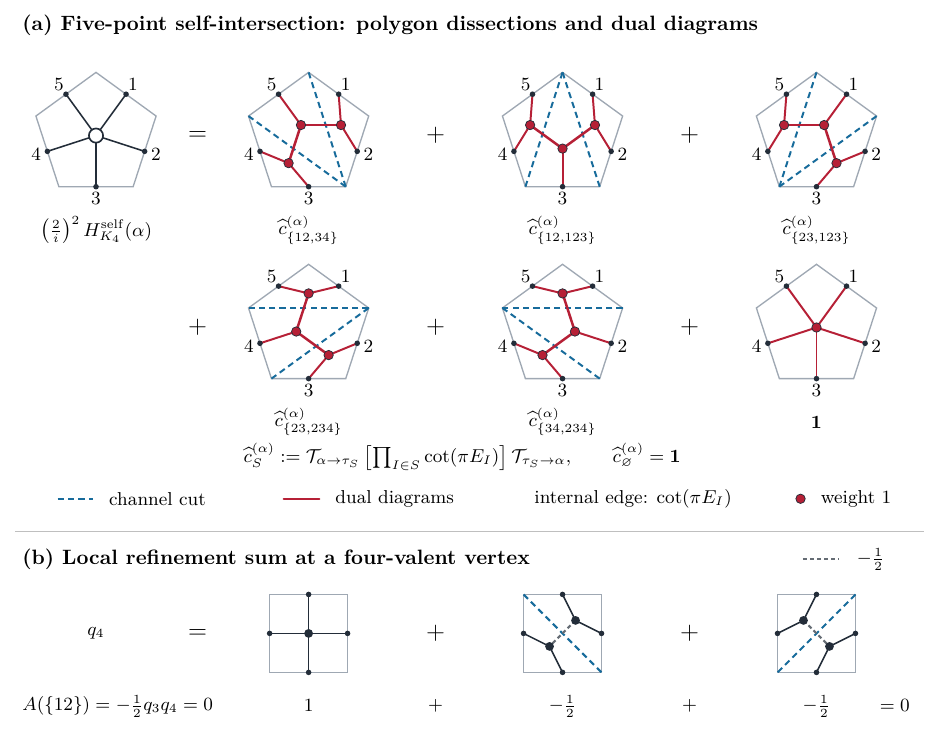}
\vspace{-5pt}
\caption{Five-point example for dual diagrams}
\label{fig:dualdiagram}
\end{figure}
For $S\in\mathscr O(P)$, this is even. Substituting
\eqref{eq:app-q-catalan} into \eqref{eq:app-refinement-factorization}
therefore gives
\begin{equation}
\begin{aligned}
(-1)^d i^{|S|}A_{n-1}(S)
&=(-1)^{d+|S|}\left(\frac{i}{2}\right)^d
\prod_{Q\in\operatorname{Cells}(S)}
\operatorname{Cat}_{(|Q|-3)/2}\\
&=\left(\frac{i}{2}\right)^d
\prod_{Q\in\operatorname{Cells}(S)}
\operatorname{Cat}_{(|Q|-3)/2}.
\end{aligned}
\end{equation}
We obtain the single-polygon rule
\begin{equation}
{
H_{K_{n-1}}^{\mathrm{self}}(\alpha)
=\left(\frac{i}{2}\right)^{n-3}
\sum_{S\in\mathscr O(P)}
\left[\prod_{Q\in\operatorname{Cells}(S)}
\operatorname{Cat}_{(|Q|-3)/2}\right]
\widehat c_S^{(\alpha)}.
}
\label{eq:app-polygon-self-cotangent}
\end{equation}
See the five-point example in \fig{fig:dualdiagram}.


\bibliographystyle{JHEP}
\bibliography{references}
\end{document}